\documentclass[a4paper,11pt]{article}
\usepackage[english]{babel}
\usepackage[utf8]{inputenc}
\usepackage{braket}
\usepackage{physics}
\usepackage{color}
\usepackage{amssymb}
\usepackage{amsmath}
\usepackage{graphicx}
\usepackage{epstopdf}
\usepackage{subfig}
\usepackage{comment}
\usepackage{mathtools}
\usepackage{ragged2e}
\usepackage{hyperref}
\hypersetup{
    colorlinks=true,
    linkcolor=blue,
    filecolor=cyan, 
    urlcolor=blue,
    citecolor=red,
} 
\usepackage[titletoc,toc,title]{appendix}
\numberwithin{equation}{section}
\usepackage{cleveref}
\usepackage[margin=2.5cm]{geometry}
\usepackage{cite}
\usepackage{epsfig}
\usepackage{float}
\usepackage{cancel}
\usepackage{amsfonts}
\usepackage{enumitem}
\usepackage[font={footnotesize,it}]{caption}
\usepackage{authblk}

\begin{document}

\title{Entanglement negativity, reflected entropy, and anomalous gravitation}

\author[1]{Debarshi Basu\thanks{\noindent E-mail:~ \href{mailto:debarshi@iitk.ac.in} {\tt debarshi@iitk.ac.in}}}

\author[1]{Himanshu Parihar\thanks{\noindent E-mail:~ \href{mailto:himansp@iitk.ac.in} {\tt himansp@iitk.ac.in}}}

\author[1]{Vinayak Raj\thanks{\noindent E-mail:~ \href{mailto:vraj@iitk.ac.in} {\tt vraj@iitk.ac.in}}}

\author[1]{Gautam Sengupta\thanks{\noindent E-mail:~ \href{mailto:sengupta@iitk.ac.in} {\tt sengupta@iitk.ac.in}}}

\affil[1]{
Department of Physics\\

Indian Institute of Technology\\ 

Kanpur 208 016, India
}

\date{}

\maketitle

\thispagestyle{empty}

\begin{abstract}

\noindent

We investigate mixed state entanglement measures of entanglement negativity and reflected entropy for bipartite states in two dimensional conformal field theories with an anomaly through appropriate replica techniques. Furthermore we propose holographic constructions for these measures from the corresponding bulk dual geometries involving topologically massive gravity in AdS$_3$ and find exact agreement with the field theory results. In this connection we extend an earlier holographic proposal for the entanglement negativity to the bulk action with a gravitational Chern-Simons term and compute its contribution to the entanglement wedge cross section dual to the reflected entropy.

\justify

\end{abstract}

\clearpage

\tableofcontents

\clearpage

\section{Introduction}
\label{sec:intro}

In the recent past the issue of quantum entanglement in extended many body systems has emerged as an exciting area for the investigation of phenomena in diverse fields from condensed matter physics to quantum gravity and black holes. In this context the characterization of entanglement in quantum field theories through the holographic AdS-CFT correspondence \cite{Susskind:1994vu, Maldacena:1997re} has attracted intense research attention over the last decade. The entanglement entropy has emerged as a reliable measure for the characterization of entanglement of bipartite pure states in these studies. A replica technique to obtain the entanglement entropy for various bipartite states in $(1+1)$-dimensional conformal field theories (CFT$_2$) was established in \cite{Calabrese:2004eu,Calabrese:2009ez,Calabrese:2009qy}. Furthermore an elegant holographic characterization of the entanglement entropy for such bipartite states in a class of CFTs was proposed in \cite{Ryu:2006bv, Hubeny:2007xt}. From these proposals the holographic entanglement entropy of a subsystem in the CFT could be expressed in terms of the area of an extremal codimension two hypersurface homologous to the subsystem. Subsequently these holographic proposals were proved in a series of works in \cite{Fursaev:2006ih, Headrick:2010zt, Casini:2011kv, Faulkner:2013yia, Lewkowycz:2013nqa, Dong:2016hjy}. 

However it is well known in quantum information theory that the entanglement entropy is not a reliable measure for the characterization of mixed state entanglement as it receives irrelevant contributions from both classical and quantum correlations. Several alternative measures to characterize mixed state entanglement has been proposed in quantum information theory most of which involve optimization over LOCC protocols and are hence difficult to compute. In this context, Vidal and Warner \cite{Vidal:2002zz} introduced a computable measure for such bipartite mixed state entanglement based on the positive partial transpose (PPT) criteria \cite{Peres:1996dw,Horodecki:1996nc} termed as entanglement negativity which was given by the trace norm of the partially transposed reduced density matrix.\footnote{Note that the entanglement negativity serves as a non-convex entanglement monotone as described in \cite{Plenio:2005cwa}.} Remarkably a suitable replica technique to compute the entanglement negativity of bipartite states in CFT$_2$ was developed in \cite{Calabrese:2012ew, Calabrese:2012nk, Calabrese:2014yza}. Furthermore in a related development another mixed state correlation measure termed reflected entropy was introduced and computed for bipartite states in CFT$_2$ through another replica technique described in \cite{Dutta:2019gen}. In a recent communication \cite{Akers:2021pvd} this was further explored in the context of random tensor networks to include novel non-perturbative effects in the R\'enyi reflected entropy spectrum.

In relation to the above developments a holographic description of such mixed state entanglement measures naturally emerged as a significant issue. This question was first addressed in \cite{Rangamani:2014ywa} where the holographic entanglement negativity for a pure vacuum state of dual CFT$_d$s was obtained. However a general holographic prescription for mixed states in CFT$_d$s remained an open issue. Subsequently, in a series of communications an elegant holographic characterization of entanglement negativity for various bipartite states in CFTs were proposed in \cite{Chaturvedi:2016rft, Chaturvedi:2016rcn, Jain:2017xsu, Chaturvedi:2017znc, Jain:2017uhe, Jain:2018bai, Malvimat:2018cfe, Malvimat:2018ood, Malvimat:2018txq, Mondal:2021kzj, KumarBasak:2020viv, Afrasiar:2021hld, Jain:2017aqk}. These proposals involved specific algebraic sums of bulk codimension two (H)RT surfaces homologous to appropriate combinations of subsystems in the dual CFT$_d$s\footnote{For applications of these holographic proposals to the black hole information loss problem, see for example \cite{KumarBasak:2021rrx}, where analogues of the Page curve for the entanglement negativity were obtained. See also \cite{Basu:2021awn, Basu:2021axf} for extensions of the above proposals to asymptotically flat spacetimes, which reproduced the field theoretic results in \cite{Malvimat:2018izs}.}. Furthermore for the AdS$_{3}$/CFT$_2$ scenario a semi-classical large central charge analysis utilizing the monodromy techniques \cite{Hartman:2013mia, Fitzpatrick:2014vua, Kulaxizi:2014nma, Malvimat:2017yaj} was established as a strong substantiation for these holographic proposals. Very recently a proof for these holographic entanglement negativity conjectures were given in \cite{KumarBasak:2020ams} based on the analysis of replica symmetry breaking saddles for the bulk gravitational path integral described in \cite{Dong:2021clv}. In this connection it should also be noted that following the gravitational path integral techniques developed in \cite{Lewkowycz:2013nqa}, a holographic duality between the reflected entropy and the minimal EWCS was established in \cite{Dutta:2019gen}. Note that the minimal cross section of the entanglement wedge\footnote{For recent developments regarding the computation of the EWCS in bulk spacetimes dual to quenched systems as well as hyperscaling violating theories, see \cite{BabaeiVelni:2019pkw,BabaeiVelni:2020wfl,Sahraei:2021wqn}.} (EWCS) has been proposed as putative dual of several quantum information measures, for example the entanglement of purification \cite{Takayanagi:2017knl, Nguyen:2017yqw}, the reflected entropy \cite{Dutta:2019gen, Akers:2021pvd} and the balanced partial entanglement \cite{Wen:2021qgx}. We should also mention here that an alternative holographic proposal for the entanglement negativity was advanced in \cite{Kudler-Flam:2018qjo, Kusuki:2019zsp} which involved the minimal area of a backreacting cosmic brane ending on the bulk entanglement wedge dual to the density matrix of the mixed state under consideration\footnote{For a covariant generalization of this alternative proposal, see \cite{KumarBasak:2021lwm}.}. This proposal was further refined in \cite {KumarBasak:2020eia} to address an outstanding issue. Note however that in the light of a recent communication \cite{Hayden:2021gno} this alternative proposal leads to a sum of the entanglement negativity and a quantity termed as the Markov gap which may be geometrically quantified in terms of the number of non trivial boundaries of the bulk EWCS.

On a separate note, in \cite{Castro:2014tta} the authors have studied the holographic characterization of entanglement entropy in $(1+1)$-dimensional conformal field theories with a gravitational anomaly (CFT$_2^a$) dual to topologically massive gravity (TMG) in asymptotically AdS$_3$ spacetime. This gravitational anomaly in such dual field theories essentially arises due to the non-conservation of the stress-energy tensor leading to unequal central charges for the left and the right moving sectors of the CFT$^a_2$. The action for the TMG in the bulk asymptotically AdS$_3$ (TMG-AdS$_3$) spacetimes involves a gravitational Chern-Simons term which modifies the shape of the worldlines of massive spinning particles propagating in the bulk geometry to that of a ribbon involving an auxiliary normal frame at each point. The Chern-Simons contribution to the entanglement entropy is then given by the boost required to propagate this auxiliary normal frame along the worldline.

As mentioned earlier, the entanglement entropy fails to correctly describe mixed state entanglement which requires the introduction of alternative entanglement or correlation measures. In this context the issue of computing such alternative measures characterizing mixed state entanglement in dual CFT$^a_2$s through appropriate replica techniques and their holographic description in the framework of the TMG-AdS$_3$/CFT$_2^a$ correspondence assumes a critical significance. In this article we address this important issue and construct suitable replica techniques to compute the entanglement negativity and the reflected entropy for various bipartite pure and mixed state configurations in dual CFT$^a_2$s. Subsequent to the field theoretic computations we turn to the holographic characterization of these mixed state entanglement measures in the framework of the TMG-AdS$_3$/CFT$_2^a$ correspondence. In particular, the holographic construction for computing the entanglement negativity for the bipartite mixed states involves a specific linear sum of the on-shell actions for massive spinning particles moving on extremal worldlines homologous to certain combinations of the intervals characterizing the mixed states.
Furthermore, we will study the effects of the gravitational anomaly in the bulk construction of the entanglement wedge dual to the density matrix of a bipartite mixed state and provide a novel prescription to compute the Chern-Simons contribution to the minimal EWCS. It is interesting to note that for a single interval at a finite temperature, as in the dual field theory, the appropriate construction of the bulk EWCS involves two large but finite auxiliary intervals sandwiching the single interval in question. Remarkably we obtain exact matches between the field theory replica technique results in the large central charge limit and the bulk holographic computation for both the measures. Interestingly we are also able to obtain the anomalous contributions from the field theory side which are dual to the contributions arising from the bulk Chern-Simons part of the action for the TMG-AdS$_3$.

The rest of the article is organized as follows. In \cref{sec:review}, we review the structure of  CFT$_2^a$ with a gravitational anomaly and a replica technique for computing the entanglement entropy in these field theories as described in \cite{Castro:2014tta}. In \cref{sec:mixed_field}, we apply the replica techniques described in \cite{Calabrese:2012ew, Calabrese:2012nk, Calabrese:2014yza, Dutta:2019gen} to compute the entanglement negativity and the reflected entropy for various bipartite pure and mixed states in such CFT$_2^a$.  Subsequently in \cref{sec:EN_geod} we provide a 
brief review of the TMG-AdS$_3$/CFT$_2^a$ correspondence and propose a holographic construction for the entanglement negativity. Following this in \cref{sec:EWCS} we describe the construction for the bulk entanglement wedge 
cross section for bipartite states in the dual CFT$_2^a$ and compare this with the reflected entropy computed in \cref{sec:mixed_field}. Finally in \cref{sec:summary}, we provide a summary of our results and comment on certain open issues. Furthermore in  \cref{sec:appendix_A}, we provide a derivation for our holographic construction for the entanglement negativity for the mixed state configuration of two adjacent intervals in the context of TMG-AdS$_3$/CFT$_2^a$ from a bulk gravitational path integral.

\section{CFTs with gravitational anomaly}
\label{sec:review}
We begin by briefly reviewing gravitational anomaly in $(1+1)$-dimensional conformal field theories (CFT$_2$) \cite{Alvarez-Gaume:1983ihn, Alvarez-Gaume:1984zlq} which arises from unequal central charges for the left and right moving sectors. The anomaly may be described through two distinct approaches. In the first the stress tensor is symmetric but not conserved and for the second we have a conserved stress tensor which is not symmetric. For the first case the anomalous divergence of the stress tensor may be expressed as \cite{Kraus:2005zm}
\begin{equation}
\nabla_\mu T^{\mu \nu}=\frac{c_L-c_R}{96 \pi} g^{\mu \nu} \epsilon^{\alpha \beta}
\partial_\alpha \partial_\rho \Gamma^\rho_{\nu \beta}.
\end{equation}
We observe from the above expression that the anomaly vanishes when the theory has equal left and right moving central charges. In the second case the stress tensor is conserved but not symmetric and the anomaly manifests itself through a broken Lorentz symmetry and in consequence the theory is rendered frame dependent. It is possible to shift between the two perspectives through the addition of a local counter term to the CFT generating functional \cite{Alvarez-Gaume:1984zlq,Bardeen:1984pm}. We will use the first approach where the stress tensor is not conserved in the following sections.

\subsection{Entanglement entropy in CFT$_2$ with gravitational anomalies}\label{sec:EE_field}
In this subsection we review the computation of the entanglement entropy for the zero and finite temperature bipartite pure and mixed state configurations of a single interval in a CFT$_2$ with a gravitational anomaly as described in \cite{Castro:2014tta}. Note that the finite temperature mixed state configuration leads to a description in the grand canonical ensemble with a chemical potential conjugate to the conserved spin angular momentum arising from the unequal central charges which is termed as the {\it angular potential }.

\subsubsection{Zero temperature}
\label{sec:EE_field_zeroT}
\begin{figure}[H]
	\centering
	\includegraphics[scale=.6]{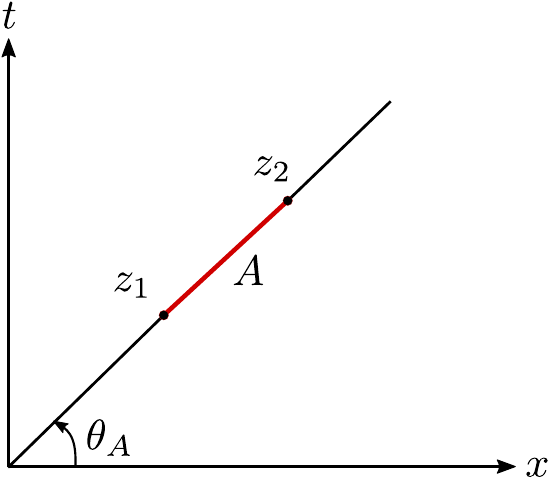}
	\caption{Schematics of a single boosted interval on a complex plane.}
	\label{fig1}
\end{figure}

The computation of the entanglement entropy in a CFT$_2$ with a gravitational anomaly follows exactly in the same fashion as for the usual scenario and involves an appropriate replica technique as described in \cite{Calabrese:2004eu, Calabrese:2009qy}. For the zero temperature configuration of a single interval it is required to consider a boosted interval described by $A\equiv [z_1,z_2] = [(x_1,t_1), (x_2,t_2)]$ and its complement $B=A^c$ denoting the rest of the system as shown in Fig. \ref{fig1}.

The entanglement entropy may then be expressed in terms of the two point twist field correlators as follows \cite{Castro:2014tta}
\begin{equation}
\begin{aligned}
\textrm{Tr}\rho_A^n =\left<\Phi_{n_e}(z_1)\Phi_{-n_e}(z_2)\right>=c_n z_{12}^{-2 h_L} \bar{z}_{12}^{-2 h_R} ,\label{Tr_rho}
\end{aligned}
\end{equation}
where $\Phi_{n_e}(z_1)$ and $\Phi_{-n_e}(z_2)$ are the twist and the anti twist fields located at the end points of the interval $A$, with conformal dimensions given as $h_L=\frac{c_L}{24}\left( n-\frac{1}{n}\right)$ and $h_R=\frac{c_R}{24}\left( n- \frac{1}{n} \right)$ which may be determined from the conformal Ward identities. Note that due to the condition $c_L \neq c_R$, the twist fields possess non-zero spin $s_n$ which is proportional to the anomaly coefficient $(c_L-c_R)$ as follows
\begin{align}
\Delta_n=\frac{c_L+c_R}{24}\left(n-\frac{1}{n}\right)\,\,,\,\, s_n=\frac{c_L-c_R}{24}\left(n-\frac{1}{n}\right)\label{mass_spin}\, ,
\end{align}
where $\Delta_n$ is the scaling dimension of the twist fields. The entanglement entropy of a single interval $A$  may now be obtained using the above expression as \cite{Castro:2014tta}
\begin{align}
S_A=-\lim_{n\to1}\partial_n\textrm{Tr}\rho_A^n=\frac{c_L}{6}\log \left(\frac{z_A}{\epsilon}\right)+\frac{c_R}{6}\log \left(\frac{\bar{z}_A}{\epsilon}\right),\label{S_EE}
\end{align}
where $z_A=z_1-z_2$ and $\epsilon$ is a UV cut-off. On using $z_A=R_A e^{i\theta_A}$ and analytically continuing to a Lorentzian signature via $z=x-t$, $\bar{z}=x+t$ we have $\theta_A=i\kappa_A$ where $\kappa$ is the boost parameter. The entanglement entropy in this case receives an additional contribution due the anomalous Lorentz boost as follows \cite{Castro:2014tta}
\begin{equation}\label{entropy-single-zero}
S_A=\frac{c_L+c_R}{6}\log\left( \frac{R_A}{\epsilon}\right)-\frac{c_L-c_R}{6}\kappa_A.
\end{equation}
In the above expression the length $R_A$ and the boost $\kappa_A$ for the boosted interval $A$ are related to the $(t, x)$-coordinates as follows
\begin{align} \label{R&K_coordinates}
R_A = \sqrt{x_{12}^2 - t_{12}^2} \, , \qquad \kappa_A = \tanh^{-1}\left( \frac{t_{12}}{x_{12}} \right).
\end{align}
The second term in \cref{entropy-single-zero} arises from the contribution from the gravitational anomaly. This reduces to the usual entanglement entropy of a single interval \cite{Calabrese:2004eu,Calabrese:2009qy} when the anomaly is absent ($c_L=c_R$).

\subsubsection{Finite temperature and angular potential}
\label{sec:EE_field_T}
For this mixed state configuration we consider a spatial interval $A\equiv [0,R_A]$ in the CFT$_2$ at a finite temperature $T=\beta^{-1}$ and with a non zero chemical potential $\Omega$ for the spin angular momentum arising from the gravitational anomaly. In this instance the CFT$_2$ with a gravitational anomaly must be described on a twisted cylinder due to the spin angular momentum. The Euclidean partition function for this CFT$_2$ following a Wick rotation is given by
\begin{equation}
\mathbb{Z}=\text{Tr}\left(e^{-\beta\,H-\beta\Omega_EJ}\right),
\end{equation}
where $H$ is the Hamiltonian, $\beta$ is the inverse temperature, $J$ is the spin angular momentum and the angular potential $\Omega_E$ is defined to be real via the standard analytic continuation $\Omega = i\Omega_E$ and we have 
\begin{equation}
H=E_R+E_L-\frac{c_L+c_R}{24}~, \quad J=E_R-E_L+\frac{c_L-c_R}{24}.
\end{equation}
The left and right moving inverse temperatures $(\beta_L,\beta_R)$ are defined in terms of $(\beta,\Omega_E)$ as
\begin{equation}\label{tempsss}
\beta_L = \beta(1+i\Omega_E)~, \quad \beta_R = \beta(1-i\Omega_E)~.
\end{equation}
Note that for the ground state on the cylinder $E_L=E_R=0$, the theory acquires a non-zero ``Casimir momentum'' $J_0$ in addition to the usual ground state energy (Casimir energy) $E_0$ as
\begin{equation}\label{gdst}
E_0 = -{c_L+c_R\over 24}~, \quad J_0={c_L-c_R\over 24}~.
\end{equation}
The CFT$_2$ on the twisted cylinder may be obtained from a Euclidean CFT$_2$ on the complex plane through the conformal transformations
\begin{equation}\label{twisted-cylinder-transf}
w=e^{2\pi z/\beta_L}, \quad \bar{w}=e^{2\pi \bar{z}/\beta_R},
\end{equation}
where $z$ and $w$ denotes the coordinate on the complex plane and the twisted cylinder respectively. Now using the transformation of the two point twist correlator under the above conformal mapping, the entanglement entropy for the mixed state of a single interval under consideration is given as \cite{Castro:2014tta}
\begin{equation}\label{entropy-single-finite}
S_A=\frac{c_L+c_R}{12}\log\left[\frac{\beta_L\beta_R}{\pi^2 \epsilon^2} \sinh \left(\frac{\pi R_A}{\beta_L}\right)\sinh \left(\frac{\pi R_A}{\beta_R}\right)\right]+\frac{c_L-c_R}{12}\log\left[\frac{\beta_L \sinh \left(\frac{\pi R_A}{\beta_L}\right)}{\beta_R \sinh \left(\frac{\pi R_A}{\beta_R}\right)}\right].
\end{equation}
The second term in the above expression quantifies the contribution due to the gravitational anomaly for $c_L\neq c_R$. In the absence of the anomaly we have $\beta_{L}=\beta_{R}$, and this reduces to the well-known expression of the entanglement entropy corresponding to the mixed state described by a single interval at a finite temperature \cite{Calabrese:2004eu,Calabrese:2009qy}. 
\section{Mixed state entanglement measures in CFT$^a_2$} \label{sec:mixed_field}

\subsection{Entanglement negativity in CFT$^a_2$}
\label{sec:negativity_field}
We begin by briefly discussing the definition of entanglement negativity in quantum information theory \cite{Vidal:2002zz}. Consider a tripartite system in a pure state consisting of the subsystems $A$, $B$ and $C$, where $AB=A \cup B$ and $C=AB^c$ being the rest of the system. For the Hilbert space $\mathcal{H}=\mathcal{H}_A \otimes \mathcal{H}_B$, the reduced density matrix for the subsystem $AB$ is defined as $\rho_{AB}=\mathrm{Tr}_{C} \,\rho $ and the partial transpose of the reduced density matrix $\rho_{A}^{T_B}$ with respect to the subsystem $B$ is given by
\begin{equation}
\mel{e^{(A)}_i e^{(B)}_j} {\rho_{AB}^{T_B}} {e^{(A)}_k e^{(B)}_l} = \mel{e^{(A)}_i e^{(B)}_l} {\rho_{AB}} {e^{(A)}_k e^{(B)}_j},
\end{equation}
where $\ket{e^{(A)}_i}$ and $\ket{e^{(B)}_j}$ are the bases for the Hilbert spaces $\mathcal{H}_A$ and $\mathcal{H}_B$. The entanglement negativity for the bipartite mixed state configuration $AB$ may then be defined as the logarithm of the trace norm of the partially transposed reduced density matrix as
\begin{equation}
\mathcal{E}(A:B) = \log \mathrm{Tr}|\rho_A^{T_B}|,
\end{equation}
where the trace norm $\mathrm{Tr}|\rho_A^{T_B}|$ is given by the sum of absolute eigenvalues of $\rho_{AB}^{T_B}$. The entanglement negativity for the bipartite states in CFT$_2$ with gravitational anomaly may be obtained through a replica technique similar to \cite{Calabrese:2012ew,Calabrese:2012nk,Calabrese:2014yza}. This involves the construction of the quantity $\mathrm{Tr} \big( \, \rho_{AB}^{T_{B}}\big)^{n}$ for even sequences of $n=n_e$ and its analytic continuation to $n_e\to 1$ which leads to the following expression 
\begin{equation}\label{replica-technique}
\mathcal{E}(A:B) = \lim_{n_e \rightarrow 1}  \log \Big[ \mathrm{Tr} \big(\,\rho_{AB}^{T_{B}}\big)^{n_e} \Big].
\end{equation}
The $\mathrm{Tr} \big(\,\rho_{AB}^{T_{B}}\big)^{n_e}$ may be expressed as a twist field correlator in the replicated CFT$^a_2$ appropriate to the mixed state configuration. 

\begin{figure}[H]
	\centering
	\includegraphics[scale=.60 ]{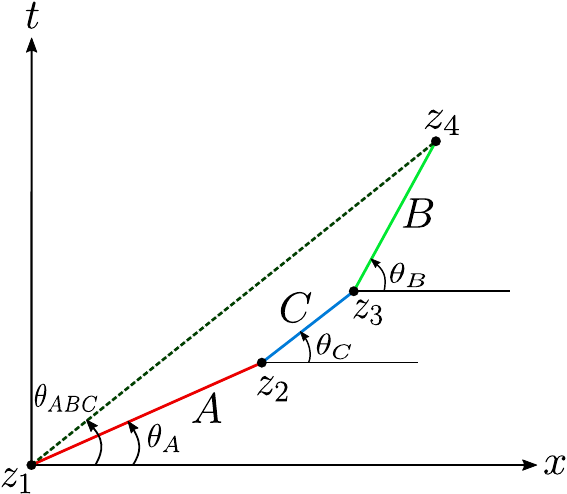}
	\caption{Schematics of two boosted disjoint intervals $A=[z_1,z_2]$ and $B=[z_3,z_4]$ on a complex plane. }
	\label{fig3}
\end{figure}

As an example for the above discussion a mixed state configuration described by two boosted disjoint intervals $A\equiv [z_1,z_2]$ and $B\equiv [z_3,z_4]$ separated by an interval $C\equiv [z_2,z_3]$ as depicted in Fig. \ref{fig3} it is possible to express the quantity $\textrm{Tr}(\rho_{AB}^{T_B})^{n_e}$ as a four point twist field correlator as follows
\begin{align}\label{four-point-twist}
\textrm{Tr}(\rho_{AB}^{T_B})^{n_e}=\left<\Phi_{n_e}(z_1)\Phi_{-n_e}(z_2)\Phi_{-n_e}(z_3)\Phi_{n_e}(z_4)\right>_{\mathbb{C} }.
\end{align}

We now proceed to compute the entanglement negativity for various bipartite states in a CFT$_2$s with gravitational anomaly ($c_L \neq c_R$) in the subsequent subsections.

\subsubsection{Single interval}
\label{sec:sing_Neg_field}
In this subsection we compute the entanglement negativity for the bipartite pure and mixed state configuration of a single interval in a CFT$_2$ in the presence of the gravitational anomaly.

\subsubsection*{Zero temperature}\label{sec:sing_Neg_zeroT}

We obtain the pure state configuration of a single interval from the two disjoint intervals through a bipartite limit described by $z_3\to z_2$, $z_4\to z_1$ where the interval $B\equiv A^c$ now describes the rest of the system. In this limit the four point twist correlator in eq. (\ref{four-point-twist}) reduces to the following two point twist correlator 
\begin{equation}\label{two-point-square}
\textrm{Tr}(\rho_A^{T_B})^{n_e}=\left<\Phi_{n_e}^2(z_1)\Phi_{-n_e}^2(z_2)\right>.
\end{equation}
The $n_e$ sheeted Riemann surface decouples into two independent $n_e/2$ sheeted Riemann surfaces in a similar manner to \cite{Calabrese:2012nk} and hence the two point correlator in eq. (\ref{two-point-square}) reduces to the following expression
\begin{equation}\label{two-point-split}
\textrm{Tr}(\rho_A^{T_B})^{n_e}=\left<\Phi_{n_e}^2(z_1)\Phi_{-n_e}^2(z_2)\right>=\left(\left<\Phi_{n_e/2}(z_1)\Phi_{-n_e/2}(z_2)\right>\right)^2.
\end{equation}
From the above equation, we find the scaling dimension of the twist fields $\Phi_{n_e}^2$ and $\Phi_{-n_e}^2$ as
\begin{align}
h_L^{(2)}=\frac{c_L}{12}\left(\frac{n_e}{2}-\frac{2}{n_e}\right),\,\,\,\,
h_R^{(2)}=\frac{c_R}{12}\left(\frac{n_e}{2}-\frac{2}{n_e}\right).
\end{align}

The entanglement negativity for the bipartite pure state configuration of a single interval at zero temperature in a CFT$^a_2$ with gravitational anomaly may then be obtained using eqs. (\ref{two-point-split}) and (\ref{replica-technique}) as follows
\begin{align}
\mathcal{E}(A)=\frac{c_L}{4}\log \left(\frac{z_A}{\epsilon}\right)+\frac{c_R}{4}\log \left(\frac{\bar{z}_A}{\epsilon}\right)+2\log c_{1/2},
\end{align}
where $z_A=z_1-z_2$, $\epsilon$ is a UV cut-off and $c_{1/2}$ is a normalization constant for the two point function. On using $z_A=R_A e^{i\theta_A}$ and analytically continuing to a Lorentzian signature via $\theta_A=i\kappa_A$ we obtain the entanglement negativity for the pure state configuration in question as follows
\begin{equation}\label{neg_sing}
\mathcal{E}(A)=\frac{c_L+c_R}{4}\log\left( \frac{R_A}{\epsilon}\right)-\frac{c_L-c_R}{4}\kappa_A+2\log c_{1/2}\,.
\end{equation}
It may be observed from the above equation, that compared to a usual CFT$_2$ \cite{Calabrese:2012nk}, the entanglement negativity receives an additional contribution arising from the anomalous Lorentz boost which is given by the second term. This result reduces to the usual entanglement negativity of a single interval at zero temperature described in  \cite{Calabrese:2012nk} for $c_L=c_R$. Note that using eq. (\ref{entropy-single-zero}), our result may be expressed as
\begin{equation}
\mathcal{E}(A)=\frac{3}{2}S_A+\textrm{const.}\,,
\end{equation}
which is expected from quantum information theory as the entanglement negativity for a pure state is given by the R\'enyi entropy of order half which is proportional to the entanglement entropy.

\subsubsection*{Finite temperature and angular potential}\label{sec:sing_Neg_T}

For this case we consider a single interval $A$ of length $R_A$ in a CFT$^a_2$ at a finite temperature $T=1/\beta$ with a conserved angular momentum $\Omega$ defined on a twisted infinite cylinder. As described in \cite{Calabrese:2014yza}, the replica manifold utilized for computing the entanglement negativity for this mixed state configuration suffers from a pathology arising due to the partial transposition over an infinite subsystem. In the present scenario of CFT$_2^a$, a similar problem arises for the infinite twisted cylinder. Following a procedure similar to that described in \cite{Calabrese:2014yza} for the entanglement negativity of a single interval at a finite temperature, we consider two adjacent large but finite auxiliary intervals of length $R$ on either side of the single interval. This configuration is then described by a four point twist field correlator as follows
\begin{equation}\label{replica-single-finite-temp}
\mathcal{E}(A)=\lim_{R\to \infty}\lim_{n_e\to1}\log \left<\Phi_{n_e}(-R)\Phi_{-n_e}^2(0)\Phi_{n_e}^2(R_A)\Phi_{-n_e}(R)\right>_{\beta_{L,R}} \,,
\end{equation}
where the subscript $\beta_{L,R}$ denotes that the four point function has to be evaluated on a twisted cylinder and $(L,R)$ in $\beta_{L,R}$ describes the left and the right moving sectors respectively. Note that in the above equation a bipartite limit $R \to \infty,\, B\equiv A^c$ has been implemented subsequent to the replica limit.
The four point twist correlator on the CFT$_2$ plane is given from \cite{Calabrese:2014yza} as follows
\begin{equation}\label{4pt-correlator-neg-T}
\left<\Phi_{n_e}(z_1)\Phi_{-n_e}^2(z_2)\Phi_{n_e}^2(z_3)\Phi_{-n_e}(z_4)\right>_\mathbb{C}=c_{n_e}c^2_{n_e/2}\left(\frac{1}{z_{14}^{2h_L}z_{23}^{2h_L^{(2)}}}\frac{\mathcal{F}_{n_e}(\eta)}{\eta^{h_L^{(2)}}}\right)\left(\frac{1}{\bar{z}_{14}^{2h_R}\bar{z}_{23}^{2h_R^{(2)}}}\frac{\bar{\mathcal{F}}_{n_e}(\bar{\eta})}{\bar{\eta}^{h_R^{(2)}}}\right), 
\end{equation}
where $\eta=\frac{z_{12}z_{34}}{z_{13}z_{24}}$ and $\bar{\eta}=\frac{\bar{z}_{12}\bar{z}_{34}}{\bar{z}_{13}\bar{z}_{24}}$ are the cross ratios and  $\mathcal{F}_{n_e}(\eta)$ and $\bar{\mathcal{F}}_{n_e}(\bar{\eta})$ are two non universal arbitrary functions. As described in \cite{Calabrese:2014yza} the non universal arbitrary functions $\mathcal{F}_{n_e}(\eta)$ and $\bar{\mathcal{F}}_{n_e}(\bar{\eta})$ at the limits $\eta,\bar{\eta}\to 1$ and $\eta,\bar{\eta}\to 0$ are given by
\begin{equation}
\mathcal{F}_{n_e}(1)=\bar{\mathcal{F}}_{n_e}(1)=1, \hspace{5mm} \mathcal{F}_{n_e}(0)=\bar{\mathcal{F}}_{n_e}(0)=C_{n_e},
\end{equation}
where $C_{n_e}$ is a non universal constant depending upon the full operator content of the theory.

We now utilize the conformal map from the CFT$_2$ plane to the twisted cylinder using eq. (\ref{twisted-cylinder-transf}) to express the four point function in the following way
\begin{equation}\label{four-point-twisted-cylinder}
\begin{aligned}
&\left<\Phi_{n_e}(-R)\Phi_{-n_e}^2(0)\Phi_{n_e}^2(R_A)\Phi_{-n_e}(R)\right>_{\beta_{L,R}} \\
&\qquad\qquad\qquad=c_{n_e}c^2_{n_e/2}\left[ \frac{\beta_L}{\pi}\sinh\left(\frac{2\pi R}{\beta_L}\right) \right]^{-2h_L} \left[ \frac{\beta_L}{\pi}\sinh\left(\frac{\pi R_A}{\beta_L}\right) \right]^{-2h_L^{(2)}}\frac{\mathcal{F}_{n_e}(\eta)}{\eta^{h_L^{(2)}}} \\
&
\qquad\qquad\qquad\qquad\qquad \times \left[ \frac{\beta_R}{\pi}\sinh\left(\frac{2\pi R}{\beta_R}\right) \right]^{-2h_R} \left[ \frac{\beta_R}{\pi}\sinh\left(\frac{\pi R_A}{\beta_R}\right) \right]^{-2h_R^{(2)}}\frac{\bar{\mathcal{F}}_{n_e}(\bar{\eta})}{\bar{\eta}^{h_R^{(2)}}}\,.
\end{aligned}
\end{equation}
Under the conformal transformation from CFT$_2$ plane to the twisted cylinder, the cross ratios in the bipartite limit ($R \to \infty$) are given as 
\begin{equation}
\lim_{R \to \infty} \eta= e^{-\frac{2\pi R_A}{\beta_L}}, \,\,\,
\lim_{R \to \infty} \bar{\eta}= e^{-\frac{2\pi R_A}{\beta_R}}.
\end{equation}
We now employ eq. (\ref{four-point-twisted-cylinder}) in (\ref{replica-single-finite-temp}) to obtain the entanglement negativity for the mixed state configuration of a single interval at finite temperature and an angular potential as follows
\begin{equation}\label{neg-single-finite}
\begin{aligned}
\mathcal{E}(A)
&=\frac{c_L}{4}\log\left[\frac{\beta_L}{\pi \epsilon}\sinh\left(\frac{\pi R_A}{\beta_L}\right)\right ]+\frac{c_R}{4}\log\left[\frac{\beta_R}{\pi \epsilon}\sinh\left(\frac{\pi R_A}{\beta_R}\right)\right ]-\frac{c_L}{4}\frac{\pi R_A}{\beta_L}-\frac{c_R}{4}\frac{\pi R_A}{\beta_R}\\
&\quad+f\left(e^{-\frac{2\pi R_A}{\beta_L}}\right)+\bar{f}\left(e^{-\frac{2\pi R_A}{\beta_R}}\right)+\text{const.}\,.
\end{aligned}
\end{equation}
Here $\epsilon$ is a UV cut-off and the arbitrary functions $f(\eta)$ and $\bar{f}(\bar{\eta})$ is given by
\begin{equation}
f(\eta)=\lim_{n_e\to 1}\log [\mathcal{F}_{n_e}(\eta)], \, \, \, \, 
\bar{f}(\bar{\eta})=\lim_{n_e\to 1}\log [\bar{\mathcal{F}}_{n_e}(\bar{\eta})].
\end{equation}
and the last term is a non universal constant for the four point function. Note that the expression in \cref{neg-single-finite} matches with the result in \cite{Chaturvedi:2016opa} for $c_L=c_R$ when the anomaly is absent. We also observe that on using eq. (\ref{entropy-single-finite}), the above eq. (\ref{neg-single-finite}) may be expressed as
\begin{equation}
\begin{aligned}
\mathcal{E}(A)=\frac{3}{2} \big[ S_A-S_A^{\textrm{th}} \big]+ f\left(e^{-\frac{2\pi R_A}{\beta_L}}\right)+\bar{f}\left(e^{-\frac{2\pi R_A}{\beta_R}}\right)+\text{const.},
\end{aligned}
\end{equation}
where $S_A$ and $S_A^{\textrm{th}}$ denote the entanglement entropy and the thermal entropy of the mixed state described by a single interval in the CFT$_2^a$. From the above equation it is observed that the universal part of the entanglement negativity described by the first term involves the elimination of the thermal entropy from the entanglement entropy which is consistent with its characterization as an upper bound on the distillable entanglement in quantum information theory whereas the other terms are non universal contributions.

\subsubsection{Two adjacent intervals}\label{sec:adj_Neg_field}
Having described the different cases for a single interval in the CFT$_2$ with a gravitational anomaly under consideration we now turn our attention to the computation of the entanglement negativity for bipartite mixed state configurations of two adjacent intervals in such CFT$_2$s.

\subsubsection*{Zero temperature}\label{sec:adj_Neg_field_zeroT}
\begin{figure}[H]
	\centering
	\includegraphics[scale=.60 ]{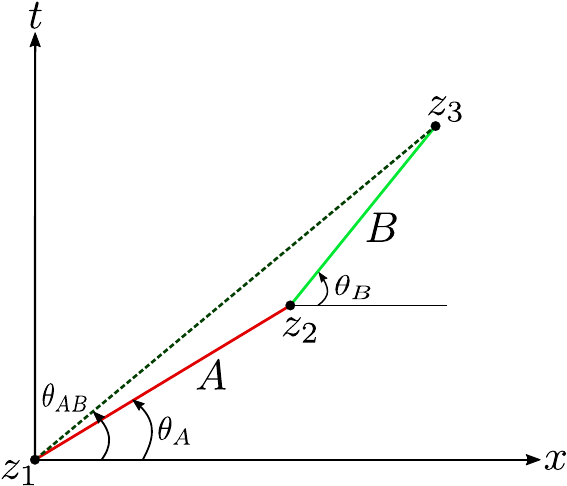}
	\caption{Schematics of tow boosted adjacent intervals $A=[z_1,z_2]$ and $B=[z_2,z_3]$. }
	\label{fig2}
\end{figure}

For the zero temperature case we consider the adjacent limit $z_3\to z_2$ for the two disjoint intervals configuration to arrive at the configuration of adjacent intervals which is depicted in Fig. \ref{fig2}. 

In this limit the four point twist correlator in eq. (\ref{four-point-twist}) reduces to a three point correlator as follows
\begin{equation}\label{three-point-correlator}
\begin{aligned}
\textrm{Tr}(\rho_{AB}^{T_B})^{n_e}&=\left<\Phi_{n_e}(z_1)\Phi_{-n_e}^2(z_2)\Phi_{n_e}(z_3)\right>\\
&=c_{n_e}^2 \mathcal{C}_{\Phi \Phi^2\Phi } \frac{1}{(z_A z_B)^{h_L^{(2)}}z_{AB}^{2h_L-h_L^{(2)}}}
\frac{1}{(\bar{z}_A \bar{z}_B)^{h_R^{(2)}}\bar{z}_{AB}^{2h_R-h_R^{(2)}}}\,,
\end{aligned}
\end{equation}
where $z_A=z_2-z_1$, $z_B=z_3-z_2$ and $z_{AB}=z_3-z_1$. Making a transition to a Lorentzian signature as $z_A=R_A e^{-\kappa_A}$, $z_B=R_B e^{-\kappa_B}$ and $z_{AB}=R_{AB} e^{-\kappa_{AB}}$ and using the weights of the twist fields and \cref{replica-technique}, we obtain the entanglement negativity for the mixed state of adjacent intervals at a zero temperature as follows
\begin{equation}\label{adj-neg-zero}
\mathcal{E}(A:B)=\frac{c_L+c_R}{8}\log \left(\frac{R_A R_B}{\epsilon \,R_{AB}}\right)-\frac{c_L-c_R}{8}(\kappa_A+\kappa_B-\kappa_{AB})+\text{const.},
\end{equation}
where $\epsilon$ is a UV cut-off.
We observe that the second term in the above equation for the entanglement negativity arises from the gravitational anomaly. Note that the above expression in eq. (\ref{adj-neg-zero}) reduces to the corresponding entanglement negativity in \cite{Calabrese:2012nk} for $c_L=c_R$.

\subsubsection*{Finite temperature and angular potential}\label{sec:adj_Neg_field_T}
For the case of a finite temperature and an angular potential as described earlier we consider the configuration of adjacent intervals $A$ and $B$ in a CFT$_2$ at finite temperature $T=1/\beta$ and chemical potential for the angular momentum $\Omega$ which is now located on a twisted cylinder. This may be obtained through the conformal map from the complex plane $z$ to the twisted cylinder $w$ as described in eq. (\ref{twisted-cylinder-transf}). The three point twist correlator transforms under the conformal transformation in the following way
\begin{equation}\label{three-point-transf}
\begin{aligned}
\left<\Phi_{n_e}(w_1,\bar{w}_1)\Phi_{-n_e}^2(w_2,\bar{w}_2)\Phi_{n_e}(w_3,\bar{w}_3)\right>_{\beta_{L,R}}&=\prod_{i=1}^{3} \left(\frac{dw_i}{dz_i}\right)^{-h_L^{(i)}}\left(\frac{d\bar{w}_i}{d\bar{z}_i}\right)^{-h_R^{(i)}}\\
&  \times \left<\Phi_{n_e}(z_1,\bar{z}_1)\Phi_{-n_e}^2(z_2,\bar{z}_2)\Phi_{n_e}(z_3,\bar{z}_3)\right>_{\mathbb{C}}\,,
\end{aligned}
\end{equation}
where $h_{L}^{(i)}\,,\,h_{R}^{(i)}$ are the conformal dimensions of the twist fields placed at $(w_i,\bar{w}_i)$. We now choose the coordinate of adjacent intervals on the cylinder as $w_1=\bar{w}_1=-R_A$, $w_2=\bar{w}_2=0$ and $w_3=\bar{w}_3=R_B$. Then the entanglement negativity for the mixed state configuration of two adjacent intervals may be computed using eq. (\ref{three-point-correlator}) in (\ref{three-point-transf}) and eq. (\ref{replica-technique}) as follows
\begin{equation}\label{adj-neg-T}
{\cal E}(A:B)=\frac{c_L}{8}\log\Bigg[\bigg(\frac{\beta_{L}}{\pi \epsilon}\bigg)\frac{\sinh{\big(\frac{\pi R_A}{\beta_{L}}\big)}\sinh{\big(\frac{\pi R_B}{\beta_{L}}\big)}}{\sinh({\frac{\pi R_{AB}}{\beta_{L}} })}\Bigg]
+\frac{c_R}{8}\log\Bigg[\bigg(\frac{\beta_{R}}{\pi \epsilon}\bigg)\frac{\sinh{\big(\frac{\pi R_A}{\beta_{R}}\big)}\sinh{\big(\frac{\pi R_B}{\beta_{R}}\big)}}{\sinh({\frac{\pi R_{AB}}{\beta_{R}} })}\Bigg],
\end{equation}
where $\epsilon$ is a UV cut-off and $R_{AB}=R_A+R_B$. Interestingly the above result matches with corresponding entanglement negativity \cite{Jain:2017uhe} in the absence of an anomaly ($c_L=c_R$).

\subsubsection{Two disjoint intervals}
\label{sec:disj_Neg_field}
In this section we focus on the bipartite mixed state configuration of two disjoint intervals in a CFT$_2$ with a gravitational anomaly ($c_L \neq c_R$).
\subsubsection*{Zero temperature}\label{sec:disj_Neg_FT}
For this case, as described earlier, we consider the configuration of two boosted disjoint intervals $A$ and $B$ as shown in Fig. \ref{fig3}. The explicit form of the four point twist correlator involved in the eq. (\ref{four-point-twist}) is not known generally as it depends on an arbitrary non universal function of the cross ratios. However in the large central charge limit when the two disjoint intervals are in proximity ($ 1/2 < \eta < 1 $), the universal part of the four point function in the $t$-channel may be extracted utilizing a monodromy technique and is given as \cite{Hartman:2013mia,Kulaxizi:2014nma,Malvimat:2018txq}
\begin{equation}\label{four-point-limit-disj}
\lim_{ n_e \to 1 } \left<\Phi_{n_e}(z_1)\Phi_{-n_e}(z_2)\Phi_{-n_e}(z_3)\Phi_{n_e}(z_4)\right>_{\mathbb{C} }
= \left ( 1 - \eta \right )^{ \hat h_L } \left ( 1 - \bar{\eta} \right )^{ \hat h_R } ,
\end{equation}
where $\eta=\frac{z_{12} z_{34}}{z_{13}z_{24}}$ is the cross ratio and $\hat h_L,\hat h_R$ are the conformal dimensions of the operator with the dominant contribution in the corresponding conformal block expansion. The dominant contribution to the four point twist correlator in eq. (\ref{four-point-limit-disj}) arises from the conformal block with the conformal dimension $h_L^{(2)}\equiv \hat h_L$ and $h_R^{(2)}\equiv \hat h_R$ and in the $n_e\to 1$ limit\footnote{Note that the negative conformal dimensions of the twist field $\Phi^2_{n_e}$ in the replica limit $n_e\to 1$ has to be understood only in the sense of an analytic continuation.}
\begin{align}
\hat h_L=-\frac{c_L}{8}~~ ,~~  \hat h_R=-\frac{c_R}{8}.
\end{align}

The entanglement negativity for the bipartite mixed state configuration of disjoint intervals in proximity in a CFT$_2$ with gravitational anomaly may then be obtained using eq. (\ref{four-point-limit-disj}) and  (\ref{replica-technique}) as
\begin{equation} \label{neg-disj-cross-ratio}
\mathcal{E}(A:B)=\frac{c_L}{8}\log\left(\frac{1}{1-\eta}\right)+\frac{c_R}{8}\log\left(\frac{1}{1-\bar{\eta}}\right).
\end{equation}
As earlier making a transition to a Lorentzian signature as $z_{ij}=R_{ij}\,e^{-\kappa_{ij}}$, where $i,j=1,2,3,4$ and $z_{AC}=z_{13}$, $z_{BC}=z_{24}$, $z_{ABC}=z_{14}$ and $z_{C}=z_{23}$, the above equation may be expressed in the following form
\begin{equation}\label{neg-disj-zero}
\mathcal{E}(A:B)=\frac{c_L+c_R}{8}\log \left(\frac{R_{AC} \,R_{BC}}{R_{ABC} \,R_C}\right)-\frac{c_L-c_R}{8}(\kappa_{AC}+\kappa_{BC}-\kappa_{ABC}-\kappa_C).
\end{equation}
Note that the above result is independent of the UV cut-off which is similar to the corresponding result for usual CFT$_2$. We also observe that the second term in the above expression arises from the gravitational anomaly and is frame dependent. Furthermore the above expression in eq. (\ref{neg-disj-zero}) matches with the corresponding result in \cite{Malvimat:2018txq} in the absence of the anomaly $(c_L=c_R$).

\subsubsection*{Finite temperature and angular potential}\label{sec:disj_Neg_field_T}
As earlier for this case we consider the configuration of  two disjoint intervals $A$ and $B$ in a CFT$_2$ at a finite temperature $T=1/\beta$ and chemical potential for the angular momentum $\Omega$ located on a twisted cylinder. Following the technique described earlier the four point twist correlator on the twisted cylinder may be obtained from the four point correlator on the complex plane through the following transformation
\begin{equation}\label{disj-transf}
\begin{aligned}
&\left<\Phi_{n_e}(w_1,\bar{w}_1)\Phi_{-n_e}(w_2,\bar{w}_2)\Phi_{-n_e}(w_3,\bar{w}_3)\Phi_{n_e}(w_4,\bar{w}_4)\right>_{\beta_{L,R}}=\prod_{i=1}^{4} \left(\frac{dw_i}{dz_i}\right)^{-h_L}\left(\frac{d\bar{w}_i}{d\bar{z}_i}\right)^{-h_R}\\
& \quad \hspace{6 cm}
\left<\Phi_{n_e}(z_1,\bar{z}_1)\Phi_{-n_e}(z_2,\bar{z}_2)\Phi_{-n_e}(z_3,\bar{z}_3)\Phi_{n_e}(z_4,\bar{z}_4)\right>_{\mathbb{C}}\,.
\end{aligned}
\end{equation}
The lengths of the disjoint intervals on the twisted cylinder maybe chosen as $w_2-w_1=R_A$, $w_3-w_2=R_C$ and $w_4-w_3=R_B$ and the entanglement negativity for this mixed state configuration may be now obtained using eqs. (\ref{four-point-limit-disj}), (\ref{twisted-cylinder-transf}), (\ref{disj-transf}) and (\ref{replica-technique}) as follows
\begin{equation}\label{neg-disj-finite}
{\cal E}(A:B)
= \frac{ c_L }{ 8 } \log \left [ \frac
{\sinh \left ( \frac{ \pi R_{AC} }{ \beta_{L} } \right )
	\sinh \left ( \frac{ \pi R_{BC} }{ \beta_{L} } \right )
}{
	\sinh \left ( \frac{ \pi R_C }{ \beta_{L} } \right )
	\sinh \left ( \frac{ \pi R_{ABC} }{ \beta_{L} } \right )
} \right ] 
+ \frac{ c_R }{ 8 } \ln \left [ \frac
{\sinh \left ( \frac{ \pi R_{AC} }{ \beta_{R} } \right )
	\sinh \left ( \frac{ \pi R_{BC} }{ \beta_{R} } \right )
}{
	\sinh \left ( \frac{ \pi R_C }{ \beta_{R} } \right )
	\sinh \left ( \frac{ \pi R_{ABC} }{ \beta_{R} } \right )
} \right ],
\end{equation}
where $R_{AC}$, $R_{BC}$ and $R_{ABC}$ are the lengths of the intervals $A\cup C$, $B\cup C$ and $A\cup B\cup C$ respectively. We note that the above result is once again cut-off independent similar to the corresponding case in usual CFT$_2$s. The above result once more matches exactly with the corresponding result in \cite{Malvimat:2018ood} when the anomaly is absent ($i.e. \,c_L=c_R$).

\subsection{Reflected entropy in CFT$^a_2$}
\label{sec:reflecetd}

We now turn our attention to another mixed state entanglement measure known as the reflected entropy which involves both classical and quantum correlations. In what follows we provide a brief review for the definition and computation of this measure in usual CFT$_2$s as described in \cite{Dutta:2019gen}. To this end it is required to consider a bipartite quantum system $A\cup B$ in a mixed state $\rho_{AB}$ and its canonical purification  in a doubled Hilbert space $\mathcal{H}_A \otimes \mathcal{H}_B \otimes \mathcal{H}_{A^\star} \otimes \mathcal{H}_{B^\star}$. This is denoted as $\ket{\sqrt{\rho_{AB}}}$ where $A^\star$ and $B^\star$ represent the CPT conjugate of the subsystems $A$ and $B$ respectively.

The reflected entropy $S_R(A: B)$ may then be defined as the von Neumann entropy of the reduced density matrix $\rho_{AA^\star}$ \cite{Dutta:2019gen} as follows
\begin{align}
S_R(A: B) \equiv S_{vN}(\rho_{ {AA^\star}})_{\sqrt{\rho_{AB}}},
\end{align}
where $\rho_{AA^\star}$ is defined as the reduced density matrix traced over $\mathcal{H}_B\otimes\mathcal{H}^\star_B$, given as
\begin{equation}
\rho_{AA^\star} = \text{Tr}_{\mathcal{H}_B\otimes\mathcal{H}^\star_B}\ket{\sqrt{\rho_{AB}}}\bra{\sqrt{\rho_{AB}}}.
\end{equation}

\begin{figure}[H]
	\centering
	\includegraphics[scale=.55 ]{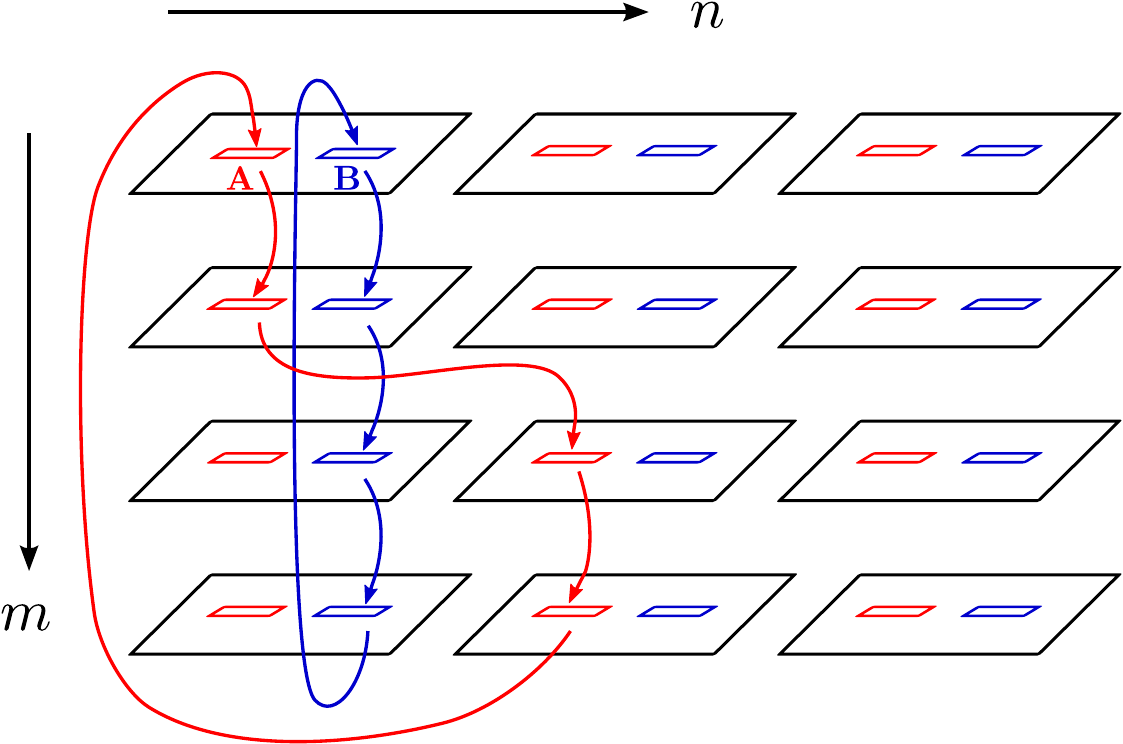}
	\caption{Structure of the replica manifold computing the R\'enyi reflected entropy between subsystems $A$ and $B$ in the state $\ket{\psi_m}$. The sewing of the individual replicas along the subsystems $A$ and $B$ are denoted by red and blue arrows corresponding to the twist fields $\sigma_{g^{}_A}$ and $\sigma_{g^{}_B}$, respectively. Figure modified from \cite{Chandrasekaran:2020qtn}.}
	\label{fig:reflected-replica-simple}
\end{figure}

Interestingly the authors in \cite{Dutta:2019gen} developed a novel replica technique to compute the reflected entropy between two subsystems $A$ and $B$ which we briefly review below. To begin with, one constructs the state $|{\rho_{AB}^{m/2}}\rangle \equiv \ket{\psi_m}$ by considering an $m$-fold replication of the original manifold where $m \in 2 \mathbb{Z}^+$. Subsequently the R\'enyi reflected entropy for this state $\ket{\psi_m}$ is computed as the R\'enyi entropy $S_n\left(AA^\star\right)_{\psi_m}$ of the reduced density matrix 
\begin{equation}
\rho_{AA^\star}^{(m)} = \text{Tr}_{\mathcal{H}_B\otimes\mathcal{H}^\star_B}\ket{{\rho_{AB}^{m/2}}}\bra{\rho_{AB}^{m/2}},
\end{equation}
which involves another replication in the R\'enyi index $n$ and results in a $nm$-sheeted replica manifold\footnote{See \cite{Dutta:2019gen,Jeong:2019xdr} for details about replica construction of the state $\ket{{\rho_{AB}^{m/2}}}$ and the sewing mechanism of such replica sheets.} as shown in \cref{fig:reflected-replica-simple}.

In the replica technique, this R\'enyi reflected entropy is given in terms of a properly weighted partition function $Z_{n,m}$ on the above replica manifold which in turn may be obtained as the correlation functions of twist operators $\sigma_{g_A}$ and $\sigma_{g_B}$ inserted at the endpoints of the intervals $A \equiv [z_1,z_2]$ and $B \equiv [z_3,z_4]$ as follows \cite{Dutta:2019gen}
\begin{equation}\label{renyi-reflected-entropy}
S_n\left(AA^\star\right)_{\psi_m}=\frac{1}{1-n}\log\frac{Z_{n,m}}{\left(Z_{1,m}\right)^n}=\frac{1}{1-n}\log\frac{\left<\sigma_{g^{}_A}(z_1)\sigma_{g_A^{-1}}(z_2)\sigma_{g^{}_B}(z_3)\sigma_{g_B^{-1}}(z_4)\right>_{\mathrm{CFT}^{\bigotimes mn}}}{\left(\left<\sigma_{g^{}_m}(z_1)\sigma_{g_m^{-1}}(z_2)\sigma_{g^{}_m}(z_3)\sigma_{g_m^{-1}}(z_4)\right>_{\mathrm{CFT}^{\bigotimes m}}\right)^n}\,.
\end{equation}
In the denominator of the above equation the partition function $Z_{1,m}$ arises from the normalization of the state $|{\rho_{AB}^{m/2}}\rangle$ and $\sigma_{g_m}$ are the twist fields at the endpoints of the intervals in $m$-replicated manifold. Having reviewed the definition and the replica technique to compute the reflected entropy for mixed states in a CFT$_2$ we now turn our attention to compute the same for various bipartite states in a CFT$_2^a$ with a gravitational anomaly in the following subsections.

\subsubsection{Two disjoint intervals} In this subsection we utilize the replica techniques described above to compute the reflected entropy for the zero and finite temperature mixed state configuration of two disjoint intervals in a CFT$_2^a$ with a gravitational anomaly.

\subsubsection*{Zero temperature}

For the zero temperature case we consider the configuration of two boosted disjoint intervals described by the intervals $A\equiv [z_1,z_2]$ and $B\equiv [z_3,z_4]$. Note that the conformal dimensions for the twist operators $\sigma_{g^{}_A}$, $\sigma_{g^{}_B}$ and $\sigma_{g^{}_B g_A^{-1}}$ for the left moving sector with a central charge $c_L$ may be written for our case of unequal central charges as follows
\begin{equation}\label{reflected-twist-field}
h^{A}_{L}=h_{L}^{B}=\frac{n \,c_{L}}{24}\left(m-\frac{1}{m}\right),
\quad h_{L}^{{B} A^{-1}}=\frac{2 \,c_{L}}{24}\left(n-\frac{1}{n}\right),
\end{equation}
with similar expressions for the right moving sector involving the central charge $c_R$. The conformal dimensions for $\sigma_{g^{}_m}$ may be obtained from \cref{reflected-twist-field} by setting $n = 1$. In the t-channel, the four point function in the numerator of  \cref{renyi-reflected-entropy} can be expanded in terms of the conformal blocks of the replica theory $\text{CFT}^{\bigotimes mn}$ as
\begin{equation}\label{four-point-reflected-expansion}
\begin{aligned}
\left<\sigma_{g^{}_A}(z_1)\sigma_{g_A^{-1}}(z_2)\sigma_{g^{}_B}(z_3)\sigma_{g_B^{-1}}(z_4)\right>_{\mathrm{CFT}^{\bigotimes mn}}&=z_{41}^{2h_L}\bar{z}_{41}^{2h_R}z_{23}^{2h_L}\bar{z}_{23}^{2h_R}
\sum_p C_p^2 \mathcal{F}_L\left(mnc_L, h_L^{}, h^{(p)}_L , 1-\eta\right) \\
&\quad\times\mathcal{F}_R\left(mnc_R, h_R^{}, h^{(p)}_R , 1-\bar{\eta}\right),
\end{aligned}
\end{equation}
where $h_{L(R)}=\frac{n \, c_{L(R)}}{24}\left(m-\frac{1}{m}\right)$, $\eta=\frac{z_{12}z_{34}}{z_{13}z_{24}}$ is the cross ratio and $\mathcal{F}$ is the Virasoro conformal block corresponding to the exchange of the primary operators with dimensions $h^{(p)}$. Note that in the above expansion $C_p$ is the OPE coefficient appearing in the three point function. The explicit closed form structure for the Virasoro conformal block is not known generally. In the following we will make use of the semi-classical limit described by
\begin{equation}
mnc_{L}\to 0, \quad \epsilon_{L}=\frac{6h_{L}}{mnc_{L}} \,\, \mathrm{and}\,\,
\epsilon_{L}^{(p)}=\frac{6h_{L}^{(p)}}{mnc_{L}} \,\,\qquad \mathrm{fixed},
\end{equation}
and similar expressions for the right moving sectors involving $h_R$ and $c_R$. It is well known that in the above semi-classical limit, the Virasoro conformal block $\mathcal{F}$ exponentiates in the following way \cite{Belavin:1984vu, Zamolodchikov1987}
\begin{equation}\label{log-F}
\begin{aligned}
\log \mathcal{F}_L\left(mnc_L, h_L^{}, h^{(p)}_L , 1-\eta\right) \approx -\frac{mnc_L}{6} f_L \left(\epsilon_L, \epsilon_L^{(p)}, 1-\eta\right), \\ 
\log \mathcal{F}_R\left(mnc_R, h_R^{}, h^{(p)}_R , 1-\eta\right) \approx -\frac{mnc_R}{6} f_R \left(\epsilon_R, \epsilon_R^{(p)}, 1-\bar{\eta}\right).
\end{aligned}
\end{equation}
In the t-channel, the dominant contribution to the four point correlator arises from the intermediate operator with the lowest conformal dimensions $h^{(p)}$ in the OPE expansion. It is given for the left moving sector as \cite{Jeong:2019xdr}
\begin{equation}
h_{L}^\text{low}=h^{(p)}_{L}=h_{L}^{B A^{-1}},\quad \epsilon_{L}^\text{low}=\epsilon_{L}^{(p)}=\frac{6 h_{L}^\text{low}}{mnc_{L}},
\end{equation}
with similar expressions for the right moving sector. The perturbative expansion of  $f_L$ in $\epsilon_L$ and $\epsilon_L^\text{low}$ can be expressed as \cite{Fitzpatrick:2014vua}
\begin{equation}\label{f_L}
f_L \left(\epsilon_L, \epsilon_L^\text{low}, 1-\eta\right)=\epsilon_L^\text{low}\log\left(\frac{1+\sqrt{\eta}}{1-\sqrt{\eta}}\right)+\mathrm{higher\, order\, terms}\,.
\end{equation}
One can also arrive at the explicit form of $C_p$ in this case in a similar fashion as described in \cite{Dutta:2019gen} as $C_p=(2m)^{-2h_L/n-2h_R/n}$. Now using eqs. (\ref{four-point-reflected-expansion}), (\ref{log-F})  and (\ref{f_L}), the reflected entropy for the mixed state of two disjoint intervals in a CFT$_{2}$ with gravitational anomaly may expressed as
\begin{align}
S_R(A:B)&=\lim_{ n \to 1 }\lim_{m \to 1 }\,S_n\left(AA^\star\right)_{\psi_m}\notag\\
&=\frac{c_L}{6}\log\left(\frac{1+\sqrt{\eta}}{1-\sqrt{\eta}}\right)+\frac{c_R}{6}\log\left(\frac{1+\sqrt{\bar{\eta}}}{1-\sqrt{\bar{\eta}}}\right)\,.\label{SR_disj}
\end{align}
We observe that the reflected entropy factorizes into the left and right moving contributions in the presence of the gravitational anomaly. Note that the above expression reduces to the corresponding reflected entropy in \cite{Dutta:2019gen} for the usual scenario ($c_L=c_R$).

\subsubsection*{Finite temperature and angular potential}
For this case we again consider the mixed state configuration of two disjoint intervals $A$ and $B$ in a CFT$_2$ now at a finite temperature $T=1/\beta$ and a chemical potential $\Omega$ for the angular momentum. In this case once again note that the CFT$^a_2$ is defined on a twisted cylinder which may be obtained from the usual complex plane utilizing eq. (\ref{twisted-cylinder-transf}). The four point twist correlator in this case transforms under this conformal map as follows
\begin{equation}\label{refl-disj-transf}
\begin{aligned}
&\left<\sigma_{g^{}_A}(w_1,\bar{w}_1)\sigma_{g_A^{-1}}(w_2,\bar{w}_2)\sigma_{g^{}_B}(w_3,\bar{w}_3)\sigma_{g_B^{-1}}(w_4,\bar{w}_4)\right>_{\beta_{L,R}}=\prod_{i=1}^{4} \left(\frac{dw_i}{dz_i}\right)^{-h_L^{(i)}}\left(\frac{d\bar{w}_i}{d\bar{z}_i}\right)^{-h_R^{(i)}}\\
& \quad \hspace{6 cm}
\left<\sigma_{g^{}_A}(z_1,\bar{z}_1)\sigma_{g_A^{-1}}(z_2,\bar{z}_2)\sigma_{g^{}_B}(z_3,\bar{z}_3)\sigma_{g_B^{-1}}(z_4,\bar{z}_4)\right>_{\mathbb{C}}\,.
\end{aligned}
\end{equation}
The reflected entropy for the mixed state of disjoint intervals may now be obtained by evaluating the four point function on a twisted cylinder using eqs. (\ref{refl-disj-transf}) and (\ref{four-point-reflected-expansion}) as follows
\begin{equation}\label{SR_disj_T}
S_R(A:B)=\frac{c_L}{6}\log\left(\frac{1+\sqrt{\xi}}{1-\sqrt{\xi}}\right)+\frac{c_R}{6}\log\left(\frac{1+\sqrt{\bar{\xi}}}{1-\sqrt{\bar{\xi}}}\right),
\end{equation}
where $\xi, {\bar \xi}$ are given by
\begin{align}
\xi = \frac {\sinh \frac{\pi w_{12}}{\beta_L} \sinh \frac{\pi w_{34}}{\beta_L}} {\sinh \frac{\pi w_{13}}{\beta_L} \sinh \frac{\pi w_{24}}{\beta_L}} ~~,~~ \bar\xi = \frac {\sinh \frac{\pi \bar w_{12}}{\beta_R} \sinh \frac{\pi \bar w_{34}}{\beta_R}} {\sinh \frac{\pi \bar w_{13}}{\beta_R} \sinh \frac{\pi \bar w_{24}}{\beta_R}},  \label{crossratio-T}
\end{align}
where $A\equiv [w_1,w_2]$ and $B\equiv [w_3,w_4]$ are the intervals on the twisted cylinder with the coordinates $w,{\bar w}$. As earlier, we observe that the reflected entropy splits into left and right moving components in the presence of the gravitational anomaly.

\subsubsection{Two adjacent intervals} We now turn our attention to the mixed state configuration of two adjacent intervals in the CFT$^a_2$.

\subsubsection*{Zero temperature}
For the zero temperature case we consider the configuration of adjacent intervals $A \equiv [z_1,z_2]$ and $B \equiv [z_2,z_3]$ which may be obtained by taking the adjacent limit $z_3\to z_2$ and relabelling $z_4 \equiv z_3$ in the disjoint interval configuration. In this adjacent limit, the R\'enyi reflected entropy may be expressed in terms of a three point twist correlator as
\begin{equation}
S_n\left(AA^\star\right)_{\psi_m}=\frac{1}{1-n}\log\frac{\left<\sigma_{g^{}_A}(z_1)\sigma_{g^{}_B g_A^{-1}}(z_2)\sigma_{g_B^{-1}}(z_3)\right>_{\mathrm{CFT}^{\bigotimes mn}}}{\left(\left<\sigma_{g^{}_m}(z_1)\sigma_{g_m^{-1}}(z_3)\right>_{\mathrm{CFT}^{\bigotimes m}}\right)^n} ,
\end{equation}
On utilizing the conformal dimensions of the twist fields from eq. (\ref{reflected-twist-field}) and the form of the three point twist correlator above, the reflected entropy for the mixed state configuration of two adjacent intervals at zero temperature may be obtained by taking the replica limit $m\to 1, n\to 1$ as follows
\begin{align}
S_R(A:B)=\frac{c_L+c_R}{6}\log \left(\frac{R_A R_B}{\epsilon \,R_{AB}}\right)-\frac{c_L-c_R}{6}(\kappa_A+\kappa_B-\kappa_{AB})+\frac{c_L+c_R}{6}\log 2,\label{SR_adj}
\end{align}
where $\epsilon$ is a UV cut-off and $(R_A, \kappa_A)$, $(R_B, \kappa_B)$ and $(R_{AB}, \kappa_{AB})$ are lengths and boosts of intervals $A$, $B$ and $A \cup B$ respectively. Note that on comparing this expression for the reflected entropy with that of a usual CFT$_2$ it is observed that the second term arises due to the presence of the gravitational anomaly.

\subsubsection*{Finite temperature and angular potential}
For this case we consider the mixed state configuration under consideration in a CFT$_2$ at a finite temperature $T=1/\beta$ with a chemical potential $\Omega$ for the conserved angular momentum. The corresponding CFT$^a_2$ is once again defined on a twisted cylinder. The end point coordinates of the adjacent intervals on the twisted cylinder are $w_1=\bar{w}_1=-R_A$, $w_2=\bar{w}_2=0$ and $w_3= \bar{w}_3 =R_B$. The transformation of the three point twist correlator under the conformal map given by eq. (\ref{twisted-cylinder-transf}) may be expressed as
\begin{equation}\label{refl-three-point-transf}
\begin{aligned}
\left<\sigma_{g^{}_A}(w_1,\bar{w}_1)\sigma_{g^{}_B g_A^{-1}}(w_2,\bar{w}_2)\sigma_{g_B^{-1}}(w_3,\bar{w}_3)\right>_{\beta_{L,R}}&=\prod_{i=1}^{3} \left(\frac{dw_i}{dz_i}\right)^{-h_L^{(i)}}\left(\frac{d\bar{w}_i}{d\bar{z}_i}\right)^{-h_R^{(i)}}\\
&  \left<\sigma_{g^{}_A}(z_1,\bar{z}_1)\sigma_{g^{}_B g_A^{-1}}(z_2,\bar{z}_2)\sigma_{g_B^{-1}}(z_3,\bar{z}_3)\right>_{\mathbb{C}}\,,
\end{aligned}
\end{equation}
On using the  above eq. (\ref{refl-three-point-transf}) and the form of the usual three point correlator in a CFT$_2$ , the reflected entropy for the mixed state of adjacent intervals may be obtained as follows
\begin{align}
S_R(A:B)&=\frac{c_L}{6}\log\Bigg[\bigg(\frac{\beta_{L}}{\pi \epsilon}\bigg)\frac{\sinh{\big(\frac{\pi R_A}{\beta_{L}}\big)}\sinh{\big(\frac{\pi R_B}{\beta_{L}}\big)}}{\sinh({\frac{\pi (R_{A}+R_B)}{\beta_{L}} })}\Bigg]
\notag\\
&+\frac{c_R}{6}\log\Bigg[\bigg(\frac{\beta_{R}}{\pi \epsilon}\bigg)\frac{\sinh{\big(\frac{\pi R_A}{\beta_{R}}\big)}\sinh{\big(\frac{\pi R_B}{\beta_{R}}\big)}}{\sinh({\frac{\pi (R_{A}+R_B)}{\beta_{R}} })}\Bigg]\notag\\
&+\frac{c_L+c_R}{6}\log 2.\label{SR_adj_T}
\end{align}
As earlier the reflected entropy decouples into left and right moving components in the presence of the gravitational anomaly.

\subsubsection{Single interval}\label{sec:S_R-single}

We now discuss the case of the bipartite state described by a single interval in a CFT$^a_2$ in this subsection.

\subsubsection*{Zero temperature}
In this case we consider the pure state configuration of a single interval $A\equiv [z_1,z_2]$ at zero temperature in a CFT$^a_2$ which can be obtained from the two disjoint intervals result by taking the limits $z_3\to z_2$, $z_4\to z_1$. The R\'enyi reflected entropy for this configuration is then given by the two point twist correlator as follows
\begin{equation}\label{reflected-two-point-single}
S_n\left(AA^\star\right)_{\psi_m}=\frac{1}{1-n}\log\left<\sigma_{g^{-1}_B g_A^{}}(z_1)\sigma_{g^{}_B g_A^{-1}}(z_2)\right>_{\mathrm{CFT}^{\bigotimes mn}}.
\end{equation}
The reflected entropy for this pure state of a single interval at zero temperature is then obtained as follows
\begin{equation}\label{S_R-single-zero}
S_R(A:A^c)=\frac{c_L+c_R}{3}\log\left( \frac{R_A}{\epsilon}\right)-\frac{c_L-c_R}{3}\kappa_A,
\end{equation}
where $\epsilon$ is a UV cut-off and $R_A$ and $\kappa_A$ are the length and boost of the interval $A$. Note that we may also obtain the reflected entropy for a single interval by using the property of the reflected entropy for a pure state i.e. $S_R(A:B)=2 S(A)$ to arrive at an identical result.

\subsubsection*{Finite temperature and angular potential}
Finally we consider the mixed state configuration of a single interval in a CFT$^a_2$ with a conserved angular momentum and at a finite temperature $T=1/\beta$. As described in the previous subsection, for the case of a single interval $A \equiv [0,R_A]$ with $B \equiv A^c$, the R\'enyi reflected entropy of order $n$ in the state $\psi_m$ appears to be given by \cref{reflected-two-point-single} where the two-point twist correlator now has to be evaluated on the twisted cylinder. Utilizing the conformal transformation from the complex plane to the twisted cylinder given in \cref{twisted-cylinder-transf}, we may obtain 
\begin{align}
	S_n\left(AA^\star\right)_{\psi_m}&=\frac{1}{1-n}\log\left<\sigma_{g^{-1}_B g_A^{}}(z_1)\sigma_{g^{}_B g_A^{-1}}(z_2)\right>_{\mathrm{CFT}_{\beta_{L,R}}^{\bigotimes mn}}\notag \\ 
	&=\left( 1+\frac{1}{n}\right) \left[\frac{c_L}{6} \log\left( \frac{\beta_L}{\pi \epsilon} \sinh \frac{\pi R_A}{\beta_L} \right) + \frac{c_R}{6} \log \left( \frac{\beta_R}{\pi \epsilon} \sinh \frac{\pi R_A}{\beta_R} \right)\right].
\end{align}
Now taking the replica limits $m \to 1, n \to 1$, this computation leads to 
\begin{align}
S_R^\text{naive}(A:B) = \frac{c_L}{3} \log\left( \frac{\beta_L}{\pi \epsilon} \sinh \frac{\pi R_A}{\beta_L} \right) + \frac{c_R}{3} \log \left( \frac{\beta_R}{\pi \epsilon} \sinh \frac{\pi R_A}{\beta_R} \right).\label{S_R-naive}
\end{align}
This resembles the expression for twice the entanglement entropy for the given single interval at a finite temperature in \cref{entropy-single-finite}. However this result leads to serious inconsistencies. In the high temperature limit $\beta_{L(R)} \to 0$, the above reflected entropy diverges linearly which is unphysical. This may be seen in the following way. For very high temperatures the state $\ket{\sqrt{\rho_{AB}}}$ reduces to a product of Bell pairs\footnote{ To see this, recall that the purified state on the doubled Hilbert space has the following structure \cite{Dutta:2019gen,Jeong:2019xdr}:	
\begin{align}
		\ket{\sqrt{\rho_{AB}}} = \sum_a \sqrt{p_a} \ket{\psi_a}_{AB} \ket{\psi_a}_{A^\star B^\star}.
\end{align}
For a thermal state with $p_a \propto e^{-\beta E_a}$, at very high temperatures $\beta \to 0$, we have 
\begin{align}
		\ket{\sqrt{\rho_{AB}}} \propto \sum_a  \ket{\psi_a}_{AB} \ket{\psi_a}_{A^\star B^\star}\,,
\end{align}
which is indeed a product Bell state.} between the mirrored regions $AB$ and $A^\star B^\star$ \cite{Dutta:2019gen}. Therefore $AB$ and $A^\star B^\star$ are maximally entangled which implies $A A^\star$ cannot be entangled with $B B^\star$ in the high temperature limit and consequently $S_R(A:B)$ should vanish. It interesting to note that a similar problem had been identified for the case of the entanglement negativity for the configuration of a single interval in a thermal CFT$_2$ in \cite{Calabrese:2014yza}, which has been utilized in subsection \ref{sec:sing_Neg_field} in the context of CFT$_2^a$.  

As described in \cite{Calabrese:2014yza} in the context of the entanglement negativity, in order to understand the pathology of the above naive computation we need to examine the structure of the replica manifold in computing the R\'enyi reflected entropy more carefully. To begin with, we recall that the finite temperature density matrix $\rho_{A}$ is defined on a cylinder of circumference $\beta$ which has a branch cut along the subsystem $A$. In the case of the R\'enyi reflected entropy for the state $\ket{\psi_m}=|\rho_{AB}^{m/2}\rangle$, the trace of the $n$-th power of the reduced density matrix $\rho^{(m)}_{AA^\star}$ computes the partition function on the replica manifold consisting of $nm$ cylinders with branch cuts along $A$ and $B$ sewed in a fashion similar to that described in \cite{Dutta:2019gen,Jeong:2019xdr}. In particular, the cuts along $B$ are always sewed vertically, while there are additional horizontal sewing along $A$ on the zeroth and $m/2$-th replica sheets, similar to that in \cref{fig:reflected-replica-simple}.  

\begin{figure}[h!]
	\centering
	\includegraphics[scale=0.5]{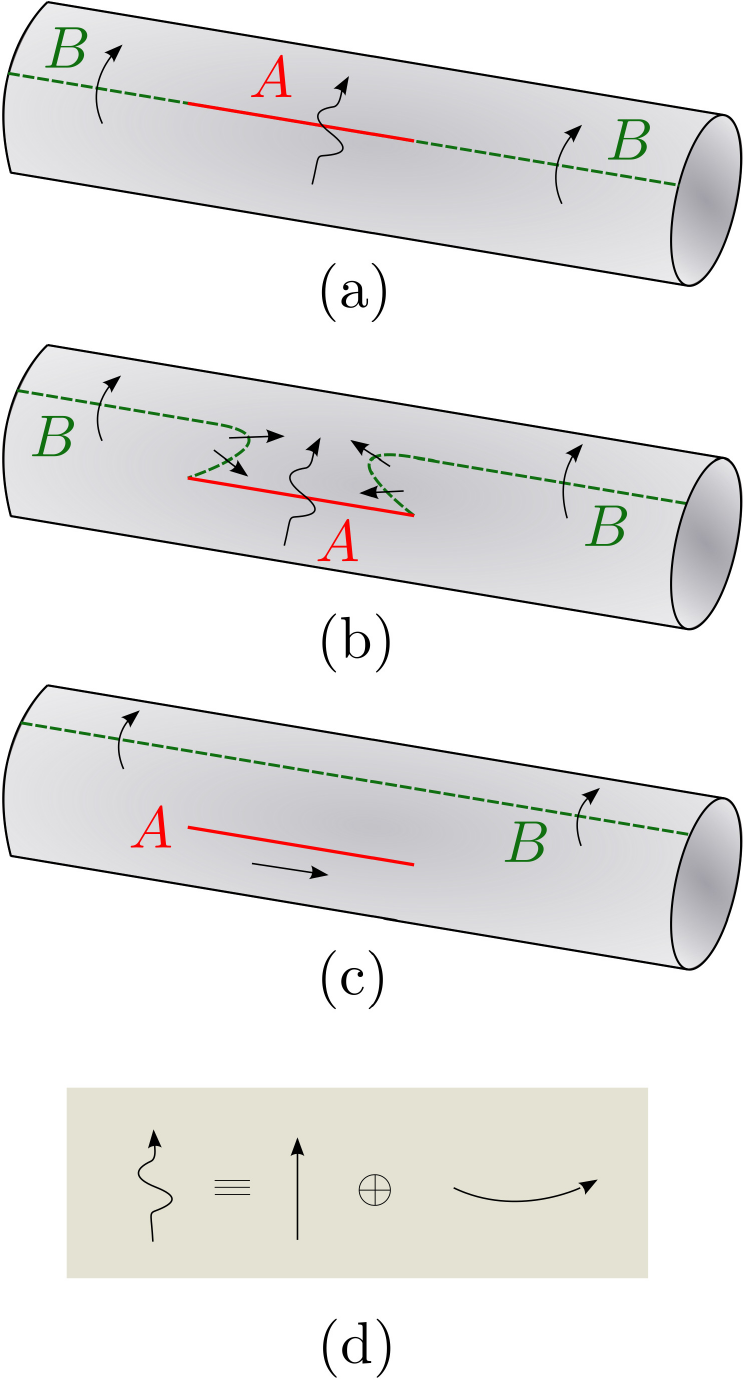}
	\caption{Schematics of the replica manifold computing $	S_n\left(AA^\star\right)_{\psi_m}$ and the reflected entropy of a single interval $A$ at finite temperature. (a) Simple arrows along $B\equiv A^c$ indicates that one passes from the $m$-th copy to the $(m+1)$-th copy through the branch cuts sewed vertically in the $m$-direction while the wiggly arrow denotes the special connection of the copies of $A$ which involves occasional sewing in the $n$-direction. (b) Deforming the cut along $B$ as indicated: part of it superimposes onto $A$ and the rest becomes an infinite cut extending along the length of the whole cylinder. (c) The merging of the red and green cuts along $A$ results in an effective cancellation of the vertical sewing of the branch cuts along $A$, leaving the sewing in the $n$-direction unaffected. This deformation procedure leads to an auxiliary infinite cut along the length of the cylinder which cannot be removed. This is the origin of the pathology in the naive computation of the reflected entropy for a single interval at finite temperature.  (d) The wiggly arrow along the subsystem $A$ denotes the sewing along both the $m$- and $n$-directions.}
	\label{fig:single-T-cylinder-reflected}
\end{figure}

For the present scenario involving a single interval $A$ and its compliment the situation is depicted in  \cref{fig:single-T-cylinder-reflected}\textcolor{blue}{(a)}, where the wiggly arrow on the subsystem $A$ denotes the non-trivial sewing procedure in both $n$ and $m$ directions as shown in \cref{fig:single-T-cylinder-reflected}\textcolor{blue}{(d)}, and the arrows on the subsystem $B$ represents the regular sewing in the $m$ direction only. The cuts along $B$ may be deformed as shown in  \cref{fig:single-T-cylinder-reflected}\textcolor{blue}{(b)} without changing the topology of the manifold. Upon deforming the cuts along $B$, its partial superimposition on $A$ effectively removes the sewing of the copies of $A$ along the $m$-direction leaving the $n$-direction unaffected as shown in \cref{fig:single-T-cylinder-reflected}\textcolor{blue}{(c)}. This amounts to branch cuts along copies of $A$ which are sewed only in the $n$-direction along with an infinite branch cut described by the green line. This infinite branch cut along $B$ which connects the different replica copies cannot be removed in a consistent manner. Therefore the structure of the replica manifold computing the reflected entropy for the single interval at the finite temperature is more complex than we had naively assumed. Although we have kept ourselves confined to the description on an ordinary cylinder for brevity, the above analysis generalizes in a straightforward fashion for the case of twisted cylinders with circumferences $\beta_L$ and $\beta_R$. This may be observed from the fact that the twisted cylinder can be interpreted as two decoupled cylinders for the left-moving and right-moving CFT modes. 

\begin{figure}[h!]
	\centering
	\includegraphics[scale=0.5]{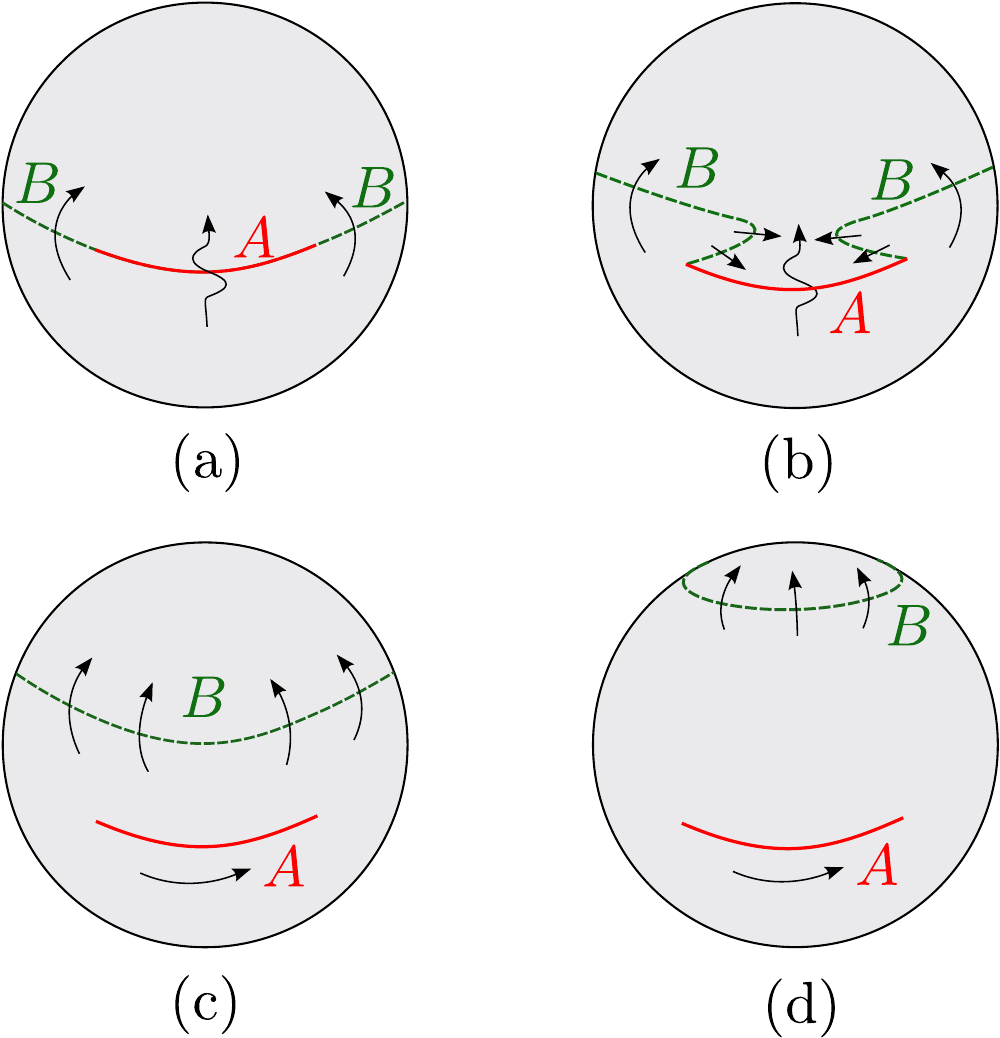}
	\caption{Schematics of the replica manifold computing $	S_n\left(AA^\star\right)_{\psi_m}$ and the reflected entropy of a single interval $A$ at zero temperature. The complex planes are topologically equivalent to spheres and we perform the same deformation procedure as described in \cref{fig:single-T-cylinder-reflected}. In this case, the green dashed line denoting the infinite branch cut upon deformation may be shrunk to a point on the north pole and is therefore eliminated. As a result, the reflected entropy of the single interval at zero temperature is correctly captured by the remaining branch cuts along $A$ which are sewed only in the $n$-direction. }
	\label{fig:single-T-reflected}
\end{figure}

The above discussion requires a critical re-examination of the naive procedure for the configuration of a single interval at zero temperature where this problem did not arise. For this purpose we recapitulate the structure of the replica manifold used to compute the R\'enyi reflected entropy in \cref{fig:single-T-reflected}. Recall that for the zero temperature case the
cylinder in \cref{fig:single-T-cylinder-reflected} has an infinite circumference that renders the geometry to that of a complex plane which is topologically equivalent to a sphere as shown in \cref{fig:single-T-reflected}\textcolor{blue}{(a)}. It is possible to perform a similar deformation of the branch cut along $B$ to superimpose over the interval $A$ as shown in \cref{fig:single-T-reflected}\textcolor{blue}{(b)}. However in this case the auxiliary infinite branch cut shown by green dashed line in \cref{fig:single-T-reflected}\textcolor{blue}{(c)} may be shrunk to a point at the north pole and can thus be eliminated. We are then only left with a branch cut along the subsystem $A$ which connects the replica sheets only in $n$-direction as depicted in \cref{fig:single-T-reflected}\textcolor{blue}{(d)}. This is reminiscent of the fact that the two point function involved in the computation of the reflected entropy of the single interval $A$ involves only the composite twist operators $\sigma_{g^{}_B g_A^{-1}}$ which correspond to hopping through the replica sheets in the $n$-direction. 

From the above discussions, it is evident that the finite temperature reflected entropy between two subsystems cannot be computed by naively mapping from the complex plane to the cylinder if an infinite part of an infinite system is involved
as was also the case for the entanglement negativity described in \cite{Calabrese:2014yza}. Therefore it is required to regularize the infinite branch cut described by the green dashed line in \cref{fig:single-T-cylinder-reflected}\textcolor{blue}{(c)}. We follow the procedure described in \cite{Calabrese:2014yza} in the context of the entanglement negativity for a single interval in a thermal CFT$_2$ by shifting the endpoints of the infinite branch cuts to finite distances.To this end, we introduce two large auxiliary intervals $B_1$ and $B_2$, each of finite length $R$, sandwiching the single interval $A$ in question and focus on the following four-point function on the twisted cylinder
\begin{align}
\left<\sigma_{g_B}(-R)\sigma_{g^{-1}_B g_A^{}}(0)\sigma_{g^{}_B g_A^{-1}}(R_A)\sigma_{g_B^{-1}}(R)\right>_{\mathrm{CFT}_{\beta_{L,R}}^{\bigotimes mn}}\,.\label{4pt-sigle-cyl}
\end{align} 
The R\'enyi reflected entropy between $A$ and $B\equiv B_1\cup B_2$ in the state $\psi_m$ is then obtained as
\begin{equation}
S_n\left(AA^\star\right)_{\psi_m}=\frac{1}{1-n}\log \frac{\left<\sigma_{g_B}(-R)\sigma_{g^{-1}_B g_A^{}}(0)\sigma_{g^{}_B g_A^{-1}}(R_A)\sigma_{g_B^{-1}}(R)\right>_{\mathrm{CFT}_{\beta_{L,R}}^{\bigotimes mn}}}{\bigg( \left<\sigma_{g_m^{}}(-R)\sigma_{g_m^{-1}}(R)\right>_{\mathrm{CFT}_{\beta_{L,R}}^{\bigotimes m}}\bigg)^n}\,,\label{Sn_R-correct-singleT}
\end{equation}
where the subscript $\beta_{L,R}$ denotes that the four point function has to evaluated on a twisted cylinder. To compute the reflected entropy of the single interval $A$, we first compute the above correlation function of twist operators normalized by a similar correlator on the $m$-replica manifold $\mathrm{CFT}_{\beta_{L,R}}^{\bigotimes m}$ and take the replica limits $m,n\to 1$. Subsequently, we take the limit $R \to \infty$ which is tantamount to the bipartite limit $B_1\cup B_2\to A^c$. As we shall see below these two limits do not commute, and we obtain a different expression from the naive one in \cref{S_R-naive}.

\begin{figure}[H]
	\centering
	\includegraphics[scale=.60 ]{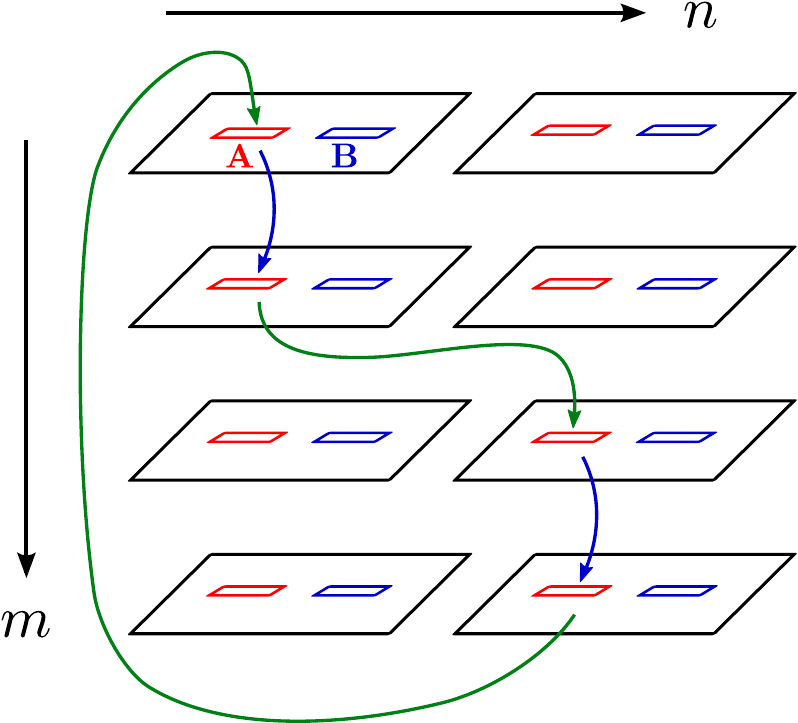}
	\caption{Replica structure corresponding to OPE of $\sigma_{g^{}_A g_B^{-1}}$ and $\sigma_{g^{}_B}$ denoted by green and blue arrows respectively.}
	\label{fig:reflected-replica-composite}
\end{figure}

Utilizing the conformal map in \cref{twisted-cylinder-transf}, we may obtain the four-point twist correlator in \cref{4pt-sigle-cyl} from the corresponding four-point function on the complex plane. However, any four-point function of primary operators on the complex plane involves an arbitrary function of the harmonic ratios $\eta\,,\,\bar {\eta}$. We would like to understand the behaviour of the four-point correlation function in the $s$- and $t$-channels described respectively by $\eta\,,\,\bar {\eta}\to 0$ and $\eta\,,\,\bar {\eta}\to 1$. To see this, we consider the following OPEs between various primaries
\begin{align}
\sigma_{g^{}_A}(z_1) \sigma_{g_A^{-1}} (z_2) = \frac{c_{nm}}{z_{12}^{2 h^{A}_{L}} \bar z_{12}^{2 h^{A}_{R}}}\,\mathbb{I} + \ldots , ~~ \sigma_{g^{-1}_B g_A^{}}(z_1)\sigma_{g^{}_B g_A^{-1}}(z_2) = \frac{\tilde{c}_{nm}}{z_{12}^{2 h_{L}^{BA^{-1}}} \bar z_{12}^{2 h^{BA^{-1}}_{R}}}\,\mathbb{I} + \ldots, ~~ z_1 \to z_2, \label{gAgA_gABgAB}
\end{align}
\begin{align}
\sigma_{g^{}_B}(z_1) \sigma_{g_{B}^{-1} g_A^{}} (z_2) = \frac{C_{B,B^{-1}A,A}}{z_{12}^{2 h^{A}_{L}} \bar z_{12}^{2 h^{A}_{R}}}\,\sigma_{g^{}_A} (z_1) + \ldots , \quad \quad z_1 \to z_2\,.\label{gA-gAB}
\end{align}
where $C_{B,B^{-1}A,A}$ is the corresponding OPE coefficient. While \cref{gAgA_gABgAB} is more or less straightforward to anticipate,  \cref{gA-gAB} requires a little inspection as the actions of $\sigma_{g^{}_B}$ and $\sigma_{g_{B}^{-1} g_A^{}}$ are seemingly independent of each other. One way to verify this is to utilize the following relations for the symmetry group elements $g_A,g_B\in S_{nm}$ \cite{Dutta:2019gen}
\begin{align}
g_A^{} = (\tau_n^{(0)})^{-1} \tau_n^{(m/2)} g_m \quad \, , \quad \quad \quad g_B^{} = g_m \quad \, , \quad \quad \quad g_{B}^{-1} g^{}_{A} = (\tau_n^{(0)})^{-1} \tau_n^{(m/2)}\,,
\end{align}
where $\tau_n^{(k)}$ are the elements of the replica symmetry group $S_{nm}$ which permutes the $m=k$-th replica sheet in the $n$-direction and $g_m$ denotes the full $m$-cyclic permutation.  
The OPE in \cref{gA-gAB} may also be visualized from the sewing procedure in the replica geometry as shown in \cref{fig:reflected-replica-composite}. Now, utilizing the structures of the OPEs in \cref{gAgA_gABgAB,gA-gAB}
it is possible to fix the form of the four point twist correlator on the complex plane with well defined cluster properties in the $s$ and $t$-channels respectively. Finally, the four point twist correlator of the twist fields $\sigma_g$ on the CFT$_2$ plane is given by
\begin{align}
\Big<\sigma_{g_B}(z_1)\sigma_{g^{-1}_B g_A^{}}(z_2)&\sigma_{g^{}_B g_A^{-1}}(z_3)\sigma_{g_B^{-1}}(z_4)\Big>_{\mathrm{CFT}^{\bigotimes mn}}\notag \\&=k_{mn}\left(\frac{1}{z_{14}^{2h^{B}_{L}}z_{23}^{2h_{L}^{{B} A^{-1}}}}\frac{\mathcal{G}_{mn}(\eta)}{\eta^{h_L^{{B} A^{-1}}}}\right)\left(\frac{1}{\bar{z}_{14}^{2h^{B}_{R}}\bar{z}_{23}^{2h_R^{{B} A^{-1}}}}\frac{\bar{\mathcal{G}}_{mn}(\bar{\eta})}{\bar{\eta}^{h_R^{{B} A^{-1}}}}\right), 
\end{align}
where $\eta=\frac{z_{12}z_{34}}{z_{13}z_{24}}$ and $\bar{\eta}=\frac{\bar{z}_{12}\bar{z}_{34}}{\bar{z}_{13}\bar{z}_{24}}$ are the cross ratios.
The non universal arbitrary functions $\mathcal{G}_{mn}(\eta)$ and $\bar{\mathcal{G}}_{mn}(\bar{\eta})$ at the limits $\eta,\bar{\eta}\to 1$ and $\eta,\bar{\eta}\to 0$ may then be determined from the OPEs in \cref {gAgA_gABgAB,gA-gAB}  as
\begin{equation}
\mathcal{G}_{mn}(1)=\bar{\mathcal{G}}_{mn}(1)=1, \hspace{5mm} \mathcal{G}_{mn}(0)=\bar{\mathcal{G}}_{mn}(0)=C_{mn},
\end{equation}
where $C_{mn}$ is a non universal constant depending upon the full operator content of the theory.

We now utilize the conformal map from the CFT$_2$ plane to the twisted cylinder using eq. (\ref{twisted-cylinder-transf}) to express the four point function on the twisted cylinder in the following way
\begin{equation}
\begin{aligned}
&\left<\sigma_{g_B}(-R)\sigma_{g^{-1}_B g_A^{}}(0)\sigma_{g^{}_B g_A^{-1}}(R_A)\sigma_{g_B^{-1}}(R)\right>_{\mathrm{CFT}_{\beta_{L,R}}^{\bigotimes mn}} \\
&\qquad\qquad\qquad=k_{mn}\left[ \frac{\beta_L}{\pi}\sinh\left(\frac{2\pi R}{\beta_L}\right) \right]^{-2h^{B}_{L}} \left[ \frac{\beta_L}{\pi}\sinh\left(\frac{\pi R_A}{\beta_L}\right) \right]^{-2h_L^{{B} A^{-1}}}\frac{{\mathcal{G}}_{mn}({\xi})}{\xi^{h_L^{{B} A^{-1}}}} \\
&
\qquad\qquad\qquad\qquad \times \left[ \frac{\beta_R}{\pi}\sinh\left(\frac{2\pi R}{\beta_R}\right) \right]^{-2h^{B}_{R}} \left[ \frac{\beta_R}{\pi}\sinh\left(\frac{\pi R_A}{\beta_R}\right) \right]^{-2h_R^{{B} A^{-1}}}\frac{\bar{\mathcal{G}}_{mn}(\bar{\xi}\,)}{\bar{\xi}^{\,h_R^{{B} A^{-1}}}}\,,\label{4pt-singleT}
\end{aligned}
\end{equation}
where $\xi\,,\,\bar \xi$ are the finite temperature cross-ratios, defined in \cref{crossratio-T}. In the bipartite limit $R \to \infty$, these are given as 
\begin{equation}
\lim_{R\to \infty} \xi= e^{-\frac{2\pi R_A}{\beta_L}}, \,\,\,
\lim_{R\to \infty} \bar{\xi}= e^{-\frac{2\pi R_A}{\beta_R}}.\label{cross-bipartite}
\end{equation}
Note that, for finite $n,m$, if we take the bipartite limit $R\to\infty$, the four-point correlator in \cref{4pt-singleT} vanishes identically. Therefore, one must take the replica limit prior to the bipartite limit as anticipated earlier.
Now using \cref{Sn_R-correct-singleT,4pt-singleT} and taking the bipartite limit $R \to \infty$ subsequent to the replica limit $n\to 1, m\to 1$, the reflected entropy for the single interval at a finite temperature and non zero angular potential may be obtained as

\begin{equation}
\begin{aligned}
S_R(A:B)&=\frac{c_L}{3} \log \left [ \frac{\beta_L}{\pi\epsilon} \sinh \frac{\pi R_A}{\beta_L} \right ] + \frac{c_R}{3} \log \left [ \frac{\beta_R}{\pi\epsilon} \sinh \frac{\pi R_A}{\beta_R} \right ] - \frac{c_L}{3} \frac{\pi R_A}{\beta_L} - \frac{c_R}{3} \frac{\pi R_A}{\beta_R}\\
&\quad+g\left(e^{-\frac{2\pi R_A}{\beta_L}}\right)+\bar{g}\left(e^{-\frac{2\pi R_A}{\beta_R}}\right)+\text{const.}\,.\label{S_R-correct-singleT}
\end{aligned}
\end{equation}
Here we have restored the UV cut-off $\epsilon$, and the arbitrary functions $g(\xi)$ and $\bar{g}(\bar{\xi})$ describing the non universal contributions are given by
\begin{equation}
g(\xi)=\lim_{n,m \to 1}\log [{\mathcal{G}}_{mn}({\xi})], \, \, \, \, 
\bar{g}(\bar{\xi})=\lim_{n,m \to 1}\log [\bar{\mathcal{G}}_{mn}(\bar{\xi})]\,.
\end{equation}
The expression in \cref{S_R-correct-singleT} is indeed different from the naive result in \cref{S_R-naive}. One interesting feature of the formula \eqref{S_R-correct-singleT} is that the linear terms proportional to the temperatures exactly cancel the high temperature divergences in \cref{S_R-naive} rendering the reflected entropy of the single interval in question finite but small at very high temperatures. Note that the reflected entropy is now dependent on the full operator content of the specific field theory under consideration through the non-universal functions $g$ and $\bar g$ whose large central charge behaviour may be extracted through the semi-classical monodromy techniques described in \cite{Fitzpatrick:2014vua,Malvimat:2017yaj,Kulaxizi:2014nma}. We leave a more careful analysis of the large central charge structure of the conformal block for future.


\section{Entanglement negativity from holographic duality}
\label{sec:EN_geod}

Having completed the field theoretic analysis of the entanglement structure for bipartite mixed states in CFT$_2^a$s, we now advance a holographic construction for the entanglement negativity in the context of the AdS/CFT correspondence for dual conformal field theories with a gravitational anomaly (CFT$^a_2$s). In this case the dual geometry is described by topologically massive gravity (TMG) in  a bulk AdS$_3$ spacetime \cite{Kraus:2005zm, Castro:2014tta}. In what follows we propose specific holographic prescriptions involving the bulk geometry described above, for the entanglement negativity of various bipartite states in the dual CFT$^a_2$s.

\subsection{Review of the setup and basic definitions}\label{sec:TMG_review}
In this subsection we briefly recapitulate the essential features of the holographic correspondence in the context of Topologically Massive Gravity (TMG) in AdS$_3$ which will be henceforth termed as TMG-AdS$_3$ where the dual conformal field theory CFT$^a_2$ admits a gravitational anomaly. The bulk action for TMG in AdS$_3$ is given by a sum of the usual Einstein-Hilbert action with the gravitational Chern-Simons (CS) term as follows \cite{Castro:2014tta, Jiang:2019qvd, Gao:2019vcc, Tachikawa:2006sz}
\begin{equation} \label{action_TMG}
S=\frac{1}{16 \pi G_N}\left[ \int \textrm{d$^3$}x\sqrt{-g}\left(R+\frac{2}{\ell^2}\right) - \frac{1}{2\mu} \int \textrm{Tr}\left(\mathbf{\Gamma} \wedge \textrm{d}\mathbf{\Gamma} +\frac{2}{3} \mathbf{\Gamma} \wedge \mathbf{\Gamma} \wedge \mathbf{\Gamma} \right) \right]\,,
\end{equation}
where the matrix-valued one-form $\mathbf{\Gamma}^{\mu}_{\nu} =\Gamma_{\rho\nu}^{\mu}dx^{\rho}$ defines the gravitational connection and $\Lambda =-\frac{2}{\ell^2}$ is the negative cosmological constant for AdS$_3$ with a 
radius $\ell$. The mass-dimension one real constant $\mu$ describes the coupling of the CS term with the Einstein-Hilbert action and the (covariant) equations of motion for the above action is given as \cite{Castro:2014tta,Jiang:2019qvd}
\begin{equation} \label{eom_TMG}
R_{\mu \nu}-\frac{1}{2}g_{\mu \nu} \left( R+\frac{2}{l^2} \right) = -\frac{1}{\mu} C_{\mu \nu},
\end{equation}
where $C_{\mu \nu}$ is the Cotton tensor \cite{Castro:2014tta,Jiang:2019qvd}. Remarkably, for a vanishing Cotton tensor the theory still admits Einstein like metrics and therefore such solutions are always locally AdS$_3$. In this article, we restrict ourselves to such locally AdS$_3$ solutions for which the Brown-Henneaux symmetry analysis leads to two copies of the Virasoro algebra with
central charges \cite{Hotta:2008yq,Compere:2008cv}
\begin{equation} \label{central_charges}
\begin{aligned}
c_L = \frac{3 l}{2 G_N} \left(1+\frac{1}{\mu} \right), \;\;\;  c_R = \frac{3 l}{2 G_N} \left(1-\frac{1}{\mu} \right)\,.
\end{aligned}
\end{equation}
This clearly indicates that the corresponding dual conformal field theory CFT$^a_2$ admits a gravitational anomaly.

As described in \cite{Castro:2014tta,Jiang:2019qvd}, for locally AdS$_3$ solutions to TMG, the holographic principle dictates that the primary operators in the CFT$^a_2$ correspond to massive spinning particles propagating along extremal worldlines in the bulk geometry. The on-shell action for such a particle of mass $m$ and spin $s$ is given by \cite{Castro:2014tta}
\begin{equation}\label{On-shell}
S_\textrm{on-shell}=\int_{\mathcal{C}} \textrm{d} \tau \left( m\,\sqrt{g_{\mu \nu} \dot X ^\mu \dot X ^\nu}+s\,\tilde{n}\cdot\nabla n\right)+S_\textrm{constraints}\,,
\end{equation}
where $\tau$ parametrizes the length along the worldline $\mathcal{C}$ of the particle, $\tilde{n}$ and $n$ are unit space-like and time-like vectors respectively, both normal to the trajectory of the particle $X ^\mu$, and $S_\textrm{constraints}$ is an action imposing these constraints through appropriate Lagrange multipliers \cite{Castro:2014tta}. These constraints leads to orthonormal triads  of the vectors $(\dot X,n,\tilde{n})$ at each point of the bulk spacetime which renders the worldlines to the shape of ribbons. The motion of such massive spinning particles is described by the Mathisson-Papapetrou-Dixon (MPD) equations which follow from the extremization of the above on-shell action \cite{Castro:2014tta, Gao:2019vcc}. Although the local minimum or the saddle point of the worldline action \cref{On-shell} is not necessary a geodesic, in locally AdS spacetimes geodesics still form one simple class of solutions to the MPD equations. In the following we will restrict to such solutions in the TMG background where such massive spinning particles moving in locally AdS spacetimes follow the geodesics.

In order to set up the holographic computations for the entanglement measures in locally AdS$_3$ spacetimes described by TMG, we first consider the phase space of AdS$_3$ solutions in the light-cone coordinates \cite{Gao:2019vcc, Jiang:2019qvd}\footnote{Note that the radial coordinate in \cite{Jiang:2019qvd} is related to the holographic coordinate $\rho$ in the present formulation as $\rho=r+\frac{T_u^2T_v^2}{4r}$.}
\begin{align}
\textrm{d}s^2= \frac{\textrm{d}\rho^2}{4(\rho^2-T_u^2T_v^2)}+2\rho\, \textrm{d}u \, \textrm{d}v + T_u^2 \, \textrm{d}u^2 + T_v^2 \, \textrm{d}v^2 \,,\label{BTZmetric}
\end{align}
with the identifications $u\sim u+2\pi$, $v\sim v+2\pi$ and the AdS$_3$ radius $l=1$. The $T_u, T_v$ in the above equation are parameters and in these coordinates the factorization of the bulk left moving and the right moving sectors described by the null coordinates $u,v$ is manifest. The case of the Poincar\'e AdS$_3$ may be obtained from the above metric by setting $T_u=T_v=0$, namely \cite{Jiang:2019qvd}:
\begin{align}
\textrm{d}s^2= \frac{\textrm{d}\rho^2}{4\rho^2}+2\rho \, \textrm{d}u \, \textrm{d}v \,. \label{Poincare_metric}
\end{align}
Similarly, the BTZ black hole may be obtained by identifying $T_u\,,\, T_v$ with the left and right moving temperatures in the corresponding dual CFT$^a_2$. In the following we will focus on the case of the Poincar\'e AdS$_3$ for brevity and postpone the discussion of the BTZ black hole till subsection \ref{sec:HEE}. 

In the above light-cone coordinates, a geodesic curve connecting two points on the asymptotic boundary ($\rho \to \infty $) with the coordinates  $\left(-\frac{\Delta u}{2},-\frac{\Delta v}{2},\infty\right)$ and $\left(\frac{\Delta u}{2},\frac{\Delta v}{2},\infty\right)$ admits of the following parametrization \cite{Gao:2019vcc}
\begin{align}
& u(\tau)=\frac{\Delta u}{2}\tanh\left(\tau+\frac{1}{2}\log(\Delta u\Delta v)\right)\,,\notag\\
& v(\tau)=\frac{\Delta v}{2}\tanh\left(\tau+\frac{1}{2}\log(\Delta u\Delta v)\right)\,,\notag\\
& \rho(\tau)=\frac{1}{2}\left(e^{\tau}+\frac{e^{-\tau}}{\Delta u\Delta v}\right)^2\,,\label{Geod_para}
\end{align}
where $\tau$ parametrizes the proper length along the geodesic. The tangent vector to the geodesic may be written as the unit vector along the $\tau$-direction \cite{Gao:2019vcc} as
\begin{align}
\dot{X}\equiv\partial_{\tau}=\frac{1/\Delta v}{\rho}\partial_u+\frac{1/\Delta u}{\rho}\partial_v+\frac{4u\rho}{\Delta u}\partial_{\rho}\label{Tangent}
\end{align}
As described earlier, for a massive spinning particle propagating in the bulk TMG-AdS$_3$ spacetime the worldline action in \cref{On-shell} consists of two parts. The first part consists of the usual geodesic length describing the intrinsic properties of the bulk which is obtained from the normalization of the tangent vector \cref{Tangent}, $\dot{X}^2=1$, indicating that the worldline of the particle has the trivial metric induced from AdS$_3$. The second part comprises of the Chern-Simons contribution due to the spin of the particle and this quantifies the extrinsic properties of the worldline. Such extrinsic properties are essentially described in terms of two mutually orthogonal vectors $n$ and $\tilde{n}$ normal to the worldline. The extrinsic curvature and torsional properties may then be studied through the change of the normal frame $(\dot{X},n,\tilde{n})$ as the worldline is traversed.

A particularly useful parametrization of the bulk vectors normal to the geodesic described by \cref{Geod_para} was given in \cite{Gao:2019vcc}. At the two endpoints of the geodesic, the boundary value of the normal vector $n$ is given by\footnote{Note that, in $(2+1)$-dimensions the other normal vector may be determined as $\tilde{n}^{\mu}= \epsilon^{\mu \nu \rho} \dot{X}_{\nu} n_{\rho}$.}
\begin{align} \label{Normal_vectors}
n_b=\pm\frac{\Delta u}{\Delta v\sqrt{2\rho_{\infty}}}\partial_u\mp\frac{\Delta v}{\Delta u\sqrt{2\rho_{\infty}}}\partial_v\,,
\end{align}
where the up sign corresponds to the left part of the geodesic with $u < 0$, and the down sign corresponds to the right part with $u > 0$, and $\rho_{\infty}$ denotes the value of the holographic coordinate at the boundary, which is UV-divergent. The specific form of these normal vectors may be determined uniquely in the following way. One first considers a parallel transported normal frame $(q,\tilde{q})$ along the worldline $\mathcal{C}$ of the particle and sets up the boundary values of the normal vector $n$ from the boundary CFT data. Finally, the actual normal vector $n$ satisfying the boundary conditions can be found through a local Lorentz rotation of the parallel transported frame. The above boundary values specify the gauge choice corresponding to the local $SO(1,1)$ rotation of the normal frame. 
\subsection{Holographic entanglement entropy in TMG-AdS$_3$} \label{sec:HEE}
In the framework of the AdS/CFT correspondence, the holographic entanglement entropy of a subsystem in the dual field theory is computed via the notion of generalized gravitational entropy \cite{Lewkowycz:2013nqa}. In this context, one performs a replication of the dual gravitational theory defined on a replica manifold $\mathcal{M}_n$ and subsequently takes the orbifold geometry $\mathcal{M}_n/\mathbf{Z}_n$ by quotienting with the $\mathbf{Z}_n$ replica symmetry. Note that this replication of the bulk is reminiscent of a similar replication of the dual field theory at the boundary of the spacetime which serves as a boundary condition to the gravitational equations of motion. In the quotient geometry $\mathcal{M}_n/\mathbf{Z}_n$, there are conical defects on the entangling surface at the boundary of the subsystem under consideration. As described earlier in subsection \ref{sec:EE_field}, in the AdS$_3$/CFT$_2$ setting, one places twist operators at the endpoints $\partial_iA$ of the boundary interval $A$ and the entanglement entropy of the subsystem is computed through the correlation function of such twist operators. In the setup of TMG in AdS$_3$, these twist operators correspond to bulk massive spinning particles of mass $m_n=\Delta_n$ and spin $s_n$ (cf. \cref{mass_spin}) moving on extremal worldlines. Utilizing the construction described in \cite{Castro:2014tta}, the two-point twist correlator may be computed in terms of the on-shell action of such massive spinning particles in the bulk, as
\begin{equation} \label{twist_correlator}
\left<\Phi_{n}(\partial_1 A) \Phi_{-n} (\partial_2 A)\right> \sim e^{- \Delta_n S_\textrm{on-shell}^\textrm{EH} - s_n \,S_\textrm{on-shell}^\textrm{CS} }\,,
\end{equation}
where $S_\textrm{on-shell}^\textrm{EH}$ and $S_\textrm{on-shell}^\textrm{CS}$ denote the on-shell actions corresponding to the Einstein-Hilbert and the Chern-Simons contributions respectively. Now using \cref{Tr_rho,S_EE}, the modified HRT formula for the entanglement entropy may be obtained as follows
\begin{equation} \label{HRT_modified}
S_{\textrm{HEE}}=\underset{\mathcal{C}}{\text{min\,\,ext}} \, \frac{\mathcal{L}_\mathcal{C}}{4 G_N} \equiv \underset{\mathcal{C}}{\text{min\,\,ext}}\, \frac{1}{4 G_N} \left( L_{\mathcal{C}} + \frac{\mathcal{T}_{\mathcal{C}}}{\mu} \right),
\end{equation}
where the extremization prescription renders the particle worldline $\mathcal{C}$ on-shell and $\mu$ describes the coupling of the CS term with the Einstein-Hilbert action. In \cref{twist_correlator} the length of the geodesic $L_{\mathcal{C}}$ and the twist $\mathcal{T}_{\mathcal{C}}$ in the ribbon-shaped worldline is given respectively by the Einstein-Hilbert and the Chern-Simons contribution to the on shell action, as \cite{Castro:2014tta,Gao:2019vcc}
\begin{align} 
&L_{\mathcal{C}}\equiv S_\textrm{on-shell}^\textrm{EH} = \int_{\mathcal{C}} \textrm{d} \tau \sqrt{g_{\mu \nu} \dot X ^\mu \dot X ^\nu} ,\notag\\
&\mathcal{T}_{\mathcal{C}}\equiv S^\textrm{CS}_\textrm{on-shell}=\int_{\mathcal{C}} \textrm{d} \tau \, \tilde{n}.\nabla n = \log \left[ \frac{q(\tau_f).n_f - \tilde q (\tau_f).n_f}{q(\tau_i).n_i - \tilde q (\tau_i).n_i} \right]\,,\label{L&T}
\end{align}
where $\tau$ parameterizes the proper length along the geodesic, $n_i\,,\, n_f$ defines the boundary values of the normal vector $n$ while $(q (\tau_i),\tilde q (\tau_i))$ and $(q (\tau_f),\tilde q (\tau_f))$ determines the initial and final parallel transported frame at the boundary.
\begin{figure}[h!]
	\centering
	\includegraphics[scale=0.8]{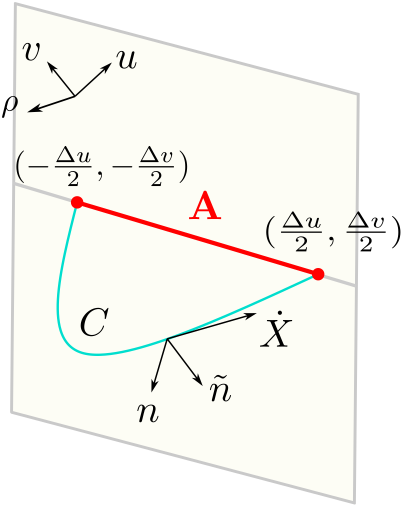}
	\caption{Extremal curve homologous to an interval $A=\left[\left(-\frac{\Delta u}{2},-\frac{\Delta v}{2}\right),\left(\frac{\Delta u}{2},\frac{\Delta v}{2}\right)\right]$ in a CFT$_{2}^a$ dual to topologically massive gravity in asymptotically AdS$_3$ spacetime. The normal frame formed by the vectors $(\dot{X},n,\tilde{n})$ gives rise to a sense of direction at every point on this extremal curve rendering it to be ribbon-shaped.  Figure modified from \cite{Jiang:2019qvd}.}
	\label{fig:Single-EE}
\end{figure}

In the following, we briefly review the computations of the holographic entanglement entropy for a single interval in the dual field theory utilizing the frameworks described in \cite{Castro:2014tta,Gao:2019vcc}. To this end, consider a boosted interval $A$ of length $R_A$ and boost parametrized by the hyperbolic boost angle $\kappa_A$ in the CFT$_2^a$ in the ground state dual to the Poincar\'e TMG-AdS$_3$ spacetime. In the symmetric setup with the interval $A=\left[\left(-\frac{\Delta u}{2},-\frac{\Delta v}{2}\right),\left(\frac{\Delta u}{2},\frac{\Delta v}{2}\right)\right]$ as depicted in \cref{fig:Single-EE}, one may choose the parallel transported vectors\footnote{Note that these normal vectors are different from those used in \cite{Castro:2014tta}. This is due to the fact that \cite{Gao:2019vcc} utilizes a different gauge choice than those made in \cite{Castro:2014tta}.} to be \cite{Gao:2019vcc}
\begin{align} 
&q = \pm \sqrt{\frac{\Delta u}{2\rho \Delta v }} \partial_u \mp \sqrt{\frac{\Delta v}{2 \rho\Delta u }} \partial_v\,,\notag\\
&\tilde{q} = - \frac{1}{\sqrt{2\rho u^2+\frac{\Delta u}{\Delta v}}}\left(u\sqrt{\frac{\Delta u}{\Delta v}} \partial_u - u\sqrt{\frac{\Delta v}{\Delta u}} \partial_v + 2\rho \sqrt{\frac{\Delta v}{\Delta u}} \partial_\rho\right), \label{q_boosted}
\end{align}
where, once again, the up sign corresponds to the left half of the geodesic with $u < 0$ and the down sign corresponds to the right half of the geodesic with $u > 0$. It is easy to check that the above parametrization satisfies the constraint equations \cite{Gao:2019vcc}
\begin{align}
q^2=-1~~,~~\tilde{q}^2=1~~,~~q\cdot\tilde{q}=q\cdot\dot{X}=\tilde{q}\cdot\dot{X}=0\,.
\end{align}
Now utilizing the boundary value of the true normal vector $n$ from \cref{Normal_vectors} as well as the auxiliary parallel transported vectors in \cref{q_boosted}, the extremal length $L_A$ and the twist $\mathcal{T}_A$ of the worldline homologous to the boosted interval $A$ in question may be obtained from \cref{L&T} to be 
\begin{subequations} \label{L&T_Boosted}
	\begin{equation} \label{L_A}
	L_A = 2 \, \log \frac{R_A}{\epsilon},
	\end{equation}
	\begin{equation} \label{T_A}
	\mathcal{T}_A = 2 \, \kappa_A,
	\end{equation}
\end{subequations}
where $\epsilon = 1 / (2 \rho_\infty)$ is a UV cut-off of the dual CFT$_2^a$. The fact that these results are exactly the same as those obtained in \cite{Castro:2014tta} should come as no surprise, since the final result for the holographic entanglement entropy should be independent of the gauge choice made. The holographic entanglement entropy for the single boosted interval is then obtained using \cref{HRT_modified} as
\begin{equation} \label{EE_sing_geod}
\begin{aligned}
S_A &= \frac{1}{2 G_N} \log \frac{R_A}{\epsilon} + \frac{1}{2 \mu G_N} \kappa_A \\
&=\frac{c_L+c_R}{6}\log \frac{R_A}{\epsilon} - \frac{c_L-c_R}{6} \kappa_A,
\end{aligned}
\end{equation}
where the Brown-Henneaux central charges given in \cref{central_charges} have been used in the last equality. This expression matches exactly with the field theory computations in \cite{Castro:2014tta}, reviewed in subsection \ref{sec:EE_field_zeroT}.

Next we move to the computation of the holographic entanglement entropy for a single interval $A$ of length $R_A$ in a thermal CFT$^a_2$ as described in \cite{Castro:2014tta} utilizing the setup of \cite{Gao:2019vcc}.  The bulk dual for such CFT$^a_2$s with inverse temperatures for the left and the right moving modes given by $\beta_L$ and $\beta_R$, is described by rotating BTZ black holes in TMG with the metric given in \cref{BTZmetric}. Similar to the zero temperature case, one may again introduce two bulk orthogonal vectors $n$ and $\tilde n$ at each bulk point normal to the worldline \cite{Castro:2014tta,Gao:2019vcc} using the parallel transported normal frame $(q, \tilde q)$. Subsequently, utilizing these vectors the length $L_A$ and the twist $\mathcal{T}_A$ of the geodesic worldline homologous to the interval $A$ may be computed using \cref{L&T} as follows \cite{Castro:2014tta}
\begin{subequations} \label{L&T_T}
	\begin{equation} \label{L_A_sing_T}
	L_A = \log \left( \frac{\beta_L \beta_R}{\pi^2 \epsilon^2} \sinh \frac{\pi R_A}{\beta_L} \sinh \frac{\pi R_A}{\beta_R} \right),
	\end{equation}
	\begin{equation} \label{T_A_sing_T}
	\mathcal{T}_A = \log \left( \frac{\beta_R \sinh \frac{\pi R_A}{\beta_R}}{\beta_L \sinh \frac{\pi R_A}{\beta_L}} \right).
	\end{equation}
\end{subequations}
The holographic entanglement entropy for the single interval in question may then by obtained using \cref{HRT_modified} to be \cite{Castro:2014tta}
\begin{equation} \label{EE_sing_T_geod}
\begin{aligned}
S_A &= \frac{1}{4 G_N} \log \left( \frac{\beta_L \beta_R}{\pi^2 \epsilon^2} \sinh \frac{\pi R_A}{\beta_L} \sinh \frac{\pi R_A}{\beta_R} \right) + \frac{1}{4 \mu G_N} \log \left( \frac{\beta_R \sinh \frac{\pi R_A}{\beta_R}}{\beta_L \sinh \frac{\pi R_A}{\beta_L}} \right) \\
&= \frac{c_L}{6} \log \left [ \frac{\beta_L}{\pi \epsilon} \sinh \frac{\pi R_A}{\beta_L} \right ] + \frac{c_R}{6} \log \left [ \frac{\beta_R}{\pi \epsilon} \sinh \frac{\pi R_A}{\beta_R} \right ],
\end{aligned}
\end{equation}
where in the last equality the Brown-Henneaux central charges in \cref{central_charges} has been utilized. The above expression matches with the corresponding field theory result eq. \eqref{entropy-single-finite} obtained in \cite{Castro:2014tta}.


\subsection{Holographic entanglement negativity for two disjoint intervals} \label{sec:EN_disj_geod}

As discussed earlier, the entanglement entropy fails to be a viable entanglement measure for bipartite mixed states and it is required to consider alternate entanglement measures for their characterization. In this context as indicated in previous sections the entanglement negativity serves as a convenient computable measure for the characterization of  mixed state entanglement and it was possible to compute this quantity directly for bipartite mixed states in CFT$^a_2$ described in subsection \ref{sec:negativity_field}. In this subsection we address the significant issue of the holographic characterization of the entanglement negativity for such conformal field theories through the framework of the TMG-AdS$_3$/CFT$_2^a$ 
correspondence.

We begin with the bipartite mixed state of two disjoint intervals in close proximity in CFT$^a_2$s  dual to (2+1)-dimensional bulk TMG-AdS$_3$ spacetimes. In this context we consider two disjoint intervals given by $A = [z_1,z_2]$ and $B = [z_3, z_4]$ in such dual CFT$^a_2$s. As described in subsection \ref{sec:disj_Neg_field} the relevant four-point twist correlator may be expressed in terms of the conformal cross-ratios. In the large central charge limit this four point correlator is then given as in \cref{four-point-limit-disj} using the monodromy analysis. From the right-hand-side of  \cref{four-point-limit-disj} using the definition of the two-point function in \cref{Tr_rho}, we observe that the four-point correlator may be factorized in the large central charge limit in terms of certain two-point twist correlators as follows
\begin{equation}
\begin{aligned}
\left< \Phi_{n_e}(z_1) \Phi_{-n_e}(z_2) \Phi_{-n_e}(z_3) \Phi_{n_e}(z_4) \right> &= \frac{ \left< \Phi_{n_e / 2}(z_1) \Phi_{- n_e / 2}(z_3) \right> \left< \Phi_{n_e / 2}(z_2) \Phi_{- n_e / 2}(z_4) \right>}{\left< \Phi_{n_e / 2}(z_1) \Phi_{- n_e / 2}(z_4) \right> \left< \Phi_{n_e / 2}(z_2) \Phi_{- n_e / 2}(z_3) \right> }\\
&\qquad\qquad\qquad\qquad\qquad\qquad\qquad+ \mathcal{O} \left( \frac{1}{c_L}, \frac{1}{c_M} \right).
\end{aligned}
\end{equation}
Subsequently using the modified holographic dictionary given in \cref{twist_correlator,L&T} and in the replica limit of $n_e \to 1$, we obtain the holographic entanglement negativity for two disjoint intervals in proximity in the following form
\begin{equation} \label{EN_disj_conjecture}
\mathcal{E}(A:B)=\frac{3}{16 G_N} \left( \mathcal{L}_{A \cup C} + \mathcal{L}_{B \cup C} - \mathcal{L}_{A \cup B \cup C} - \mathcal{L}_C \right),
\end{equation}
where $\mathcal{L}_X$ corresponds to interval $X$ in the dual field theory and is as defined in \cref{HRT_modified}. It is interesting to note that, similar to the AdS$_3$/CFT$_2$ case as described in \cite{Malvimat:2018ood, Malvimat:2018txq}, the above mentioned proposal for the holographic entanglement negativity for two disjoint intervals may be expressed in terms of the holographic mutual information on utilizing the HRT formula in \cref{HRT_modified}, as 
\begin{align} \label{EN_disj_mutual}
	\mathcal{E}(A:B)=\frac{3}{4} \Big(\mathcal{I}(A \cup C: B) - \mathcal{I}(B:C) \Big).
\end{align}
In the following subsections we will utilize the above holographic proposal in \cref{EN_disj_conjecture} to obtain the holographic entanglement negativity for two disjoint intervals in proximity in CFT$_2^a$s at zero temperature as well at finite temperature dual to TMG-AdS$_3$ geometries.
\begin{figure}[h!]
	\centering
	\includegraphics[scale=0.7]{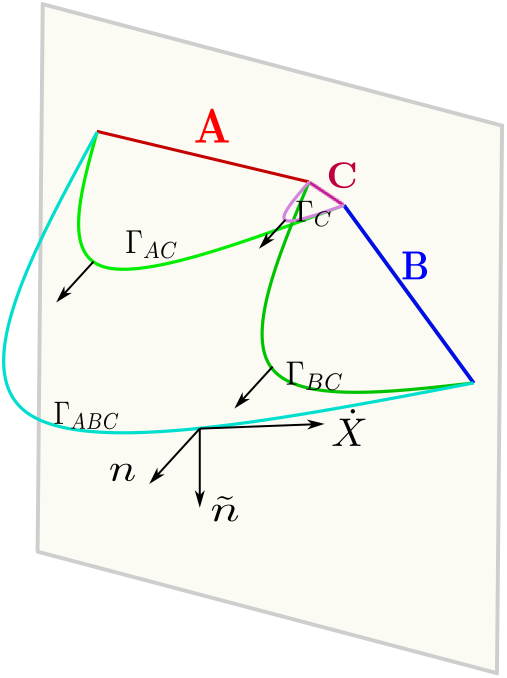}
	\caption{Schematics of the holographic construction for computing the entanglement negativity for two disjoint intervals $A$ and $B$ in a CFT$_2^a$ dual to topologically massive gravity in asymptotically  AdS$_3$ spacetimes. The normal frames to the extremal curves are depicted through the black arrows.}
	\label{disjTMGgeod}
\end{figure} 

\subsubsection{Poincar\'e TMG-AdS$_3$} \label{sec:EN_disj_boosted_geod}
In this subsection we consider two disjoint boosted intervals $A$ and $B$ with lengths $R_A$ and $R_B$ and boosts $\kappa_A$ and $\kappa_B$ respectively in a CFT$_2^a$ in its ground state dual to a bulk TMG-AdS$_3$ spacetime as depicted in \cref{disjTMGgeod}. The interval separating $A$ and $B$ is labelled $C$ here with a length $R_C$ and boost $\kappa_C$. The lengths and twists of geodesic worldlines homologous to these intervals in the dual field theory are given in \cref{L&T_Boosted}. Utilizing these expressions for the lengths and twists in our proposal described in \cref{EN_disj_conjecture}, we may obtain the holographic entanglement negativity for the mixed state configuration of the two disjoint intervals (in proximity)  in question as
\begin{equation} \label{EN_disj_geod}
\begin{aligned}
\mathcal{E}(A:B) &= \frac{3}{8 G_N} \left[ \log \frac {R_{AC} R_{BC}} {R_{ABC} R_C} + \frac{1}{\mu} \left( \kappa_{AC} + \kappa_{BC} - \kappa_{ABC} - \kappa_C \right) \right]\\
&=\frac{c_L+c_R}{8} \log \frac {R_{AC} R_{BC}} {R_{ABC} R_C} - \frac{c_L-c_R}{8} \left( \kappa_{AC} + \kappa_{BC} - \kappa_{ABC} - \kappa_C \right),
\end{aligned}
\end{equation}
where $(R_{AC},\kappa_{AC})$, $(R_{BC},\kappa_{BC})$ and $(R_{ABC},\kappa_{ABC})$ correspond to the lengths and the boosts for intervals $A \cup C$, $B \cup C$ and $A \cup B \cup C$ respectively in the dual CFT$_2^a$. We have also used the Brown-Henneaux central charges given in \cref{central_charges} in the last equality above. Note that the above result is cut-off independent and similar to the results described in \cite{Malvimat:2018ood, Malvimat:2018txq} for the usual AdS/CFT framework in the absence of any anomaly and in \cite{Basu:2021awn, Basu:2021axf} in the context of flat-space holography. Interestingly our result matches exactly with the universal part of the corresponding field theory result in \cref{neg-disj-zero} in the large central charge limit which serves as a strong consistency check for our proposal.

\subsubsection{Rotating BTZ black holes} \label{sec:EN_disj_T_geod}

We now consider two disjoint intervals $A$ and $B$ of lengths $R_A$ and $R_B$ with an interval $C \subseteq (A \cup B)^c$ of length $R_C$ separating $A$ and $B$ in a thermal CFT$^a_{2}$ defined on twisted cylinders of circumferences $\beta_L$ and $\beta_R$. The corresponding bulk dual for this mixed state configuration in the thermal CFT$^a_2$ is described by a rotating planar BTZ black hole in TMG-AdS$_3$ spacetime. As in the previous subsection we obtain the holographic entanglement negativity for this mixed state configuration using the length and the twist of the geodesic worldline homologous to an interval in a thermal CFT$^a_2$ given in \cref{L&T_T} and utilizing our proposal in \cref{EN_disj_conjecture} as
\begin{equation} \label{EN_disj_T_geod}
\begin{aligned}
\mathcal{E}(A:B) = & \frac{c_L}{8} \log \left( \frac {\sinh \frac{\pi R_{AC}}{\beta_L} \sinh \frac{\pi R_{BC}}{\beta_L}} {\sinh \frac{\pi R_{ABC}}{\beta_L} \sinh \frac{\pi R_{C}}{\beta_L}} \right) + \frac{c_R}{8} \log \left( \frac {\sinh \frac{\pi R_{AC}}{\beta_R} \sinh \frac{\pi R_{BC}}{\beta_R}} {\sinh \frac{\pi R_{ABC}}{\beta_R} \sinh \frac{\pi R_{C}}{\beta_R}} \right),
\end{aligned}
\end{equation}
where $R_{AC}$, $R_{BC}$ and $R_{ABC}$ correspond to the length of the intervals $A \, \cup \, C$, $B \cup C$ and $A \cup B \cup C$ in the dual CFT$_2^a$ respectively and we have used the Brown-Henneaux central charges given in \cref{central_charges}. As earlier we observe that the above result is cut-off independent  similar to the usual AdS$_3$/CFT$_2$ scenario without any anomaly \cite{Malvimat:2018txq, Malvimat:2018ood}. Once again our result matches with the universal part of the corresponding field theory result obtained in \cref{neg-disj-finite} in the large central charge limit which constitutes a strong consistency check.



\subsection{Holographic entanglement negativity for two adjacent intervals} \label{sec:EN_adj_geod}
\begin{figure}[h!]
	\centering
	\includegraphics[scale=0.7]{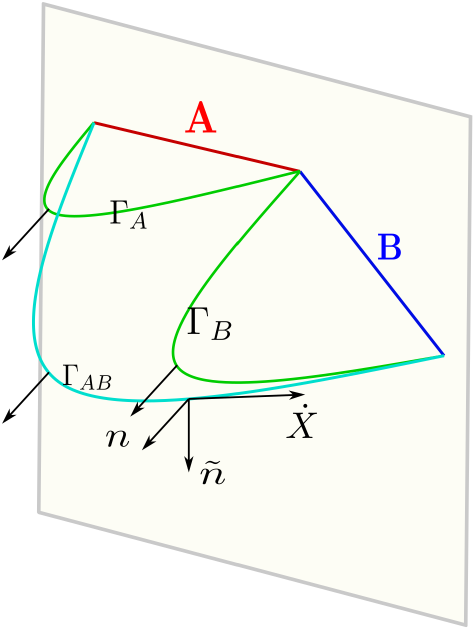}
	\caption{Schematics of the holographic construction for computing the entanglement negativity for two adjacent intervals $A$ and $B$ in a CFT$_2^a$ dual to topologically massive gravity in asymptotically  AdS$_3$ spacetimes. The normal frames to the extremal curves are depicted through the black arrows.}
	\label{adjTMGgeod}
\end{figure}

Having described the holographic entanglement negativity for two disjoint intervals in CFT$^a_{2}$s under consideration, we now proceed to compute the same for bipartite mixed states involving two adjacent intervals. To this end, we consider two adjacent intervals $A = [z_1, z_2]$ and $B = [z_2, z_3]$ in the dual CFT$_2^a$ as depicted in \cref{adjTMGgeod}. As described earlier the entanglement negativity for this configuration involves a three-point twist correlator given in \cref{three-point-correlator}. In the large central charge limit the dominant universal part may be expressed in terms of certain two-point twist correlators in the dual CFT$_2^a$ as follows
\begin{equation}
\begin{aligned}
\left< \Phi_{n_e} (z_1) \Phi^2_{- n_e} (z_2) \Phi_{n_e} (z_3) \right> = & \frac{\left< \Phi_{n_e / 2} (z_1) \Phi_{- n_e / 2} (z_2) \right> \left< \Phi_{n_e / 2} (z_2) \Phi_{- n_e / 2} (z_3) \right>}{\left< \Phi_{n_e / 2} (z_1) \Phi_{- n_e / 2} (z_3) \right>} \left< \Phi_{n_e} (z_1) \Phi_{- n_e} (z_3) \right> \notag \\
&\qquad\qquad\qquad\qquad\qquad\qquad\qquad\qquad\qquad\qquad+ \mathcal{O} \left( \frac{1}{c_L}, \frac{1}{c_M} \right).
\end{aligned}
\end{equation}
Utilizing  the modified holographic dictionary in \cref{twist_correlator,L&T}, in the replica limit $n_e \to 1$ we obtain the holographic entanglement negativity for the mixed state of two adjacent intervals as follows
\begin{equation} \label{EN_adj_conjecture}
\mathcal{E}(A:B)=\frac{3}{16 G_N} \left( \mathcal{L}_A + \mathcal{L}_B - \mathcal{L}_{A \cup B} \right)\equiv \frac{3}{4}\mathcal{I}(A:B),
\end{equation}
where $\mathcal{L}_X$ is related to the length and the twist of the geodesic worldline homologous to the interval $X$ in the dual field theory as given in \cref{HRT_modified}. In the following subsections we proceed to compute the holographic entanglement negativity for the mixed state configuration of two adjacent intervals in zero and a finite temperature CFT$^a_2$s utilizing the above proposal \cref{EN_adj_conjecture}.

\subsubsection{Poincar\'e TMG-AdS$_3$} \label{sec:EN_adj_boosted_geod}

For the first case we consider two adjacent boosted intervals $A$ and $B$ of lengths $R_A$ and $R_B$ and boosts $\kappa_A$ and $\kappa_B$ respectively, in a zero temperature CFT$_2^a$ dual to a bulk Poincar\'e TMG-AdS$_3$ spacetime. As earlier utilizing the length and the twist of a geodesic worldline homologous to a boosted interval given in \cref{L&T_Boosted}, we may compute the holographic entanglement negativity for the mixed state configuration  in question using our proposal in \cref{EN_adj_conjecture} as
\begin{equation} \label{E_adj_geod}
\begin{aligned}
\mathcal{E}(A:B) &= \frac{3}{8 G_N} \left[ \log \frac {R_A R_B} {\epsilon \, R_{AB}} + \frac{1}{\mu} \left(\kappa_A + \kappa_B - \kappa_{AB} \right) \right]\\
&=\frac{c_L+c_R}{8} \log \left(\frac {R_A R_B} {\epsilon \, R_{AB}}\right) - \frac{c_L-c_R}{8} \left( \kappa_A + \kappa_B - \kappa_{AB} \right),
\end{aligned}
\end{equation}
where $\epsilon$ is a UV cut-off and $R_{AB}$ and $\kappa_{AB}$ correspond to the length and the boost of the interval $A \cup B$ in the dual CFT$_2^a$. We have also used the Brown-Henneaux central charges given in eq. \eqref{central_charges} in the last equality. The above expression for the holographic entanglement negativity for the mixed state of two adjacent intervals in the CFT$_2^a$ vacuum dual to the Poincar\'e TMG-AdS$_3$ spacetime matches exactly with the universal part of the corresponding field theory result described earlier in \cref{adj-neg-zero}.


\subsubsection{Rotating BTZ black holes} \label{sec:EN_adj_T_geod}

Next we consider two adjacent intervals $A$ and $B$ of length $R_A$ and $R_B$ respectively in a thermal CFT$^a_{2}$ defined on a twisted cylinder with circumferences given by the inverse temperatures $\beta_L$ and $\beta_R$. The bulk dual in this case is described by a rotating planar BTZ black hole in the TMG-AdS$_3$ spacetime. The length and the twist of the geodesic worldline homologous to an interval in such field theories are given in \cref{L&T_T}. We may now obtain the holographic entanglement negativity for the  mixed state configuration of two adjacent intervals in the dual  CFT$^a_{2}$ using our proposal in \cref{EN_adj_conjecture} as follows
\begin{equation} \label{E_adj_T_geod}
\begin{aligned}
\mathcal{E}(A:B) = & \frac{c_L}{8} \log \left( \frac{\beta_L}{\pi \epsilon} \frac {\sinh \frac{\pi R_A}{\beta_L} \sinh \frac{\pi R_B}{\beta_L}} {\sinh \frac{\pi R_{AB}}{\beta_L}} \right) + \frac{c_R}{8} \log \left( \frac{\beta_R}{\pi \epsilon} \frac {\sinh \frac{\pi R_A}{\beta_R} \sinh \frac{\pi R_B}{\beta_R}} {\sinh \frac{\pi R_{AB}}{\beta_R}} \right),
\end{aligned}
\end{equation}
where $\epsilon$ is a UV cut-off and $R_{AB}=R_A+R_B$ corresponds to the length of the interval $A \cup B$ in the dual CFT$^a_2$ and the Brown-Henneaux central charges given in eq. \eqref{central_charges} have been utilized in the above expression. Once again we observe that our result matches exactly with the corresponding field theory result obtained in \cref{neg-disj-finite}.



\subsection{Holographic entanglement negativity for a single interval} \label{sec:EN_sing_geod}
Finally, we proceed to the holographic characterization of the entanglement negativity for the pure and mixed state configurations of a single interval at zero and a finite temperature in the dual CFT$^a_{2}$s.


\subsubsection{Poincar\'e TMG-AdS$_3$} \label{sec:EN_sing_boosted_geod}

In this case, we consider the pure vacuum state of a boosted interval $A$ of length $R_A$ and boost $\kappa_A$ in a CFT$^a_2$ dual to a bulk Poincar\'e TMG-AdS$_3$ geometry. As described in subsection \ref{sec:sing_Neg_zeroT}, the entanglement negativity for such a state in the dual field theory involves  two point twist correlators. Utilizing the modified holographic dictionary in \cref{twist_correlator}, the required twist correlator may be expressed as
\begin{equation} \label{EN_sing_geod}
\left<\Phi^2_{n_e}(z_1)\Phi^2_{-n_e}(z_2)\right> =\left(\left<\Phi_{n_e/2}(z_1)\Phi_{-n_e/2}(z_2)\right>\right)^2=  e^{- 2\Delta_{n_e/2} L_A - 2\, s_{n_e/2} {\mathcal{T}_A}},
\end{equation}
where $L_A$ and $\mathcal{T}_A$ denote the length and the twist of the geodesic worldline homologous to the interval $A$ in the dual CFT$_2^a$. Using \cref{EN_sing_geod}, we may now obtain the holographic entanglement negativity for the pure state of a single boosted interval in question as 
\begin{equation} \label{E_sing_geod}
\begin{aligned}
\mathcal{E}(A) &= \frac{c_L+c_R}{4} \log \frac{R_A}{\epsilon} - \frac{c_L-c_R}{4} \kappa_A,
\end{aligned}
\end{equation}
where we have used the Brown-Henneaux central charges in \cref{central_charges} and the expressions for the length $L_A$ and and the twist $\mathcal{T}_A$ in \cref{L&T_Boosted}. The holographic entanglement negativity obtained above matches exactly with the corresponding field theory result in \cref{neg_sing}. Also note that the above expression for the holographic entanglement negativity may be re-written as
\begin{equation}
\mathcal{E}(A) = \frac{3}{2} S_A,\label{Instructive_sing}
\end{equation}
where $S_A$ is the holographic entanglement entropy for the single interval in question given in \cref{EE_sing_geod}. This is in conformity with quantum information theory expectations as the entanglement negativity for a pure state is given by the R\'enyi entropy of order half which in this case is $\frac{3}{2}S_A$.


\subsubsection{Rotating BTZ black holes} \label{sec:EN_sing_T_geod}
\begin{figure}[h!]
	\centering
	\includegraphics[scale=1.0]{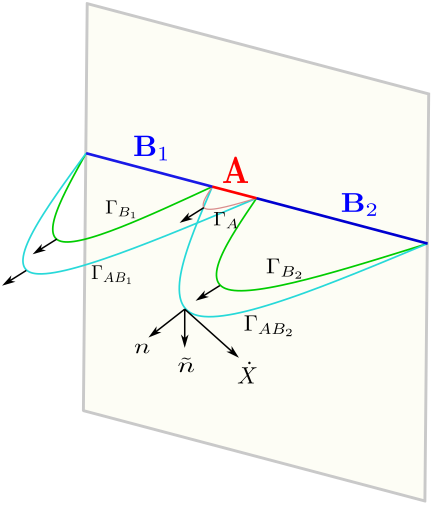}
	\caption{Schematics of the holographic construction for computing the entanglement negativity for a single interval in a thermal CFT$_2^a$ dual to topologically massive gravity in a rotating BTZ black hole. The normal frames to the extremal curves are depicted through the black arrows.}
	\label{singTMGgeod}
\end{figure}

Finally we consider the scenario of a single interval at a finite temperature in a CFT$_2^a$ defined on a twisted cylinder with circumferences given by inverse temperatures $\beta_L$ and $\beta_R$. However as discussed in the corresponding field theory analysis in subsection \ref{sec:sing_Neg_T} and the holographic constructions in
\cite{KumarBasak:2020eia, Basu:2021awn, Basu:2021axf} we require to consider the single interval $A=[w_2, w_3]$ of length $R_A$ sandwiched between two large but finite auxiliary intervals $B_1=[w_1, w_2]$ and $B_2=[w_3, w_4]$ of lengths $R$ on either sides on a constant time slice as depicted in \cref{singTMGgeod}. We perform the computation for this setup involving the finite auxiliary intervals and ultimately implement the bipartite limit $B = B_1 \cup B_2 \to A^c$ to 
restore the original configuration of a single interval in a thermal CFT$_2^a$. 

As seen in subsection \ref{sec:sing_Neg_T}, the field theory computation of the entanglement negativity employs a four-point twist correlator. In \cref{4pt-correlator-neg-T}, this twist correlator is expressed in terms of the cross-ratios, the coordinates of the intervals and certain non-universal functions. However in the large central charge limit, the dominant contribution arises from the universal part of the field theory result. Now, using the definition of a two-point function in the usual CFT$_{2}$ given in \cref{Tr_rho}, in the large central charge limit, we observe that the four-point twist correlator in question can be expressed as
\begin{equation} \label{four_point_sing_T_split}
\begin{aligned}
&\left< \Phi_{n_e}(w_1) \Phi_{-n_e}^2(w_2) \Phi_{n_e}^2(w_3) \Phi_{-n_e}(w_4) \right> \\
&=\left( \left< \Phi_{n_e/2} (w_2) \Phi_{-n_e/2} (w_3) \right> \right)^2 \left< \Phi_{n_e} (w_1)\Phi_{n_e} (w_4) \right>\frac{\left< \Phi_{n_e/2} (w_1) \Phi_{-n_e/2} (w_2) \right> \left< \Phi_{n_e/2} (w_3) \Phi_{-n_e/2} (w_4) \right>}{\left< \Phi_{n_e/2} (w_1) \Phi_{-n_e/2} (w_3) \right> \left< \Phi_{n_e/2} (w_2) \Phi_{-n_e/2} (w_4)  \right>} \\
&\qquad\qquad\qquad\qquad\qquad\qquad\qquad\qquad\qquad\qquad\qquad\qquad\qquad\qquad\qquad\qquad\qquad+ \mathcal{O} \left( \frac{1}{c_L} , \frac{1}{c_M} \right).
\end{aligned}
\end{equation}
Utilizing the holographic dictionary in \cref{twist_correlator,L&T}, it is possible to express the above four-point twist correlator in terms of the lengths and the twists for the bulk geodesic worldlines homologous to appropriate combinations
of the intervals in the dual CFT$^a_2$. Finally implementing the bipartite limit $R \to \infty$ subsequent to the replica limit $n_e \to 1$, we obtain the holographic entanglement negativity for the single interval $A$ in question as follows
\begin{equation} \label{EN_sing_T_conjecture}
\begin{aligned}
\mathcal{E}(A) = \lim_{B_1\cup B_2 \to A^c} \frac{3}{16 G_N} \left( 2 \mathcal{L}_A + \mathcal{L}_{B_1} + \mathcal{L}_{B_2} - \mathcal{L}_{A \cup B_1} - \mathcal{L}_{A \cup B_2} \right),
\end{aligned}    
\end{equation}
where $\mathcal{L}_X$ corresponds to interval $X$ in the dual field theory and is given in \cref{HRT_modified}. It is important to note here that the order of the application of the two limits, namely, the replica limit $n_e \to 1$ and the bipartite limit $R \to \infty$, is important and they do not commute. Interestingly, utilizing the modified HRT formula in \cref{HRT_modified}, we may rewrite the above expression for the holographic entanglement negativity in terms of the holographic mutual information between various subsystems involved, as
\begin{align} \label{EN_sing_mutual}
\mathcal{E}(A)=\lim_{B_1\cup B_2 \to A^c}\frac{3}{4} \Big(\mathcal{I}(A: B_1) + \mathcal{I}(A:B_2) \Big),
\end{align}
which conforms to the earlier findings in \cite{Jain:2017uhe,Jain:2017aqk} in the context of AdS$_3$/CFT$_2$.

The length and the twist of a generic geodesic worldline homologous to an interval in the dual thermal CFT$_2^a$ are given in \cref{L&T_T}. Using these in \cref{EN_sing_T_conjecture} we may obtain the holographic entanglement negativity for the mixed state of a single interval in a thermal CFT$_2^a$ dual to the rotating planar BTZ black hole in TMG-AdS$_3$ spacetime as
\begin{equation} \label{E_sing_T_geod}
\begin{aligned}
\mathcal{E}(A) = \frac{c_L}{4} \log \left [ \frac{\beta_L}{\pi} \sinh \frac{\pi R_A}{\beta_L} \right ] + \frac{c_R}{4} \log \left [ \frac{\beta_R}{\pi} \sinh \frac{\pi R_A}{\beta_R} \right ] - \frac{c_L}{4} \frac{\pi R_A}{\beta_L} - \frac{c_R}{4} \frac{\pi R_A}{\beta_R},
\end{aligned}
\end{equation}
where we have utilized the Brown-Henneaux central charges given in \cref{central_charges}. We note here that again our result matches exactly with the universal part of the corresponding field theory result in eq. \eqref{neg-single-finite} in the large central charge limit which once more serves as a strong consistency check for our holographic construction. 

Interestingly, we observe that the above result may also be expressed in the following way as
\begin{equation}
\mathcal{E}(A) = \frac{3}{2} ( S_A - S_A^\text{th} ).
\end{equation}
where $S_A$ is the entanglement entropy for the interval $A$ in the thermal CFT$_2^a$ as given in \cref{EE_sing_T_geod} and  $S_A^\text{th}$ is the thermal contribution to the entanglement entropy which  
is subtracted. This illustrates that the entanglement negativity provides an upper bound to the distillable entanglement as described in quantum information theory.

\section{EWCS in TMG-AdS$_3$/CFT$^a_2$}
\label{sec:EWCS}
In this section we will provide a construction for the bulk entanglement wedge cross section (EWCS) for a subregion in the 
CFT$^a_2$ and investigate the effects of the gravitational anomaly on the structure of the entanglement wedge.
In the following we begin with the evaluation of the Chern-Simons contribution to the bulk minimal EWCS and subsequently utilize the same to provide an alternative holographic characterization for the entanglement of 
bipartite pure and mixed states in the dual CFT$^a_2$ described by two disjoint, adjacent and a single interval 
configurations dual to bulk Poincar\'e TMG-AdS$_3$ and the BTZ black hole geometries. In this context, we recall that in the usual AdS/CFT scenario the holographic reflected entropy has been shown to be twice the minimal EWCS in \cite{Dutta:2019gen}. In this work, we extend this duality in the context of the TMG-AdS$_3$/CFT$_2^a$ scenario and obtain the reflected entropy for the various bipartite states in the dual CFT$_2^a$ from the bulk EWCS and compare with the corresponding field theory replica technique results.

\begin{figure}[h!]
	\centering
	\includegraphics[scale=0.9]{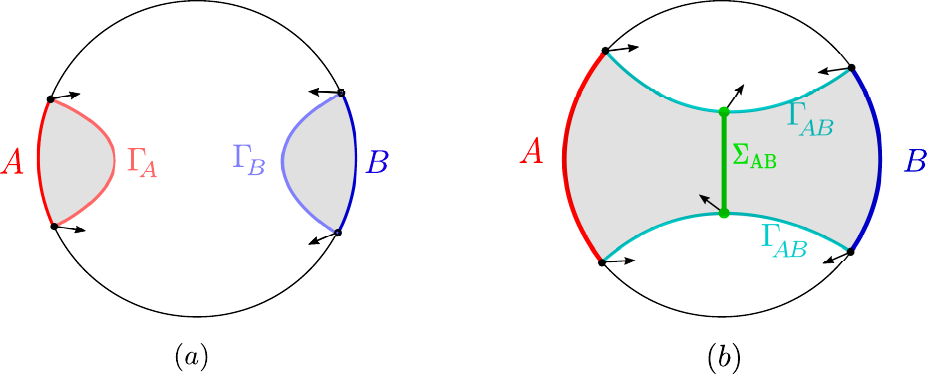}
	\caption{Schematics of the bulk entanglement wedge corresponding to two disjoint intervals $A$ and $B$ in the dual CFT$^a_2$ and the candidate extremal curves. The black arrows denote the bulk vectors normal to the extremal curves. (a) Disconnected entanglement wedge with trivial cross section. (b) Connected entanglement wedge bounded by the hypersurfaces $A\cup B\cup \Gamma_{AB}$.}
	\label{Wedge-generic}
\end{figure}

For this purpose we consider two generic disjoint subsystems $A$ and $B$ in the dual CFT$^a_2$ and as described in \cite{Castro:2014tta}, the holographic entanglement entropy for this configuration is given in terms of the areas (lengths) of the codimension-two extremal HRT surfaces (geodesics) homologous to the subsystem $A\cup B$, namely, $\Gamma_{A} \,,\,\Gamma_{B}$ and $\Gamma_{AB}$. The entanglement wedge dual to the reduced density matrix $\rho_{AB}$ is defined as the codimension-one region of the bulk spacetime bounded by the union of the HRT surfaces homologous to $A\cup B$ and the subsystems $A$ and $B$ themselves \cite{Takayanagi:2017knl}, as shown by the shaded regions in \cref{Wedge-generic}. For small subsystems $A$ and $B$, if they are separated enough, the entanglement entropy is computed through the combination of the disconnected HRT surfaces $\Gamma_{A}$ and $\Gamma_{B}$ and consequently the entanglement wedge is disconnected with a trivial cross-section (\cref{Wedge-generic}\textcolor{blue}{(a)}). On the other hand, when the subsystems are large enough so that the entanglement entropy is obtained through the extremal surface $\Gamma_{AB}$ as depicted in \cref{Wedge-generic}\textcolor{blue}{(b)}, one obtains a connected entanglement wedge $\Xi_{AB}$ bounded by the union of the hypersurfaces $A\cup B\cup \Gamma_{AB}$ \cite{Takayanagi:2017knl,Nguyen:2017yqw}, namely 
\begin{align}
\partial\,\Xi_{AB}\equiv A\cup B\cup \Gamma_{AB}\,.
\end{align}

As described earlier for the dual CFT$^a_2$ the bulk action includes a gravitational Chern-Simons term which requires the construction of timelike vectors at each point in the bulk which are constrained to be normal to the extremal worldlines of massive spinning particles. In this case the bulk entanglement wedge admits of extra gauge degrees of freedom arising from these timelike vectors which requires gauge fixing conditions obtained through the choice of appropriate local frames. In this case to define the minimal cross section of the entanglement wedge, we first divide the geodesic $\Gamma_{AB}$ in two segments as \cite{Takayanagi:2017knl,Nguyen:2017yqw}
\begin{align}
\Gamma_{AB}=\Gamma_{AB}^{(A)}\cup \Gamma_{AB}^{(B)}\,,\label{EW-partition}
\end{align}
and subsequently construct the extremal curve $\Sigma_{AB}$ homologous to the segment $\tilde{\Gamma}^{(A)}\equiv A\cup \Gamma_{AB}^{(A)}$ in the entanglement wedge \cite{Takayanagi:2017knl,Nguyen:2017yqw}. The entanglement wedge cross section is then defined as the minimal length of the curve sought out from all the candidate $\Sigma_{AB}$s, where the minimization is performed over all possible partitions in \cref{EW-partition}. In the present scenario of TMG in asymptotically AdS$_3$ spacetimes dual to anomalous CFT$_2$s, this minimal length picks up contributions from both the Einstein-Hilbert as well as the Chern-Simons part of the gravitational action. The familiar Einstein-Hilbert contribution is just given by the usual length of the minimal curve $\Sigma_{AB}$ as \cite{Takayanagi:2017knl,Nguyen:2017yqw}
\begin{align}
E_W^{\textrm{EH}}=\underset{\Gamma_{AB}^{(A)}\subset \Gamma_{AB}}{\text{min\,\,ext}} \, \left[\frac{L\left(\Sigma_{AB}\right)}{4 G_N}\right]\,.\label{EW_EH}
\end{align}
As described earlier, the effect of the gravitational Chern-Simons term is to broaden the particle worldlines in the shape of ribbons and traversing through the length of such a ribbon a torsion is experienced. This torsion in turn twists the ribbon and the Chern-Simons contribution to the EWCS is given in terms of the difference in the twists of the ribbon shaped worldline $\Sigma_{AB}$ at its two ends.
Therefore, similar to the computation of the holographic entanglement entropy  in \cite{Castro:2014tta}, the Chern-Simons contribution to the EWCS may be obtained by extremizing the boost $\mathcal{T}$ required to drag an auxiliary orthonormal frame through the length of $\Sigma_{AB}$ as
\begin{align}
E_W^{\textrm{CS}}=\underset{\Gamma_{AB}^{(A)}\subset \Gamma_{AB}}{\text{min\,\,ext}} \, \left[\frac{\mathcal{T}\left(\Sigma_{AB}\right)}{4\mu G_N}\right]\,,\label{EW_CS}
\end{align}
where, once again, the extremization is performed over all possible partitions in \cref{EW-partition}. Note that in the above definition, the coupling constant $\mu$ of the CS term appears in the denominator which ensures that the Chern-Simons contribution also carries the dimensions of length. 

With the above bulk construction of the minimal EWCS given in \cref{EW_EH,EW_CS}, we now propose following \cite{Dutta:2019gen} that the holographic reflected entropy is given by twice the total entanglement wedge cross section as
\begin{align}
	S_R(A:B)=2 E_W(A:B)\equiv\underset{\Gamma_{AB}^{(A)}\subset \Gamma_{AB}}{\text{min\,\,ext}} \, \left[\frac{1}{2 G_N}\left(L\left(\Sigma_{AB}\right)+\frac{\mathcal{T}\left(\Sigma_{AB}\right)}{\mu}\right)\right]\,.\label{S_R-holo}
\end{align}

In the following, we will compute the minimal EWCS including the Chern-Simons contribution in \cref{EW_CS} for various bipartite state configurations in the dual conformal field theory with a gravitational anomaly. Furthermore, we will examine the proposed holographic duality between the reflected entropy and the EWCS in \cref{S_R-holo} in the presence of topologically massive gravity in AdS$_3$ and find perfect agreement with the field theoretic computations in \cref{sec:reflecetd}.


\subsection{Two disjoint intervals} \label{sec:EW_disj}
We begin by computing the minimal EWCS corresponding to the mixed state configuration of two disjoint intervals in the CFT$_2^a$. The dual geometries involve topologically massive gravity in asymptotically AdS$_3$ spacetimes. A schematics of the entanglement wedge corresponding to the setup is sketched in \cref{Wedge-disj}. As described above, the computation of the minimal EWCS involves an Einstein-Hilbert contribution as well as a topological Chern-Simons contribution. In the following we will compute the minimal EWCS for two disjoint intervals in the ground state of a CFT$_2^a$ as well as for a thermal CFT$_2^a$ defined on a twisted cylinder. 

\subsubsection{Poincar\'e AdS$_3$} \label{sec:EW_disj_boosted}
\begin{figure}[h!]
	\centering
	\includegraphics[scale=0.8]{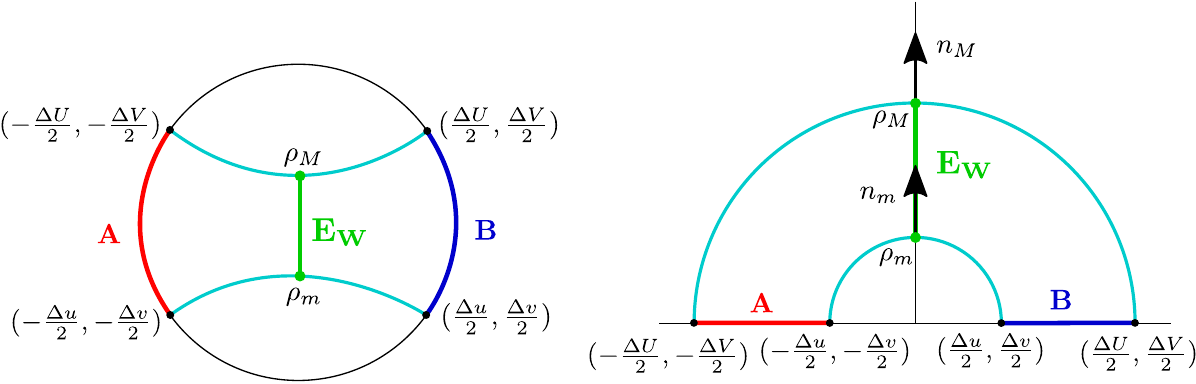}
	\caption{EWCS for two symmetrically placed disjoint intervals $A=\left[\left(-\frac{\Delta U}{2},-\frac{\Delta V}{2}\right),\left(-\frac{\Delta u}{2},-\frac{\Delta v}{2}\right)\right]$ and $B=\left[\left(\frac{\Delta u}{2},\frac{\Delta v}{2}\right),\left(\frac{\Delta U}{2},\frac{\Delta V}{2}\right)\right]$  in a CFT$_2^a$ dual to topologically massive gravity in Poincar\'e AdS$_3$.}
	\label{Wedge-disj}
\end{figure}

In this subsection we compute the minimal entanglement wedge cross-section corresponding to two boosted disjoint intervals $A$ and $B$ in the ground state of a CFT$_2^a$. The dual gravitational theory is described by TMG in Poincar\'e AdS$_3$ spacetime with the metric given in \cref{Poincare_metric}. To proceed we recall from the discussion in subsection \ref{sec:TMG_review} that in the presence of the gravitational Chern-Simons term the bulk picture is modified in terms of the inclusion of timelike vectors $n$ at each bulk site. Moreover these timelike vectors are constrained to be normal to the worldlines of massive spinning particles. As described earlier, we are interested in situations where the massive spinning particles in the bulk follow geodesics. For a geodesic worldline in Poincar\'e AdS$_3$ spacetime connecting two boundary points $\left(-\frac{\Delta u}{2},-\frac{\Delta v}{2},\infty\right)$ and $\left(\frac{\Delta u}{2},\frac{\Delta v}{2},\infty\right)$ a particularly useful parametrization of the normal vectors is given by \cite{Gao:2019vcc}
\begin{align}
	n=\pm\frac{\Delta u^2\sqrt{\rho^2-\frac{2\rho}{\Delta u\Delta v}}}{\rho\sqrt{2\Delta u^2\Delta v^2\rho-(\Delta u+\Delta v)^2}}\partial_u&\mp\frac{\Delta v^2\sqrt{\rho^2-\frac{2\rho}{\Delta u\Delta v}}}{\rho\sqrt{2\Delta u^2\Delta v^2\rho-(\Delta u+\Delta v)^2}}\partial_v\notag\\
	&+\frac{2\rho(\Delta u-\Delta v)}{\sqrt{2\Delta u^2\Delta v^2\rho-(\Delta u+\Delta v)^2}}\partial_{\rho}\,.
	\label{Normal_n}
\end{align}
The turning point ($u=0$) of the above geodesic corresponds to $\tau_m=\frac{1}{2}\log\frac{1}{\Delta u\Delta v}$. Utilizing \cref{Geod_para}, we obtain $\rho_m=\frac{2}{\Delta u\Delta v}$ and using \cref{Normal_n,Tangent} the normal frame has the following form\footnote{Note that the imaginary component of the normal vector is required for the normalization $n_m^2=-1$ and is an artefact of the gauge choice made here.}
\begin{align}
	\dot{X}_m=\frac{\Delta u}{2}\partial_u+\frac{\Delta v}{2}\partial_v~~,~~n_m=\frac{4i}{\Delta u\Delta v}\partial_{\rho}\,.\label{Xm_nm}
\end{align}
We now consider two symmetrically placed disjoint intervals $A=\left[\left(-\frac{\Delta U}{2},-\frac{\Delta V}{2}\right),\left(-\frac{\Delta u}{2},-\frac{\Delta v}{2}\right)\right]$ and $B=\left[\left(\frac{\Delta u}{2},\frac{\Delta v}{2}\right),\left(\frac{\Delta U}{2},\frac{\Delta V}{2}\right)\right]$ of equal length in ground state of the dual CFT$_2^a$ as shown in \cref{Wedge-disj}. 
Purely from the symmetry of the geometry, the minimal cross section of the corresponding entanglement wedge is given by the extremal curve (geodesic in the present setting) connecting the turning points $(0,0,\rho_m)$ and $(0,0,\rho_M)$ of the two geodesics computing the entanglement entropy $S_{A\cup B}$ of the composite system $A\cup B$. 
As usual, the contribution from the Einstein-Hilbert action computes the length of the extremal curve connecting the two turning points  
\begin{align}
	E_W^{\textrm{EH}}&=\frac{1}{4G_N}\int_{\tau_M}^{\tau_m}\textrm{d}\tau \sqrt{g_{\mu \nu} \dot X ^\mu \dot X ^\nu}\notag\\
	&\equiv \frac{\tau_m-\tau_M}{4G_N}=\frac{1}{8G_N}\log\left(\frac{\Delta U\Delta V}{\Delta u\Delta v}\right)\,.\label{EH_Poincare}
\end{align}
where we have used the expressions for the proper time at the two turning points. Now writing $\Delta u=2r\,e^{-2\kappa}$, $\Delta v=2r\,e^{2\kappa}$ and $\Delta U=2R\,e^{-2K}$, $\Delta V=2R\,e^{2K}$, \cref{EH_Poincare} reduces to
\begin{align}
	E_W^{\textrm{EH}}=\frac{1}{4G_N}\log\left(\frac{R}{r}\right)=\frac{c_L+c_R}{6}\log\left(\frac{R}{r}\right)\,,
\end{align} 
where in the last step, we have made use of the Brown-Henneaux relation \cref{central_charges}. 

In a similar fashion, the Chern-Simons contribution to the minimal EWCS may be obtained by the boost required to drag the normal frame generated by the orthonormal triad $(\dot{X},n,\tilde{n})$ from one turning point to another as
\begin{align}
	E_W^{\textrm{CS}}=\frac{1}{4\mu G_N}\int_{\tau_M}^{\tau_m}\textrm{d}\tau\, \tilde{n}.\nabla n=\frac{1}{4\mu G_N}\log \left[ \frac{q(\tau_m).n_m - \tilde q (\tau_m).n_m}{q(\tau_M).n_M- \tilde q (\tau_M).n_M} \right]\,.\label{EW_Poincare_CS}
\end{align}
The values of the parallel transported normal vectors at the turning point of a geodesic line connecting the boundary points $\left(-\frac{\Delta u}{2},-\frac{\Delta v}{2},\infty\right)$ and $\left(\frac{\Delta u}{2},\frac{\Delta v}{2},\infty\right)$ may be obtained from \cref{q_boosted} as
\begin{align}
	q (\tau_m)=\pm\frac{\Delta u}{2}\partial_u\mp\frac{\Delta v}{2}\partial_v~~,~~\tilde q (\tau_m)=\frac{4}{\Delta u^2}\partial_{\rho}\,.\label{qm}
\end{align}
Now using \cref{Xm_nm,qm}, we obtain the Chern-Simons contribution to the minimal EWCS for our setup of two symmetrically placed disjoint intervals from \cref{EW_Poincare_CS} as
\begin{align}
	E_W^{\textrm{CS}}&=\frac{1}{4\mu G_N}\log\left(\frac{\Delta v/\Delta u}{\Delta V/\Delta U}\right)\notag\\
	&=\frac{1}{4\mu G_N}\left(\kappa-K\right)=\frac{c_L-c_R}{6}\left(K-\kappa\right)\,,\label{EW-CS-disj-boosted}
\end{align}
where once again we have made use of the Brown-Henneaux relation \cref{central_charges}.

Next we will rewrite the minimal EWCS obtained above in a more formal notation utilizing the CFT$_2^a$ cross-ratios. To proceed we first note that for the present setup of two symmetrically placed boosted disjoint intervals $A\equiv[z_1,z_2]$ and $B\equiv[z_3,z_4]$ of equal length in the dual CFT$_2^a$, the (complex) cross-ratio is given by
\begin{align}
	\eta=\frac{z_{12}z_{34}}{z_{13}z_{24}}=\frac{(Z-z)^2/4}{(Z+z)^2/4}\,,
\end{align} 
where $z$ denotes the length of the interval $C$ sandwiched between $A$ and $B$, while $Z$ denotes the length of the composite system $A\cup B\cup C$. In terms of the proper lengths and boost angles corresponding to these subsystems, we have
\begin{align}
	\frac{Z}{z}\equiv\frac{R\,e^{-2K}}{r\,e^{-2\kappa}}=\frac{1+\sqrt{\eta}}{1-\sqrt{\eta}}\,.
\end{align}
Therefore, the total EWCS may be expressed in terms of the cross-ratios reminiscent of the boundary intervals as
\begin{align}
	E_W&=\frac{1}{4G_N}\log\left(\frac{R}{r}\right)+\frac{1}{4\mu G_N}\left(\kappa-K\right)\notag\\
	&=\frac{1}{4G_N}\log\left|\frac{1+\sqrt{\eta}}{1-\sqrt{\eta}}\right|+\frac{1}{4\mu G_N}\textrm{argh}\left(\frac{1+\sqrt{\eta}}{1-\sqrt{\eta}}\right)\,.\label{EW_crossratio}
\end{align}
where the hyperbolic argument for a complex variable analytically continued to Lorentzian signature $z=x-t=Re^{-\kappa}$, is defined through 
\begin{align}
	\text{argh}(z)\equiv\kappa=\tanh^{-1}\left(\frac{t}{x}\right)\,.
\end{align}
Equation \eqref{EW_crossratio} provides a comprehensive expression for the minimal EWCS for two disjoint intervals in the ground state of a CFT$_2^a$ dual to TMG in Poincar\'e AdS$_3$. Upon utilizing \cref{central_charges,S_R-holo}, the holographic reflected entropy matches exactly with the field theoretic result (in the large $c$ limit) corresponding to the present configuration of two boosted disjoint intervals given in \cref{SR_disj}. This serves as a strong consistency check for our construction of the bulk minimal EWCS. Furthermore, we note that the above expression is reminiscent of the expectations from the dual field theory. Recall that in the absence of the Chern-Simons term in the gravitational action (in the absence of anomaly in the dual CFT) the minimal EWCS for two disjoint intervals was given in terms of the CFT data as \cite{Takayanagi:2017knl} 
\begin{align}
	E_W=\frac{c}{6}\log\left(\frac{1+\sqrt{\eta}}{1-\sqrt{\eta}}\right)\,.\label{EW_TU}
\end{align}
In the presence of a gravitational anomaly, the left and right moving sectors possess different central charges and a natural generalization of \cref{EW_TU} reads
\begin{align}
	E_W&=E_W^{(L)}+E_W^{(R)}\notag\\
	&=\frac{c_L}{12}\log\left(\frac{1+\sqrt{\eta}}{1-\sqrt{\eta}}\right)+\frac{c_R}{12}\log\left(\frac{1+\sqrt{\bar{\eta}}}{1-\sqrt{\bar{\eta}}}\right)\,.\label{EW_CFT}
\end{align}
Performing the Lorentzian continuation is tantamount to hyperbolic arguments for complex quantities and therefore we reproduce \cref{EW_crossratio} from \cref{EW_CFT} upon utilizing the Brown-Henneaux relations \cref{central_charges}.


\subsubsection{Rotating BTZ black holes} \label{sec:EW_disj_T}
We now proceed to the computation of the entanglement wedge cross-section for two disjoint intervals $A\equiv [w_1, w_2]$ and $B\equiv [w_3,w_4]$ of lengths $R_A$ and $R_B$ respectively, in a thermal CFT$^a_2$ defined on a twisted cylinder with circumferences $\beta_L$ and $\beta_R$. The bulk dual for such field theories is described by rotating BTZ black holes in TMG-AdS$_3$ spacetime whose metric is given in \cref{BTZmetric}. In principle we could follow the similar recipe as in the previous case of Poincar\'e AdS$_3$ spacetime by computing the Einstein-Hilbert and the Chern-Simons contributions to the EWCS separately. This would involve a similar parametrization of the geodesics and the normal vectors which may be found in \cite{Gao:2019vcc}. However in the present article we follow a different approach where we utilize the fact that the EWCS in \cref{EW_crossratio} is written in terms of the dual field theory data and subsequently use the modified cross-ratios for the finite temperature case:
\begin{align} \label{cross_ratio_T}
	\xi = \frac {\sinh \frac{\pi w_{12}}{\beta_L} \sinh \frac{\pi w_{34}}{\beta_L}} {\sinh \frac{\pi w_{13}}{\beta_L} \sinh \frac{\pi w_{24}}{\beta_L}} ~~,~~ \bar\xi = \frac {\sinh \frac{\pi \bar w_{12}}{\beta_R} \sinh \frac{\pi \bar w_{34}}{\beta_R}} {\sinh \frac{\pi \bar w_{13}}{\beta_R} \sinh \frac{\pi \bar w_{24}}{\beta_R}},  
\end{align}
where the transformations from the complex plane to the twisted cylinder are given by $w=e^{2\pi z/\beta_L}$ and $\bar{w}=e^{2\pi \bar{z}/\beta_R}$. Therefore the expression for the EWCS we obtain for the mixed state configuration in question is given by
\begin{align}
E_W = \frac{c_L}{6}\log\left(\frac{1+\sqrt{\xi}}{1-\sqrt{\xi}}\right)+\frac{c_R}{6}\log\left(\frac{1+\sqrt{\bar{\xi}}}{1-\sqrt{\bar{\xi}}}\right)\,.\label{EW_disj_T} 
\end{align}
Once again the holographic reflected entropy computed through \cref{S_R-holo} matches perfectly with the field theory result in \cref{SR_disj_T} (obtained in the large $c$ limit) upon using the finite temperature cross-ratios in \cref{cross_ratio_T}. This once more serves as a consistency check for our holographic proposal.

\subsection{Two adjacent intervals} \label{sec:EW_adj}
Having computed the minimal EWCS for various bipartite mixed states involving two disjoint intervals in a CFT$_2^a$, we now move on to analyze the mixed state configuration of two adjacent intervals for all the previous cases. Interestingly, all of the results in this subsection may be obtained through a suitable adjacent limit of the corresponding results for the setup of two disjoint intervals in subsection \ref{sec:EW_disj}.

\subsubsection{Poincar\'e AdS$_3$} \label{sec:EW_adj_boosted}
\begin{figure}[h!]
	\centering
	\includegraphics[scale=1.0]{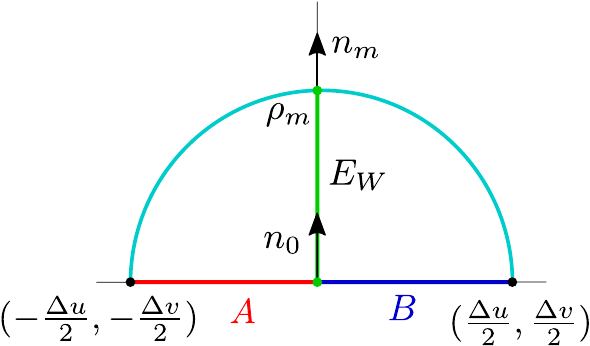}
	\caption{EWCS for two symmetrically placed adjacent intervals $A=\left[\left(-\frac{\Delta u}{2},-\frac{\Delta v}{2}\right),(0,0)\right]$ and $B=\left[(0,0),\left(\frac{\Delta u}{2},\frac{\Delta v}{2}\right)\right]$ in a CFT$_2^a$ dual to topologically massive gravity in Poincar\'e AdS$_3$.}
	\label{Wedge-adj}
\end{figure}

In this subsection we compute the minimal EWCS for two adjacent intervals
$A\equiv[z_1,z_2]$ and $B\equiv[z_2,z_3]$ in the vacuum state of  CFT$^a_2$  whose bulk dual is described by 
the Poincar\'e TMG-AdS$_3$ geometry. Once again, we start with symmetric intervals of equal length in the $u$-$v$ plane, $A=\left[\left(-\frac{\Delta u}{2},-\frac{\Delta v}{2}\right),(0,0)\right]$ and $B=\left[(0,0),\left(\frac{\Delta u}{2},\frac{\Delta v}{2}\right)\right]$ as shown in \cref{Wedge-adj}. For this symmetric setup the entanglement wedge is bounded by the extremal curve (geodesic) connecting the endpoints of the subsystem $A\cup B$ and the boundary intervals $A$ and $B$ themselves. Purely from geometric arguments the minimal cross-section is then given by the geodesic connecting the common boundary of $A$ and $B$, and the turning point of the geodesic computing the entanglement entropy of the composite subsystem $A\cup B$. 

The Einstein-Hilbert contribution to the minimal EWCS is obtained from the length of the extremal geodesic between these two points as
\begin{align}
E_W^{\textrm{EH}}=\frac{1}{4G_N}\left(\tau_{\infty}-\tau_m\right)=\frac{1}{8G_N}\log\left(\frac{\Delta u\Delta v}{\epsilon^2}\right)=\frac{1}{4G_N}\log\left(\frac{2R}{\epsilon}\right)\,.
\end{align}
where $\tau_{\infty}=\log\frac{1}{\epsilon}$ denotes the proper time required to reach the boundary along the geodesic and we have used $\Delta u=2R\,e^{-2\kappa}$, $\Delta v=2R\,e^{2\kappa}$, where $R$ is the length of either of the subsystems $A$ and $B$ and $\kappa$ is the corresponding boost.

In a similar manner to the analysis in subsection \ref{sec:EW_disj_boosted}, the Chern-Simons contribution to the minimal EWCS for the present setup of two boosted adjacent intervals may be obtained through the boost required to drag the orthonormal frame between the endpoints of the extremal geodesic as
\begin{align}
E_W^{\textrm{CS}}=\frac{1}{4\mu G_N}\int_{\tau_\infty}^{\tau_m}\textrm{d}\tau \,\tilde{n}.\nabla n=\frac{1}{4\mu G_N}\log \left[ \frac{q(\tau_m).n_m - \tilde q (\tau_m).n_m}{q(\tau_\infty).n_0- \tilde q (\tau_\infty).n_0} \right]\,.\label{EW_Poincare_CS_adj}
\end{align}
In the above expression, $q(\tau_m)$ and $\tilde q (\tau_m)$ are the parallel transported normal vectors at the turning point of the geodesic, and are given in \cref{qm}. The normal vector $n_0$ and parallel transported frame at the common boundary of $A$ and $B$ are given by
\begin{align}
n_0=\pm\frac{i}{\epsilon}\partial_{\rho}~~,~~q (\tau_\infty)=\epsilon\left(\pm\partial_u\mp\partial_v\right)~~,~~\tilde q (\tau_\infty)=-\epsilon\partial_{\rho}\,.\label{q0}
\end{align}
Therefore, substituting \cref{qm,q0} in \cref{EW_Poincare_CS_adj}, the Chern-Simons contribution is evaluated to be
\begin{align}
E_W^{\textrm{CS}}=\frac{1}{4\mu G_N}\log\left(\frac{\Delta v}{\Delta u}\right)=\frac{\kappa}{2\mu G_N}\,,\label{EW_adj}
\end{align}
where, once again we have used $\Delta u=2R\,e^{-2\kappa}$ and $\Delta v=2R\,e^{2\kappa}$. In terms of the cross-ratio of the dual CFT$_2^a$
\begin{align}
\zeta=\frac{1}{1-\eta}=\frac{z_{12}z_{23}}{\epsilon\,z_{13}}\equiv\frac{R}{\epsilon}e^{-2\kappa}\,.\label{zeta_adj}
\end{align}
the expression for the complete minimal EWCS maybe rewritten as
\begin{align}
E_W=\frac{1}{4G_N}\log |2\zeta| +\frac{1}{4\mu G_N}\kappa\,.\label{EW-adj-crossratio}
\end{align}
In \cref{zeta_adj}, $R$ denotes the length of either of the subsystems $A$ and $B$ and $\kappa$ denotes the corresponding boost. Therefore, once again the minimal EWCS may be expressed in the form \cref{EW_crossratio} via a trivial redefinition of the UV cut-off $\epsilon$. Now utilizing \cref{zeta_adj} and the conformal symmetry of the dual field theory we may obtain the total minimal EWCS for two generic adjacent intervals $A\equiv[z_1,z_2]$ and $B\equiv[z_2,z_3]$ as
\begin{align}
E_W^{\textrm{adj.}}= \frac{1}{4 G_N} \log \left(\frac {R_{12} R_{23}} {\epsilon R_{13}}\right) - \frac{1}{4 \mu G_N} \left(\kappa_{12} + \kappa_{23} - \kappa_{13} \right) + \frac{1}{4 G_N} \log 2\,.
\end{align}
Now utilizing the Brown-Henneaux central charges in \cref{central_charges}, the holographic reflected entropy for the present configuration matches exactly with that obtained in \cref{SR_adj} using the replica technique in the dual field theory. Interestingly, we may also obtain the above expression by taking an appropriate adjacent limit
\begin{align}
 R_{23}\equiv\epsilon~~,~~ \kappa_{23}\equiv0\,,
\end{align}
of the result for two disjoint intervals in \cref{EW_CFT}. This provides yet another consistency check of our bulk construction of the minimal EWCS.

\subsubsection{Rotating BTZ black holes} \label{sec:EW_adj_T}
Next we move on to the computation of the minimal EWCS for two adjacent intervals $A\equiv[w_1,w_2]$ and $B\equiv[w_2,w_3]$ in a thermal CFT$_2^a$ dual to a rotating BTZ black hole in the bulk TMG-AdS$_3$ geometry. The computation essentially follows a similar analysis as in subsection \ref{sec:EW_disj_T}. As described above, we can alternatively obtain the minimal EWCS in the present situation of two adjacent intervals by taking a suitable adjacent limit of the corresponding disjoint intervals result in \cref{EW_disj_T} as
\begin{align}
 E_W^{\textrm{adj.}} = & \frac{c_L}{12} \log \left( \frac{\beta_L}{\pi \epsilon} \frac {\sinh \frac{\pi R_A}{\beta_L} \sinh \frac{\pi R_B}{\beta_L}} {\sinh \frac{\pi \left(R_A+R_B\right)}{\beta_L}} \right) + \frac{c_R}{12} \log \left( \frac{\beta_R}{\pi \epsilon} \frac {\sinh \frac{\pi R_A}{\beta_R} \sinh \frac{\pi R_B}{\beta_R}} {\sinh \frac{\pi \left(R_A+R_B\right)}{\beta_R}} \right)  + \frac{c_L + c_R}{12} \log 2\,,\label{EW_adj_T}
\end{align}
where we have chosen the coordinate of the endpoints of the adjacent intervals on the cylinder to be $w_1=\bar{w}_1=-R_A$, $w_2=\bar{w}_2=0$ and $w_3=\bar{w}_3=R_B$. Once again, the holographic reflected entropy matches exactly with the corresponding field theoretic result obtained through the replica technique in \cref{SR_adj_T}. 

\subsection{Single interval} \label{sec:EW_sing}
Finally we focus on bipartite states involving a single interval in CFT$^a_2$s with bulk dual TMG-AdS$_3$ geometries.
In particular, we will first compute the minimal EWCS corresponding to the pure state of a single interval in the vacuum state of the anomalous CFT$_2$. Next we will consider the mixed state configuration described by a single interval in a thermal CFT$_2^a$ with a finite chemical potential defined on a twisted cylinder. The computation for the minimal EWCS for this configuration is subtle and requires a more careful analysis. 

\subsubsection{Poincar\'e AdS$_3$} \label{sec:EW_sing_boosted}
We start with the simplest pure state configuration of a single boosted interval $A$ in the ground state of a CFT$_2^a$ whose dual gravitational theory involves TMG in Poincar\'e AdS$_3$ spacetime. The minimal EWCS for this pure state is trivially equal to the entanglement entropy for the single interval and therefore is obtained simply from the modified HRT formula in \cref{HRT_modified} as
\begin{align}
E_W\equiv S_A=\frac{1}{2 G_N} \log \frac{R_A}{\epsilon} + \frac{1}{2 \mu G_N} \kappa_A\,,
\end{align}
where $R_A=\sqrt{x_A^2-t_A^2}$ is the length of the boosted interval $A$ and $\kappa_A=\tanh^{-1}\left(\frac{t_A}{x_A}\right)$ is the boost angle. Utilizing the Brown-Henneaux central charges in \cref{central_charges}, the holographic reflected entropy computed through \cref{S_R-holo} matches exactly with the corresponding field theory answer \cref{S_R-single-zero} for the pure state configuration considered here.


\subsubsection{Rotating BTZ black holes} \label{sec:EW_sing_T}
Finally, we consider the bipartite mixed state configuration described by a single interval $A \equiv [0,R_A]$ and its compliment $B=A^c$ in a CFT$_2^a$ at a finite temperature and finite chemical potential defined on a twisted cylinder. The left and right moving CFT modes involve two different temperatures $\beta_{L,R}$ as defined in \cref{tempsss}. 

As described in \cite{KumarBasak:2020eia, Basu:2021awn} in the context of the usual AdS/CFT and flat-space holography respectively, the construction of the minimal EWCS for this case is subtle and we propose a similar construction in the case of TMG-AdS$_3$/CFT$_2^a$. As described before in subsection \ref{sec:sing_Neg_T}, the mixed state of a single interval $A$ at finite temperature is correctly analysed by sandwiching it between two adjacent large but finite auxiliary intervals $B_1$ and $B_2$ of length $R$ and subsequently implementing the bipartite limit
$B_1\cup B_2\to A^c$.
Therefore we start with the tripartite pure state corresponding to $A\cup B_1\cup B_2$. For the adjacent intervals $A,B_i\,,\,i=1,2$, we have from \cref{EW_adj_T,EE_sing_T_geod} the following equality
\begin{align}
E_W\left(A:B_i\right)=\frac{1}{2}\mathcal{I}\left(A:B_i\right)+ \frac{c_L + c_R}{12} \log 2\,,\label{EW_I}
\end{align}
where $\mathcal{I}\left(A:B_i\right)$ is the holographic mutual information between $A$ and $B_i$. We now utilize the following 
inequality valid for tripartite states
\begin{align}
	 E_W(A:B_1B_2)\leq E_W(A:B_1)+E_W(A:B_2)\,,\label{EW-Inequalities}
\end{align}
and obtain an upper bound on the minimal EWCS for the present configuration.
Using \cref{EW-Inequalities,EW_I,EW_adj_T}, the upper bound on the minimal EWCS may be obtained, upon taking the bipartite limit $R\to \infty$, as follows
\begin{align}
E_W &=\lim_{B_1\cup B_2 \to A^c} \left(E_W(A:B_1)+E_W(A:B_2)\right) \notag \\
&= \lim_{ R \to \infty }\left[\frac{c_L}{6} \log \left( \frac{\beta_L}{\pi \epsilon} \frac {\sinh \frac{\pi R_A}{\beta_L} \sinh \frac{\pi R}{\beta_L}} {\sinh \frac{\pi \left(R_A+R\right)}{\beta_L}} \right) + \frac{c_R}{6} \log \left( \frac{\beta_R}{\pi \epsilon} \frac {\sinh \frac{\pi R_A}{\beta_R} \sinh \frac{\pi R}{\beta_R}} {\sinh \frac{\pi \left(R_A+R\right)}{\beta_R}} \right)\right]+\frac{c_L + c_R}{6} \log 2\notag\\
&=\frac{c_L}{6} \log \left [ \frac{\beta_L}{\pi} \sinh \frac{\pi R_A}{\beta_L} \right ] + \frac{c_R}{6} \log \left [ \frac{\beta_R}{\pi} \sinh \frac{\pi R_A}{\beta_R} \right ] - \frac{c_L}{6} \frac{\pi R_A}{\beta_L} - \frac{c_R}{6} \frac{\pi R_A}{\beta_R}+\frac{c_L + c_R}{6} \log 2\,.
\end{align}
Remarkably, utilizing \cref{central_charges} the above expression matches with half of the universal part of the reflected entropy for the mixed state configuration of a single interval at finite temperature, obtained in \cref{S_R-correct-singleT}. Note that the additive constant is contained within the non-universal functions $g,\bar{g}$ in \cref{S_R-correct-singleT} and may be extracted through a large central charge analysis of the corresponding conformal block as discussed in subsection \ref{sec:S_R-single}. 




\section{Summary} \label{sec:summary}

To summarize, in this article we have obtained the entanglement negativity and the reflected entropy for various bipartite pure and mixed state configurations in a CFT$^a_2$ with a gravitational anomaly. For this purpose we utilized a  replica technique to compute these mixed state correlation measures for various bipartite states described by
a single interval, two adjacent intervals and two disjoint intervals (in proximity) in the vacuum state of  CFT$_2^a$s and also for thermal CFT$_2^a$s with an angular potential. It is observed that the gravitational anomaly introduces a non-trivial dependence on the choice of coordinates and the observables are sensitive to such choices. The entanglement negativity as well as the reflected entropy involves an additional contribution due to the gravitational anomaly and is hence frame dependent. We note that, in the absence of the gravitational anomaly, our results reduce to the corresponding results in the literature for the usual AdS/CFT scenario.

Interestingly, we have observed that similar to the case of the entanglement negativity discussed in \cite{Calabrese:2014yza}, a naive computation of the reflected entropy for a single interval at a finite temperature leads to inconsistent results. The origin of this inconsistency is the non-trivial sewing of the different copies of subsystems in the replica manifold for the R\'enyi reflected entropy which leads to an infinite branch cut. Similar to the case of the entanglement negativity, this may be rectified through the introduction of large but finite auxiliary intervals adjacent to the single interval on either side to compute the reflected entropy and subsequently implementing an appropriate bipartite limit.

Following the field theory replica constructions we have advanced a holographic proposal for the
entanglement negativity for various bipartite pure and mixed state configurations in CFT$^a_2$s with a gravitational anomaly dual to bulk topologically massive gravity (TMG) in asymptotically AdS$_3$ geometries.
The bulk three dimensional action for the TMG-AdS$_3$ geometries involve a gravitational Chern-Simons term in addition to the usual Einstein-Hilbert term. In this context, we have extended the earlier holographic entanglement negativity proposals to accommodate the effect of the Chern-Simons term in the bulk action. Accordingly, for bipartite states described by two disjoint, adjacent and a single interval in a CFT$_2^a$, the holographic constructions involve algebraic sums of the on-shell actions of massive spinning particles moving on extremal worldlines in the dual bulk geometry, homologous to certain appropriate combinations of the intervals. The holographic entanglement negativity obtained using these constructions exactly reproduce the corresponding replica technique results in the large central charge limit. 

Subsequently we have described a construction for the EWCS in the bulk TMG-AdS$_3$ geometries dual to CFT$_2^a$s
and proposed a prescription to compute the Chern-Simons contribution to the EWCS. Remarkably the holographic reflected entropy thus obtained from the bulk EWCS exactly matches with corresponding replica technique results in the large central charge limit. This serves as a strong consistency check of our holographic construction for the reflected entropy from the bulk EWCS. Finally, in appendix \ref{sec:appendix_A} we have provided a heuristic proof of the holographic entanglement negativity proposal for the case of two adjacent intervals in a CFT$_2^a$ utilizing Euclidean gravitational path integral techniques.

Our results for the field theory replica technique computations and the corresponding holographic constructions for the entanglement negativity and the reflected entropy for bipartite states in CFT$_2^a$s with a gravitational anomaly dual to bulk TMG-AdS$_3$ geometries, described in this article provides an elegant and consistent framework to address the issue of mixed state entanglement in these interesting field theories and leads to several interesting insights and future directions for investigations. One such future direction would be to study mixed state entanglement measures in the Chern-Simons formulation of $(2+1)$-dimensional topologically massive gravity theories \cite{Ammon:2013hba}. The entanglement entropy has been studied in the TMG-AdS$_3$/CFT$_2^a$ setting in \cite{Ammon:2013hba} and in flat holographic setting in \cite{Bagchi:2014iea} utilizing a factorized Wilson line prescription. It will be interesting to extend this Chern-Simons formulation to provide holographic constructions for mixed state entanglement and correlation measures such as the entanglement negativity, the reflected entropy and the entanglement wedge. We hope to return to these interesting issues in the near future.

\section{Acknowledgement}
GS would like to thank Koushik Ray and the Indian Association for the Cultivation of Science (IACS), Kolkata, India, 
where part of this work was completed, for their warm hospitality and a stimulating research environment.

\appendix
\section{Derivation of holographic entanglement negativity in TMG-AdS$_3$/CFT$_2^a$} \label{sec:appendix_A}
In this appendix, we provide a heuristic gravitational path integral derivation of the holographic construction for the entanglement negativity from \cref{sec:EN_geod}. For brevity, we focus on the mixed state configuration of two adjacent intervals $A$ and $B$ in the CFT$_2^a$. To begin with, we note that the entanglement negativity\footnote{The entanglement negativity was called as the logarithmic negativity in \cite{Dong:2021clv}. Here we stick with the more common nomenclature in the literature to avoid any confusion.} $\mathcal{E}$ for a bipartite mixed state $\rho_{AB}$, may be obtained from a replica technique as an even analytic continuation of the R\'enyi  generalization of the entanglement negativity $\mathcal{N}^{(k)}$ in the following way \cite{Calabrese:2012ew,Calabrese:2012nk,Dong:2021clv}:
\begin{align}
\mathcal{E}(A:B)=\lim_{ n \to 1/2 } \log\mathcal{N}^{(2n)}\left(\rho_{AB}\right)
\end{align} 
The R\'enyi entanglement negativity of order $2n$ may be computed as the properly normalized partition function on the corresponding replica manifold. The replica manifold $\mathcal{B}_{2n}^{A,B}$ is constructed as the $2n$-fold branched cover of the original boundary manifold $\mathcal{B}_1$, where the individual copies are sewed cyclically along $A$ and anti-cyclically along $B$ \cite{Calabrese:2012ew,Calabrese:2012nk}. For a dual CFT$_2^a$, utilizing the holographic duality, the replica partition functions may be calculated in terms of the on-shell action of the bulk replica  TMG-AdS$_3$ geometry, denoted as $\mathcal{M}_{2n}$. Note that the asymptotic boundary of $\mathcal{M}_{2n}$ constitutes the boundary replica manifold $\mathcal{B}_{2n}^{A,B}$. 
Following \cite{Dong:2021clv}, the appropriate bulk saddle-point geometry may be found through the so-called replica symmetry breaking mechanism, where in the bulk one breaks the replica symmetry partially while respecting the full replica symmetry in the boundary. In the present case of topologically massive gravity in asymptotically AdS$_3$ spacetimes, we construct the replica non-symmetric saddle $\mathcal{M}_{2n}^{A,B\,(\text{nsym})}$ by extending the cutting and gluing procedure as described in \cite{Dong:2021clv,KumarBasak:2020ams}, which breaks the replica symmetry group of $\mathbb{Z}_{2n}$ to that of $\mathbb{Z}_{n}$ in the bulk. By employing the holographic duality in TMG-AdS$_3$/CFT$_2^a$, the replica partition function on the boundary manifold is obtained from the bulk saddle-point geometry as
\begin{align}
\mathbf{Z}[\mathcal{B}_{2n}^{A,B}]
=e^{-I_{\text{grav}}[\mathcal{M}_{2n}^{A,B\,(\text{nsym})}]}\,,
\end{align}
where $I_{\text{grav}}[\mathcal{M}_{2n}^{A,B\,(\text{nsym})}]$ is the on-shell action of the replica non-symmetric saddle $\mathcal{M}_{2n}^{A,B\,(\text{nsym})}$. Therefore, the R\'enyi entanglement negativity of order $2n$ is given by
\begin{align}
\mathcal{N}^{(2n)}(A:B)&=\frac{\mathbf{Z}[\mathcal{M}_{2n}^{A,B}]}{\left(\mathbf{Z}[\mathcal{M}_1]\right)^{2n}}\nonumber\\
&=e^{-I_{\text{grav}}[\mathcal{M}_{2n}^{A,B\,(\text{nsym})}]+2n\,I_{\text{grav}}[\mathcal{M}_1]}\,,
\end{align}
where $I_{\text{grav}}[\mathcal{M}_1]$ is the on-shell action of the bulk asymptotically AdS$_3$ geometry dual to the original CFT$_2^a$ manifold.
Next, we consider the quotient geometry $\hat{\mathcal{M}}_{2n}^{A,B\,(\text{nsym})}$ in the bulk by quotienting through the remnant $\mathbb{Z}_{n}$ symmetry  
\begin{align}
\hat{\mathcal{M}}_{2n}^{A,B\,(\text{nsym})}=\mathcal{M}_{2n}^{A,B\,(\text{nsym})}/\mathbb{Z}_n\,.
\end{align}
The quotient manifold has conical defects  $\Gamma_{A_1}^{(n)}$ and $\Gamma_{B_2}^{(n)}$, at the loci of the fixed points of the residual replica symmetry \cite{Dong:2021clv} with conical deficit angles 
\begin{align*}
\Delta\phi_n=2\pi\left(1-\frac{1}{n}\right)\,.
\end{align*}
The on-shell action of the bulk replica manifold may now be obtained from that of the quotient bulk as
\begin{align*}
I_{\text{grav}}[\mathcal{M}_{2n}^{A,B\,(\text{nsym})}]\equiv n\,I_{\text{grav}}\left(\mathcal{M}_{2}^{AB},\Gamma_{A_1}^{(n)},\Gamma_{B_2}^{(n)}\right)\,,
\end{align*}
and therefore, the R\'enyi negativity between subsystems $A$ and $B$ is given by
\begin{align}
 \log\,\mathcal{N}^{(2n)}(A:B)
&=-n\left[I_{\text{grav}}\left(\mathcal{M}_{2}^{AB},\Gamma_{A_1}^{(n)},\Gamma_{B_2}^{(n)}\right)-2\,I_{\text{grav}}[\mathcal{M}_1]\right]\,,\label{Renyi_Neg_gen}
\end{align}
To compute the on-shell action of the quotient bulk geometry $I_{\text{grav}}\left(\mathcal{M}_{2}^{AB},\Gamma_{A_1}^{(n)},\Gamma_{B_2}^{(n)}\right)$, we need to consider the contributions coming from the codimension-2 cosmic branes homologous to $A$ and $B$, which are situated at $\Gamma_{A_1}^{(n)}$ and $\Gamma_{B_2}^{(n)}$. As described in \cref{sec:EN_geod}, in the case of topologically massive gravity in the asymptotically AdS$_3$ bulk, we have massive spinning probe particles propagating along these backreacting cosmic branes. We can comprehensively determine the on-shell action of the quotient bulk in terms of the effective on-shell actions of the massive spinning particles on these cosmic branes as
\begin{align}
I_{\text{grav}}\left(\mathcal{M}_{2}^{AB},\Gamma_{A_1}^{(n)},\Gamma_{B_2}^{(n)}\right)=2\,I_{\text{grav}}[\mathcal{M}_1]&+\frac{\mathcal{L}^{(1/2)}(\Gamma_{AB})}{4G}\notag\\&+\left(1-\frac{1}{n}\right)\frac{\mathcal{L}^{(n)}(\Gamma_{A})+\mathcal{L}^{(n)}(\Gamma_{B})}{4G}\,,\label{QuotientAction}
\end{align}
where $\mathcal{L}^{(n)}(\Gamma_{X})$ is related to the length $L^{(n)}\left(\Gamma_{X}\right)$ and the twist $\mathcal{T}^{(n)}\left(\Gamma_{X}\right)$ of the backreacted codimension-2 cosmic brane homologous to subsystem $X$ as
\begin{align}
n^{2} \frac{\partial}{\partial n}\left(\frac{n-1}{n} \mathcal{L}^{(n)}\left(\Gamma_{X}\right)\right)&=L^{(n)}\left(\Gamma_{X}\right)+\frac{\mathcal{T}^{(n)}\left(\Gamma_{X}\right)}{\mu}\label{AreaCB1}\,.
\end{align}
Therefore, utilizing \cref{Renyi_Neg_gen,QuotientAction} we obtain for the R\'enyi entanglement negativity as
\begin{align}
\log\,\mathcal{N}^{(2n)}(A:B)=-n\frac{\mathcal{L}^{(1/2)}(\Gamma_{AB})}{4G}-(n-1)\frac{\mathcal{L}^{(n)}(\Gamma_{A})+\mathcal{L}^{(n)}(\Gamma_{B})}{4G}\,.
\end{align}
Now taking the $n\to 1/2$ limit, the entanglement negativity between $A$ and $B$ is given by the R\'enyi mutual information of order half as
\begin{align}
\mathcal{E}(A:B)&=\frac{\mathcal{L}^{(1/2)}(\Gamma_{A})+\mathcal{L}^{(1/2)}(\Gamma_{B})-\mathcal{L}^{(1/2)}(\Gamma_{AB})}{8G}\equiv \frac{1}{2}\mathcal{I}^{(1/2)}(A:B)\,.
\label{Neg_gen}
\end{align}
Finally, we use the fact that in the framework of TMG-AdS$_3$/CFT$_2^a$ the effects of the backreaction can be conveniently absorbed into the multiplicative factor $\mathcal{X}_2=\frac{3}{2}$ \cite{Hung:2011nu,Rangamani:2014ywa,Kudler-Flam:2018qjo,Kusuki:2019zsp} and therefore
\begin{align}
\mathcal{L}^{(1/2)}(\Gamma_{A})=\mathcal{X}_2\,\mathcal{L}(\Gamma_{A})=\frac{3}{2}\left(L(\Gamma_{A})+\frac{\mathcal{T}(\Gamma_{A})}{\mu}\right)\,,
\end{align}
leading to our holographic proposal for the entanglement negativity for two adjacent intervals $A$ and $B$, given in \cref{EN_adj_conjecture}. The holographic construction for the entanglement negativity for the other bipartite states corresponding to two disjoint intervals and a single interval may also be obtained from a gravitational replica construction employing the replica non-symmetric saddle in a similar fashion, as described in \cite{KumarBasak:2020ams}.

Finally we note that it will be interesting to explore the holographic duality between the reflected entropy and the minimal EWCS in the framework of TMG-AdS$_3$/CFT$_2^a$ from a gravitational path integral perspective, similar to that in \cite{Dutta:2019gen,Akers:2022max}. 
\newpage

\begin{center}
	\Large\textbf{Erratum}
\end{center}
In the computation of the Chern-Simons (CS) contribution to the entanglement wedge cross-section (EWCS) computed in subsections \ref{sec:EW_disj_boosted} and \ref{sec:EW_adj_boosted}, we found two compensating errors. These are listed below:
\begin{enumerate}
	\item The proposal for the CS contribution to the EWCS in \cref{EW_CS} involves the extremization over the boost $\mathcal{T}$ required to transport an auxiliary orthonormal frame through the length of the EWCS $\Sigma_{AB}$ as follows
	\begin{align}
	E_W^{\textrm{CS}}&=\underset{\Gamma_{AB}^{(A)}\subset \Gamma_{AB}}{\text{min\,\,ext}} \, \left[\frac{\mathcal{T}\left(\Sigma_{AB}\right)}{4\mu G_N}\right]=\frac{1}{4\mu G_N}\log\left[\frac{q_f\cdot n_f-\tilde{q}_f\cdot n_f}{q_i\cdot n_i-\tilde{q}_i\cdot n_i}\right]\,\tag{1},\label{EW_CS2}
	\end{align}
	where $n$ is the auxiliary timelike normal vector, $(q,\tilde{q})$ are the parallel transported normal vectors along the EWCS $\Sigma_{AB}$ and the subscripts $i$ and $f$ denote the endpoints of the EWCS. 
	
	In the calculations, instead of the normal vectors to the EWCS, we incorrectly made use of the normal vectors to the RT surfaces, given in \cref{Normal_n}. The correct normal vector to the EWCS is given in \cref{n_m} of this erratum.
	
	\item Furthermore, there was a typographical error in \cref{q_boosted}, which was carried forward in the computations of the EWCS in \cref{sec:EWCS}.  	
\end{enumerate}

These two errors compensated each other to give the final result for the EWCS in the correct form. These mistakes were overlooked by the authors in the original article as the final result matched with the field theoretic replica computations for the dual reflected entropy. 

The vector $n$ normal to the EWCS may be obtained by noting that the tangent to the (Ryu-Takayanagi) RT surface at the position $\rho_m$ serves as the other normal $\tilde{n}$ to the EWCS\footnote{We thank Prof. Qiang Wen for pointing out this crucial issue.}\cite{Wen:2022jxr}. We have obtained the correct normal vectors to the EWCS through the procedure outlined in \cite{Wen:2022jxr}. After incorporating the proper normal vectors to the EWCS and fixing the typographical errors, we found that the final expressions for the CS contribution to the EWCS remain unchanged. The relevant corrections are listed below:
\begin{itemize}
	\item In subsection \ref{sec:HEE}, the second line in \cref{q_boosted} should be modified to
	\begin{align} 
	\tilde{q} = - \frac{1}{\sqrt{2\rho u^2+\frac{\Delta u}{\Delta v}}}\Big(u\sqrt{\frac{\Delta u}{\Delta v}} \partial_u + u\sqrt{\frac{\Delta v}{\Delta u}} \partial_v - 2\rho \sqrt{\frac{\Delta u}{\Delta v}} \partial_\rho\Big)\,.\tag{2} \label{qt}
	\end{align}

	\item In subsection \ref{sec:EW_disj_boosted}, \cref{Wedge-disj} should be modified as follows
	\begin{figure}[H]
		\centering
		\includegraphics[scale=.7 ]{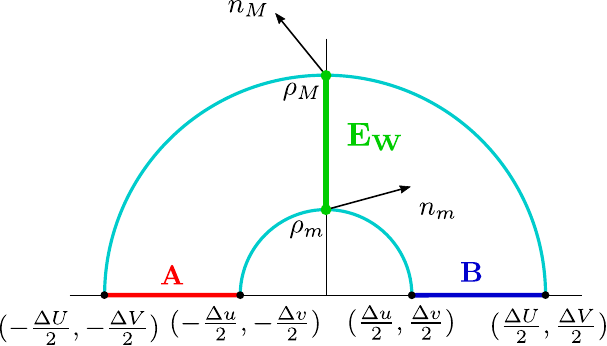}
		\caption{Updated version of \cref{Wedge-disj}, properly depicting the schematics of the vector $n$ normal to the EWCS.}
		\label{fig:disj-corr}
	\end{figure}
	
	\item In subsection \ref{sec:EW_disj_boosted}, the sentence after \cref{EW_Poincare_CS} should be modified to: ``The values of the parallel transported vectors normal to the EWCS $\Sigma_{AB}$ at the turning point of a geodesic line connecting the boundary points  $\left(-\frac{\Delta u}{2},-\frac{\Delta v}{2},\infty\right)$ and $\left(\frac{\Delta u}{2},\frac{\Delta v}{2},\infty\right)$ may be expressed as"
	
	\item Eq. \eqref{qm} in subsection \ref{sec:EW_disj_boosted} should be modified to
	\begin{align}
	& q (\tau_m)=\frac{1}{\sqrt{2\rho}}\left(\partial_u+\partial_v\right)\,~~,~~\tilde q (\tau_m)=\frac{1}{\sqrt{2\rho}}\left(\partial_u-\partial_v\right)\,.\label{q_m}\tag{3}
	\end{align}
	
	\item After \cref{qm}, the following should be added:  The auxiliary normal vector $n_m$ at the endpoint of the EWCS $\Sigma_{AB}$, labelled as $\rho_m$ in \cref{Wedge-disj}, is given by \cite{Wen:2022jxr} 
	\begin{align}
	n_m=\frac{1}{\sqrt{2\rho_m}}\left(-\sqrt{\frac{\Delta u}{\Delta v}} \partial_u 
	+ \sqrt{\frac{\Delta v}{\Delta u}} \partial_v\right)\,.\label{n_m}\tag{4}
	\end{align}
	
	\item Eq. \eqref{EW-CS-disj-boosted} in subsection \ref{sec:EW_disj_boosted} should be modified to
	\begin{align}
	E_W^{\textrm{CS}}&=\frac{1}{4\mu G_N}\log\left(\sqrt{\frac{\Delta v/\Delta u}{\Delta V/\Delta U}}\right)=\frac{1}{2\mu G_N}\left(\kappa-K\right)=\frac{c_L-c_R}{3}\left(K-\kappa\right)\,.\tag{5}
	\end{align}

	\item In subsection \ref{sec:EW_disj_boosted}, the coefficient of the second term in the first equality of \cref{EW_crossratio} should be modified to $\frac{1}{2 \mu G_N}$.
\end{itemize}
Similar to the corrections described above for the disjoint intervals, the case of adjacent intervals described in subsection \ref{sec:EW_adj_boosted} need the following modifications:
\begin{itemize}
	\item In subsection \ref{sec:EW_adj_boosted}, \cref{q0} should be modified to
	\begin{align}
	&	q (\tau_{\infty})=\frac{1}{\sqrt{2\rho_{\infty}}}\left(\partial_u+\partial_v\right)\,~~,~~n_0=\tilde q (\tau_{\infty})=\frac{1}{\sqrt{2\rho_{\infty}}}\left(\partial_u-\partial_v\right)\,.\label{qm2}\tag{6}
	\end{align}
	
	\item In subsection \ref{sec:EW_adj_boosted}, the coefficient of the logarithm in the first equality of \cref{EW_adj} should be modified to $\frac{1}{2 \mu G_N}$.
	
	\item In subsection \ref{sec:EW_adj_boosted}, the coefficient of the second term of \cref{EW-adj-crossratio} should be modified to $\frac{1}{2 \mu G_N}$.
\end{itemize}

The authors would like to thank Prof. Qiang Wen and Haocheng Zhong for highlighting these crucial issues and suggesting a resolution.


\bibliographystyle{utphys}
\bibliography{reference}

\end{document}